\documentclass[a4paper,12pt,twoside]{report}

\usepackage[margin=25mm]{geometry}
\usepackage{graphicx}
\usepackage{times,cite,url,fancyhdr}
\usepackage{hyperref}
\usepackage[co411,draft]{pera}
\usepackage[caption=false,font=footnotesize]{subfig}

\begin{document}

\title{Power Analysis Based Side Channel Attack}

\author{Hasindu Gamaarachchi \\ Harsha Ganegoda}
\date{\today}

\maketitle
\urlstyle{sf}
\thispagestyle{empty}

\pagenumbering{roman} 
\begin{abstract}
  \textit{Side channel attacks} break the secret key of a cryptosystem using channels such as sound, 
  heat, time and power consumption which are originally not intended to leak such information.
  \textit{Power analysis} is a branch of side channel attacks where power consumption data
  is used as the side channel to attack the system. First using a device like an oscilloscope power traces are collected when the cryptographic device is doing the cryptographic operation. Then those traces are statistically analysesd using methods such as Correlation Power Analysis (CPA) to derive the secret key of the system. Being possible to break Advanced Encryption Standard (AES) in few minutes, 
  power analysis attacks have become a serious security issue
  for cryptographic devices such as smart card.
  
  As the first phase of our project, we build a testbed for doing research on power analysis attacks. Since power analysis is a practical type of attack in order to do any research, a testbed is the first requirement. 
  Our testbed includes a PIC microcontroller based
  cryptographic device, power measuring circuits and a digital oscilloscope in the hardware side. In software side it includes algorithms running on the microcontroller, oscilloscope automation scripts and analysis programs. 
  We verify the functionality of the testbed by attacking AES in time less than 10 minutes.
  Since building a test bed is a complicated process, having a pre-built testbed would save the time of future researchers.

  The second phase of our project is to attack the latest cryptographic algorithm called \textit{Speck} which has been released by National Security Agency (NSA)
  for use in embedded systems. So far, Speck has not been attacked using power analysis. In spite it has lot of differences to AES making impossible to  directly use the power analysis approach used for AES, we introduce novel approaches to break Speck in less than an hour. Therefore, We practically show that even though the algorithm is very new still
  it is vulnerable to power analysis.

  The third phase of the project is to work on countermeasures. After getting familiar with the current state of art,
  we select few already introduced countermeasures and practically attack them on our testbed to do a comparative analysis.
  Meanwhile, we try to form our own countermeasures and to improve existing countermeasures.
  Under circuit based countermeasures, the existing idea of implementing power line filters is practically implemented and tested. We show that it is not safe enough.
  We try few of our own circuit based ideas as well, to evaluate how good they are as countermeasures. But unfortunately none of them are good enough.
  Under software based countermeasures existing methods called random instruction injection and randomly shuffling Sboxes are implemented and tested. We show that those countermeasures are good enough for their simplicity and cost. But we identify the possible threat due to the problem of generating a good seed for the pseudo random algorithm running on the microcontroller. We address this issue by using a hardware based  true random generator that amplifies a random electrical signal and samples to generate a proper seed.

  \vfill
\end{abstract}

  \tableofcontents
  \listoffigures
  \listoftables

  \chapter{\label{c:intro}Introduction}
  Encryption is a process where a message is encoded such that it is only understandable by the intended parties. 
  The input message known as the plain text is transformed to an output text known as cipher text using an encryption 
  algorithm based on a parameter called the key. The cipher text must be again converted back to plain text is order to 
  be read and this process known as decryption can be only done by a party who knows the secret key. 
  Various encryption algorithms exist out there while AES (Advanced Encryption Standard), DES (Data Encryption Standard) 
  and blowfish are some common examples.
 
\pagenumbering{arabic} 
\setcounter{page}{1}

  \section{\label{s:cryptanalysis}Cryptanalysis}

  A cryptosystem is a pair of algorithms including the encryption algorithm and the decryption algorithm that does 
  encryption and decryption based on a key. As the security of such a system fully depends on the secrecy of the secret key, 
  if somehow the key is compromised then any unauthorized party will be able to decrypt the cipher text. 
  Analysing a cryptosystem to find a weakness that would leak the secret key is called cryptanalysis \cite{forouzan}. 
  Currently various attacks exist which can be used to derive the key of a cryptosystem in an unauthorized fashion. 

  The most trivial of such attacks is the brute force attack. In a brute force attack all the possibilities for 
  the key are generated and are tried one by one until it matches the correct key. Brute force attacks fall 
  into the category of traditional attacks. As modern encryption algorithms consist of very large keys as large as
  128 bits or 256 bits, the number of permutations is huge that this type of attack is not practically feasible due to 
  the huge computational time required. On the other hand another category of attacks called modern attacks exist which use 
  mathematical approaches. Differential cryptanalysis and linear cryptanalysis are such methods and these methods are more 
  effective than traditional attacks.

  \section{\label{s:side_channel_attacks}Side channel attacks}
  In addition to the two types of attacks described in \autoref{s:cryptanalysis}, now a very new type of attacks called side channel
  attacks have become popular \cite{Prouff201408}. These attacks use the weakness in the physical implementation of the cryptosystem to derive the secret key. 
  For example a system can give out sound, heat, light or other electromagnetic radiation when under operation. 
  These usually can have a relationship with what is happening inside a system and an attacker can make use of those 
  things to determine the secrets inside a system. Such channels that leak out information are called side channel attacks. 
  If these side channels are used to attack a cryptosystem then they are called side channel attacks.

   \begin{figure}[htb]
    \begin{center}
      \includegraphics[width=10cm]{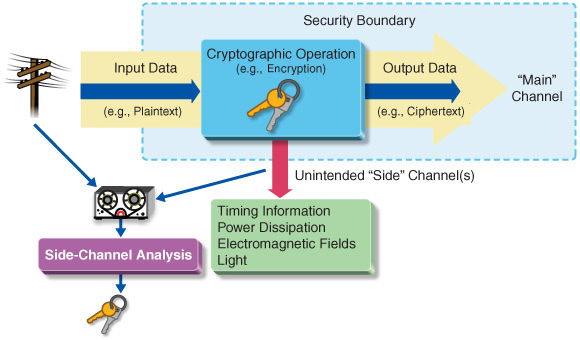}
    \end{center}
    \caption{\label{f:sidechannel}Side channel attack \cite{sidechannelpic}}
  \end{figure}

  Consider \autoref{f:sidechannel}. Plain text is given as input to the system and the system runs the encryption to output cipher text. 
  The key used in the system is unknown to the outside. But the system leaks out information using various side channels 
  unintendedly. These side channel data are collected and they are analysed together with the input data to derive the secret key. 
  This same process can be done in an decryption operation as well where now the input would be cipher text and output 
  would be plain text.  
  
  \section{\label{s:Power_analysis}Power analysis}

  Power analysis is a branch of side channel attacks where the side channel used is the power consumption. 
  In electronic devices, the instantaneous power consumption is dependent on the data that is being processed 
  in the device as well as the operation performed by that device\cite[pp.~6-12]{mangard}.  Therefore by analysing the power consumed by a 
  device when it is doing encryption or decryption the key can be deduced. 
 
  \begin{figure}[htb]
    \begin{center}
      \includegraphics[width=8cm]{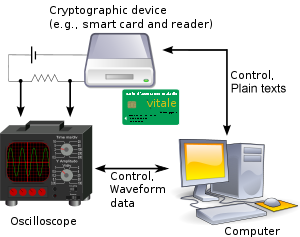}
    \end{center}
    \caption{\label{f:poweranalysis}Power analysis attack  \cite{poweranalysispic}}
  \end{figure} 

  Consider \autoref{f:poweranalysis}. The computer inputs a set of known plain texts to the cryptographic device 
  which does the encryption. While the device performs the encryption the oscilloscope measures the power consumptions. 
  For several hundreds of plain text samples power traces  are obtained and then they are analysed on a computer 
  using an algorithm such as Simple Power Analysis (SPA), Differential Power Analysis (DPA) \cite{kocher} or Correlation Power Analysis (CPA) \cite{cpabrier}
  to derive the secret key of the system. As the plain text in this case is known to the attacker this is a known plain text attacks.
 
 \section{\label{s:CMOS}Power consumption of CMOS circuits} 

  CMOS circuits are made up of pairs of p type and n type MOSFET (Metal Oxide Semiconductor Field Effect Transistors). 
  Microprocessors, microcontrollers, static RAM and many other digital logic circuits in the modern world use CMOS 
  (Complementary Metal Oxide Semiconductor) technology. Therefore for power analysis which involves power consumption 
  of a cryptographic device, focusing on power consumption of CMOS circuits is important. 
 
  Instantaneous power consumption of a circuit is given by the following equation. Here $p$ is the instantaneous power, 
  $v$ is the instantaneous voltage and $i$ is the instantaneous current.

  \[
   p = vi
  \]

  As the voltage given to a circuit is equal to the power supply voltage and because it is always constant, 
  the above equation can be now written as follows.  Here $V_{DD}$ is the voltage of the power supply. Therefore by measuring the instantaneous current, 
  the instantaneous power consumption can be deduced.
  
  \[
   p = V_{DD}i
  \]

  The power consumption of a CMOS circuits is made up of two components as static power and dynamic power. 
  Static power is due to the leakage current of transistors and therefore it is dependent 
  on the design of the circuit. On the other hand, dynamic power is due to the switching of 
  transistors and therefore it is dependent on the data being processed and the operation being done. 
  As power analysis utilizes the relationship between power consumption and the data being processed, 
  dynamic power is the relevant one. As the static power is mostly constant, the variation
  in the total power is solely due to dynamic power and therefore the total power consumption can be directly 
  used for an attack.   
 
  \begin{figure}[htb]
    \begin{center}
      \includegraphics[width=10cm]{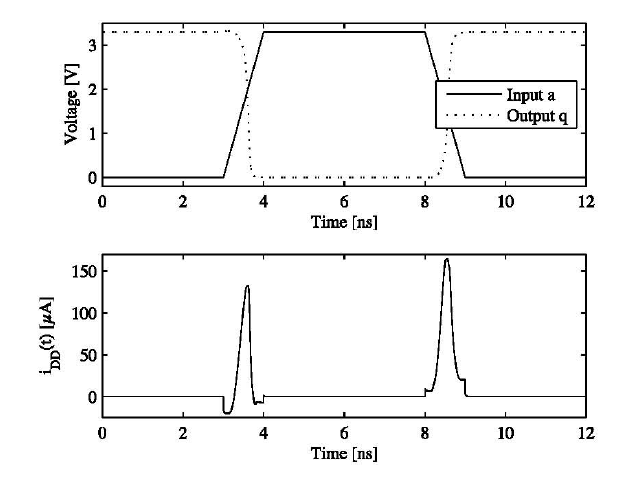}
    \end{center}
    \caption{\label{f:cmos}Dynamic power consumption during switching in a CMOS inverter\cite[p.~32]{mangard}}
  \end{figure} 
  
  Dynamic power is consumed when switching occurs in transistors. 
  That is if a CMOS cell changes from 0 to 1 or from 1 to 0, switching occurs in transistors 
  and power is consumed. \autoref{f:cmos} is for a CMOS inverter. Upper graph there shows the input signal 
  and the output signal of the CMOS inverter while the below graph shows the instantaneous current. 
  As you can see there are two transitions in the input signal where at first it goes from 0 to 1 and 
  then from 1 to 0. At these moments where switching takes place peaks are visible in the instantaneous current drawn 
  by the circuit.    

  These clear changes in the current is due to the current drawn to charge the
  capacitance in the transistors and the temporary  short circuit current that is drawn during
  switching. In devices such as a microcontrollers, the data bus which is long and connected to 
  many components have a quite big capacitance. Therefore writing to the data bus during a memory 
  read or write causes a significant power consumption. When attacking software based encryption using 
  power analysis this is the point which is mostly made use of.
  
  \section{\label{s:powermodel}Power model}
  
  A power model is used to deduce the power consumption of a circuit mathematically using the information 
  we have about the circuit. Hamming distance model is a very simple such power model. Even though it 
  is not the most accurate, for our cause it is quite suitable. 

  In the data bus explained in \autoref{s:CMOS} if the initial value on the bus was $v0$ and 
  the value after the change is $v1$ then the hamming distance can be simply found by counting the 
  number of 0 to 1 and 1 to 0 transitions that happen. If the data bus is four bits and if $v0=1010$ and $v1=0011$ 
  only two bits have changed (first and the last) and therefore the hamming distance is equal to 2. 
  
  The hamming distance between two values $v0$ and $v1$ can be calculated by the following equation where $HD$ 
  stands for Hamming Distance and $HW$ stands for Hamming Weight. Hamming weight is the number of bits 
  which are equal to 1.

  \[
   HD(v0,v1) = HW(v0 \ XOR \  v1)
  \]

  As the significant power consumption in the CMOS circuit occurs during transitions, this model 
  works fine.  Most microcontrollers usually set the data bus to 0 or 1 before writing a value to it.
  In a data bus where all the bits are set to 0 first, hamming distance is just the hamming weight
  of the value written to the bus afterwards( $HW(v \ XOR \ 0) = HW(v)$ ). If it is a data bus where all bits
  are initially set to 1 then the hamming distance is equal to hamming weight of the value written 
  subtracted by the number of bits in the bus. ( $HW(v \ XOR \ 1)=Num\_bits-HW(v)$ ). Therefore for most 
  microcontrollers hamming weight can be directly used as the power model.
 
 \section{\label{s:cpa}Correlation Power Analysis (CPA)}
  Correlation Power Analysis (CPA) as mentioned above is an algorithm used to do power analysis. 
  It is a statistical type of attack and uses the Pearson correlation coefficient to correlate data. 
  CPA is more recent and when compared to other algorithms such as DPA has numerous advantages such as 
  requirement of less number of power traces\cite{cpabrier}.
  
  As the first step of the attack, an intermediate value generated during the cryptographic operation 
  (encryption or decryption) which is a function of a variable data value and part of the key 
  must be selected. This function is known as the selection function. In the equation below $I$ is the intermediate
  value. The variable data sample which is generally plain text or cipher text is shown by $d$. $k$ is part of the key
  which will be called as the subkey after this point.

  \[
   I = f(d,k)
  \]

 Then power measurements must be done when the device is doing the cryptographic operation. 
 Here power traces must be obtained for several hundred (or thousands depending on the system)
 of samples for $d$. These obtained traces are the real power consumption values. Imagine that we 
 used $N$ number of plain text samples to get $N$ number of power traces. 
 Each power trace would have $M$ number of sampling points which correspond to 
 power consumption at each sampled moment in time.
 
 After that, the intermediate values for each of the variable data samples used and 
 the power consumption for those intermediate values must be calculated using a power
 model such as hamming distance model. This calculation must be repeated for all possible values 
 for the subkey considered. These calculated data are the hypothetical power consumption values.
 
  Finally the real power consumption values and the hypothetical power 
  consumption values per each subkey must be compared to each other to find out which subkey has the 
  highest correlation. This comparison is done using the statistical method called Pearson correlation 
  coefficient given by the equation below\cite{cpabrier}. In the equation, the $j^{th}$ sample point of the $i^{th}$ power trace has 
  been written as $W_{i,j}$. The hypothetical power consumption value for the $i^{th}$ plain text (or cipher text) 
  with respect to the appropriate subkey has been written as $H_{i}$.  An estimate for the correlation for $j^{th}$ 
  sampling point for a certain subkey can be found using the equation. 
  
  \[
     \hat\rho = \frac{ N\sum_{i=0}^{N} W_{i,j}H_{i} - \sum_{i=0}^{N} W_{i,j}\sum_{i=0}^{N} H_{i} }
       { \sqrt{N\sum_{i=0}^{N} W_{i,j}^{2} - (\sum_{i=0}^{N} W_{i,j})^{2}} \; \sqrt{N\sum_{i=0}^{N} H_{i}^{2} - (\sum_{i=0}^{N} H_{i})^{2}} }
  \]	

 Pearson correlation coefficient is a value between -1 and 1. If the value is 1 that means the 
 two data sets compared (hypothetical power consumption values and real power consumption values)
 have the best correlation. If 0 that means there is no correlation at all. -1 
 means they are correlated at best but the relationship is such that the values in the data sets are 
 inversely proportional. As the two data sets used here to find the correlation coefficient is the 
 hypothetical power consumption data and the real power consumption data, the correlation coefficient 
 depicts how much they are related. Therefore the maximum correlation coefficient would be obtained when 
 the hypothetical power consumption values for the samples are calculated using the correct key.
 
 The fact that the analysis is done separately on separate subkeys is the one that reduces the complexity
 of a correlation power analysis attack to a practically feasible value. 
 For example consider a 128 bit key. A 128 bit key has 2\textsuperscript{128} possibilities
 and checking such number of possibilities is not feasible even on a super computer. This is why a brute force approach
 is not feasible for most encryption algorithms out there today.
 Now consider a CPA attack where a subkey is size of a byte.
 Here the 128 bit key is divided into 8 bit portions to make 16 bytes.
 A byte which is 8 bits just have 256 possibilities and when all 16 keybytes are considered
 only 256x16 possibilities must be computed. This number of computations
  is practically feasible.
 
 \section[Countermeasures]{\label{s:countermeasures_intro}Countermeasures against power analysis attacks}
 
 Countermeasures are the techniques that are implemented with the ultimate goal of preventing power analysis attacks.
 But unfortunately in the security world there are no such hundred percent perfect countermeasures. Therefore the objective of a 
 countermeasure against power analysis is to make an attack extremely difficult. When the time and the cost 
 for an attack is large enough such that it would be infeasible for an attacker, such a countermeasure can be considered good enough.
 
  \begin{figure}[htb]
  	\begin{center}
  		\includegraphics[width=8cm]{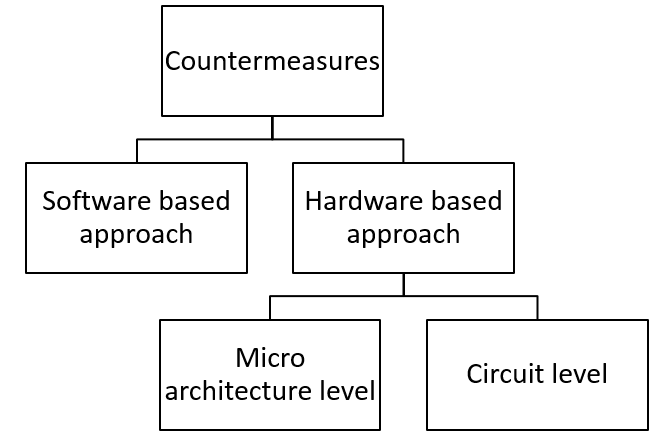}
  	\end{center}
  	\caption{\label{f:counter}Classification of countermeasures}
  \end{figure}  
 
 Countermeasures against power analysis attacks can be categorized in to two broad categories as hardware based countermeasures
 and software based countermeasures as shown in \autoref{f:counter}. Hardware based countermeasures as the names suggests is the addition of new components or modification of
 existing components of the cryptosystem, in order to minimize the leakage of secret data as power. Hardware countermeasures
 can be further classified into two categories namely circuit level countermeasures and micro architecture level countermeasures as shown in \autoref{f:counter}.
 Circuit level countermeasures involve addition of components to the circuit or modification of the circuit at a macro level. For example adding a 
 power line filter to the cryptographic device or connecting a noise source are circuit level countermeasures. Micro architecture level countermeasures involve addition of components or modification of components at logic gate level. For example, adding additional transistors
 to a CMOS gate such that power consumption becomes different, is such micro architecture level countermeasure.
 Software based countermeasures on the other hand do modifications to the program code  that runs on the microcontroller of the cryptographic device. 
 For example inserting some false instructions randomly at the middle of the encryption to change the power consumption pattern is a software based countermeasure.
 
 Due to the massive threat imposed by power analysis attacks, a lot of research is already conducted to form various countermeasures.
 Currently there are many such introduced countermeasures and are elaborated in \autoref{c:relatedwork}.
 
 \section{\label{s:project}The project and its objectives}
 Embedded cryptographic devices such as smart cards are very much vulnerable to power analysis attacks\cite[pp.~1-5]{mangard}.  
 For example, the key of certain smart cards can be obtained in a small amount of time as 90 minutes by devising 
 a power analysis attack\cite{lowcosteducation}. Embedded devices such as smart cards are widely used for authentication 
 mechanisms. Further even ATM (Automated Teller Machine) cards and SIM (Subscriber Identity Module) cards are also smart cards.
 Therefore as this type of attacks are a serious risk for the security, researchers are proposing 
 various countermeasures.
 
 To find countermeasures or do any other power analysis based research on the first place a 
 testbed is needed. That is a system comprising of a cryptosystem, power measurement 
 system and necessary software to do the analysis is needed. Building such a system involves a 
 difficult process which is time consuming. While it needs a quite good knowledge on electronics, 
 computer interfacing, mathematics and programming knowledge, the process needs spending lot of time 
 to do troubleshooting. Anyone who is interested in power analysis going through this same process 
 is a wastage of time. But if a testbed is already available one can directly start on countermeasures or similar 
 research rather than going through the troubles of building the testbed. Moreover 
 if such an already built system is available anyone who has a little electronic knowledge also can use it. 
 
 The first phase of our project is to build such a testbed for power analysis based side channel attacks. 
 As described before, the objective is to help researchers who are interested on countermeasures to save them 
 from the testbed setup time and difficultly. And also we need this testbed to continue the next phases of 
 the projects for this semester as well as the next semester. The cryptographic device of the testbed 
 is built using a reprogrammable microcontroller such as PIC or AVR. Therefore this device does software based 
 encryption opposed to hardware based encryption which in that case uses direct logic circuits or FPGA 
 (Field Programmable Gate Array). Then we attack AES using the created testbed to check whether our testbed works properly. 

 The second phase is to do cryptanalysis on an encryption algorithm called Speck. 
 Speck is a very recent encryption algorithm that was released by the NSA in June 2013. 
 This belongs to the type of ciphers called block ciphers which AES also belongs, but the
 speciality is this is a very lightweight cipher targeted mainly for embedded systems. Therefore it 
 is expected to be famous among embedded devices in the future and therefore it must be tested to 
 check whether this algorithm is vulnerable to power analysis to determine the necessity of countermeasures. 
 Up to now according to best of our knowledge there is no such power analysis based cryptanalysis done on Speck.
 Therefore in the second phases we check whether Speck is vulnerable to power analysis attacks. This phase 
 also involves implementation of Speck for 8 bit microcontrollers, due to the fact that 
 such an implementation currently is not available. 
 
 Currently there is a lot of available countermeasures against power analysis as we explain in \autoref{c:relatedwork}. But 
 at one point any of them would fail. Performance and powerfulness of computers as well as devices such as digital
 oscilloscopes are rapidly increasing and therefore the effectiveness of old countermeasures decrease. A countermeasure which appears
 to take infeasible amount of time to be attacked today might become feasible in the future with the improvement of technology.
 Further, the effectiveness of a countermeasure would compromise factors such as power, performance, device size and cost. For example adding new 
 hardware components will increase power consumption and the device size, which is not preferred. Integrating a software countermeasure would reduce the performance and efficiency.
 Meanwhile any countermeasure would definitely increase the cost of a cryptographic device. Due to all these reasons countermeasures against power analysis is still a very open research area. Therefore the third  phase of our project is to work on countermeasures against power analysis. First we study the current state of the art of hardware based countermeasures. Then we
 would practically test some selected circuit based countermeasures on the testbed we have created
 and evaluate their effectiveness. Here we limit testing of hardware based countermeasures only to circuit based countermeasures
 as micro architecture level ones would need access to advanced facilities such as fabrication. 
 Next we study the current state of art of software countermeasures and then some selected software
 countermeasures would be tested and evaluated. 
 Finally we improve a selected countermeasure with the goal of contributing towards the security of embedded
 devices. Meanwhile we would also test some new ideas to verify whether they are good enough as countermeasures or not.
 
  The rest of the report is organized as follows. 
  In \autoref{c:relatedwork} we discuss about the related work. 
  In \autoref{c:testbed} we discuss the process of building of the testbed and in \autoref{c:aes} we 
  discuss how AES algorithm was attacked and its results. 
  Then in \autoref{c:speck} how the implementation of Speck algorithm for PIC was done is explained.
  In \autoref{c:speckattack} the attack on Speck is discussed while emphasizing the
  new approaches and mechanisms needed while presenting the results.
  In \autoref{c:hardware} we elaborate some selected circuit based countermeasures and present the results
  from the tests we did for verifying their effectiveness. Several new circuit based ideas tested by us are also presented
  with evaluation results on their performance.
  In \autoref{c:software} we elaborate some selected software based countermeasures and show how effective they are, based on 
  the test we carry out. Improvements done by us to a selected software based countermeasure is also elaborated.
  Finally we conclude the report in \autoref{c:conclusion}.

 \chapter{\label{c:relatedwork}Related work}

\section{Power analysis and Testbeds}

\begin{figure}[htb]
    \begin{center}
      \includegraphics[width=6cm]{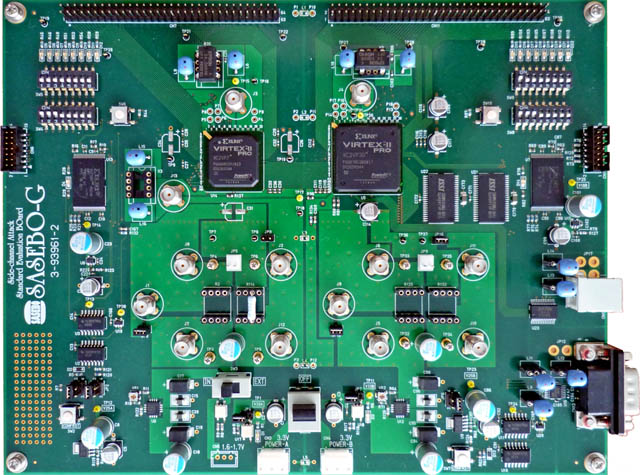}
    \end{center}
    \caption{\label{f:sasebo}A SASEBO board}
  \end{figure}

 National Institute of Advanced Industrial Science and Technology in Japan 
 has produced a side channel attack evaluation board called SASEBO board for 
 power analysis\cite{saseboboard}. The image of a board is shown in \autoref{f:sasebo}. 

 This is a commercially available board but this being built using an FPGA does hardware based encryption. 
 Power analysis on hardware based encryption and software based encryption has differences and 
 therefore for those who do research on software based encryption this board cannot be used. 
 On the other hand most cryptodevices such as smart cards use microcontrollers in their design 
 due to the low cost and hence that is software based encryption. The cryptographic device of our
 testbed does software based encryption and therefore it would be useful for anyone interested in 
 software based encryption. 
 
  Martinasek et al.\cite{martinasek} has created a testbed for Differential Power Analysis where the 
  cryptographic device is based on a PIC microcontroller. But the method they have used to 
  interface the cryptographic device to the computer is RS232. Today most computers such as laptops 
  does not have RS232 interfaces and therefore using it with such a computer is not straight forward. 
  In our setup we have interfaced the cryptographic device using Universal Serial Bus (USB) and because 
  almost all computers today are equipped with USB ports our device can be more easily used. 
  Further for measuring power they have used a current probe. A current probe is much costly and our 
  laboratory does not have such probes. On the other hand our power measurement setup measures voltage 
  rather than the current and therefore ordinary oscilloscope probes can be used.
  
  Petrvalsky et al.\cite{petrvalsky2013differential} has created the cryptographic device for their testbed using an 8051 based 
  processor for Differential Power Analysis. They have used a USART for communication and again
  the use of RS-232 causes the problems discussed in the previous work. To measure power they use 
  two different setups. One method is for ordinary oscilloscope probes while the other method is for
  differential probes which are costly. They claim that the method that does not use differential probes 
  require more power traces to derive the key than the method that use differential probes.
  In our work we introduce a new power measurement method that uses ordinary probes but yet needs 
  similar number of traces the method that used the differential probe required. This power measurement 
  method is not found in any publication available to us.
  
  Petrvalsky et al.\cite{petrvalsky2014differential} has also created a testbed based on an ARM based processor and 
  there as well the same USART have been used. In this setup for power measurement a differential 
  probe is required. This type of probes are much costly than an ordinary oscilloscope and they are 
  not available in our lab. Our power measurement setup as described above uses ordinary oscilloscope probes.
  They have alternately suggested a method to use two ordinary oscilloscope probes instead of the 
  differential probe but then with another probe for the trigger a total of three probes are needed. 
  Oscilloscopes normally come with two probes and even if three probes were available then the waves 
  from the first probe has to be subtracted from the second one which produces an unnecessary overhead. 
  Our power measurement setup only require two probes, one for power measurement and the other for the 
  trigger.
  
  Tepanek et al.\cite{lowcosteducation} designed an AVR microcontroller based power analysis testbed and the 
  cost for the testbed including the testbed is claimed to be \$2300. The upper margin 
  of their oscilloscope at that time is \$2000 and therefore the cost for the cryptographic device 
  with the power measurement circuits and relevant interfacing circuits is about \$300. This is a bit 
  higher cost for us and therefore we have come up with a setup which is less costly than that. Another
  issue in this setup as similar to the previous ones is the use of serial port for interfacing.
  
  Moreover all the publications above that discuss power analysis on software based encryption 
  does not discuss any detailed information about the testbed they have created. They neither include 
  any steps or the methodology they followed in building the testbed. Therefore those testbeds 
  are not easily reproducible. 
  
  Another issue with the above mentioned publications is that they all use AES to do power analysis. 
  Apart from them there are many other works related to power analysis but they are also mostly about AES or DES. 
  According to best of our knowledge currently there is no work that discuss about power analysis on 
  the Speck algorithm. Therefore the fact that whether the algorithm is vulnerable to power analysis 
  attacks is yet unknown. In the second phase of our project we do power analysis on Speck.
  
  Brier et al.\cite{cpabrier} introduced the Correlation Power Analysis algorithm for analysing the power 
  traces to derive the secret key. As explained in \autoref{s:cpa} we make use of this algorithm for the 
  power analysis attack. 
  
  Correlation Power Analysis algorithm has a large time complexity and hence a generic 
  implementation that runs on the CPU takes a considerable amount of time. This makes 
  it difficult due to the fact that large amount of time has to be waited to get the result for an attack. 
  Therefore we use an implementation of Correlation Power Analysis on 
  CUDA (Compute Unified Device Architecture) done for a previous project during semester 5
  which harness the power of thousands of threads in NVIDIA GPUs 
  to accelerate the algorithm more than 1000 times faster\cite{cpacuda}. As that implementation was 
  for analysing power traces for hardware based encryption, couple of changes were done to port 
  it for software based encryption. As this was for attacking AES for the attack on Speck it has 
  to be further modified to comply with that algorithm.
  
  \section{\label{relatedcounter}Countermeasures}
  
  Work related to circuit level hardware countermeasures are very few. Sprunk \cite{sprunk1995clock} in his invention uses a clock that
  outputs a stream of random clock pulses. When the clock is unpredictable, the moment at which
  a certain instruction would run is also unpredictable. Therefore the obtained power traces would be misaligned. The misalignment of
  power traces makes it necessary to collect and analyse large number of power traces which makes the attack more time consuming.
  But many communication protocols such as USB and RS232 need a stable clock and therefore usage of an unpredictable clock
  is a disadvantage is such situations. 
  
  Kocher et al. \cite{kocher} proposed introducing noise to the power line which is also a circuit level countermeasure. When noise is manually added using some noise generating
  device, it decreases the signal to noise ratio (SNR) and hence the number of required power traces increase. 
  Mangard \cite[pp.~167-175]{mangard} also proposed several circuit level countermeasures. One technique is filtering the power line.
  When an appropriate filter is connected to the cryptographic device, the exploitable power consumption patterns are attenuated
  making the attack difficult. Mangard \cite[pp.~167-175]{mangard} also proposed skipping clock pulses. When the clock signal is passed
  through a filter which randomly skips clock pulses, the alignment of the power traces are broken making the attack difficult. 
  According to best of our knowledge, most of these countermeasures are just proposed but not practically tested. 
  Therefore we practically implement some of these countermeasures and attack them to check how effective they are.
  
  On the other hand there are lot of work regarding micro architecture level hardware countermeasures. For example
  Waddle and Wagner \cite{waddle2005fault} introduced a method called dual rail logic to achieve constant power consumption for every clock cycle. This eliminate the correlation between power traces and the data and operation being processed.
  But devices with micro architecture level countermeasures are very costly. Also the power is consumed equally all the time increasing the power consumption. Further additional logic gates increases the device size as well. Working on micro architecture level countermeasures would 
  need technologies such as fabrication and hence we are not focusing on these countermeasures.
  
  On software countermeasures as well there are a lot of works. \autoref{t:softcounter} work summarizes some selected
  software based countermeasures. 

  \begin{table}
  	\begin{center}
  		\begin{tabular}{|p{6cm}|p{3cm}|p{5cm}|}\hline
  			\textbf{Method} & \textbf{Introduced by} & \textbf{Issues} \\\hline
  			Injecting dummy instructions & Kocher et al. \cite{kocher} & Cannot prevent SPA \\\hline
  			Injecting valid instructions & Ambrose et al. \cite{ambrose2007rijid} & Decreased performance\\\hline
  			Randomly shuffling Sbox operations & Mangard et al. \cite{mangard} & Limited by the number of Sboxes \\\hline
  			Dividing the standard Sbox into two different Sbox & Goubin and Patarin \cite{goubin1999and} & Susceptible to 2nd-order DPA\\\hline
  			Software balancing by processing words containing both the data bits and their complements & Daemenand and Rijmen \cite{daemen1999resistance}
  			& Susceptible to signal processing analysis \\\hline
  			Doubling the data width, to include both the original data and its complementary value & Arora et al. \cite{arora2013double} &
  			Increased power consumption \\\hline
  		\end{tabular}
  		\caption{\label{t:softcounter}Results for ground resistor method}
  	\end{center}
  \end{table}

 In dummy instruction injection method, instructions such as NOP (No Operation) are randomly inserted to the encryption code.
 This breaks the alignment of power traces and hence the number of power traces required for the attack increases.
 But NOP instructions which do not do any work would have a power consumption pattern that is easily distinguishable.
 Therefore using techniques such as SPA the traces can be visually inspected to detect the NOPs.
 
 The method proposed by Ambrose et al. \cite{ambrose2007rijid} inserts real instructions instead of NOPs. This makes visual
 inspection of power traces a challenge. But inserting extra instructions is a compromise with power consumption. But random instruction
 insertion is quite simple to implement on any microcontroller. Therefore this method is less costly and easily feasible but yet
 provides considerable level of security when compared many other countermeasures. But the effectiveness of the countermeasure has
 been only tested on a simulator. In simulators ideal conditions are assumed when modeling the real systems. But in real systems
 such ideal conditions do not exist and hence the effects may be different, Therefore we practically implement the random instruction insertion countermeasure on our testbed and attack it to evaluate the effectiveness on a real system. 
 
 The method called randomly shuffling Sbox operations introduced at \cite{mangard} proposes to randomly change the order 
 at which is the Sbox operations are done. Since the number of Sboxes in an algorithm is limited, the number
 of required power traces cannot be indefinitely increased like in random instruction injection. But as no additional operations
 are performed unnecessary computations and power consumption do not happen. We test this countermeasure as well, in our testbed.

 \chapter{\label{c:testbed}The testbed}
  
  As discussed in the introduction to do any power analysis attack research 
  first of all a testbed is needed.  Creating the testbed was a great challenge as 
  any source that properly explained the steps in a systematic fashion could not be found. 
  It was more challenging as it required knowledge in various areas such as microcontroller 
  programming, oscilloscope automation, microcontroller interfacing and electronics where they had 
  to be explored and learnt. Further, lot of failures occurred when building the system and there was 
  no proper source that explained the steps to troubleshoot and therefore most problems were solved 
  mostly by experimentation. In this chapter we discuss the methodology that was followed in constructing 
  the testbed while elaborating the difficulties encountered and how they were solved. 
  The testbed construction has been divided into the following sections although they are 
  tightly combined, for the convenience of proceeding the explanation.
  
  \section{\label{s:cryptographic_device}Cryptographic device}
 
  \subsection{\label{s:AESonPIC}AES on PIC microcontroller}
  
  Cryptographic device is responsible for carrying out the cryptographic operation which is encryption, 
  decryption or both. This device mimics the real device under attack such as a smart card, but for power
  analysis research as various modifications are to be tested on the system, using a custom cryptographic 
  device over an already built smart card gives the opportunity to do any modification. 
  As explained in the introduction, the scope of our project was to do research on software based
  encryption. Therefore the cryptographic device was built using a microcontroller. 
  Due to the availability of devices such as programmers and also compilers with us PIC microcontrollers 
  by Microchip company was decided to be used. 
  
\begin{figure}[htb]
    \begin{center}
      \includegraphics[width=6cm]{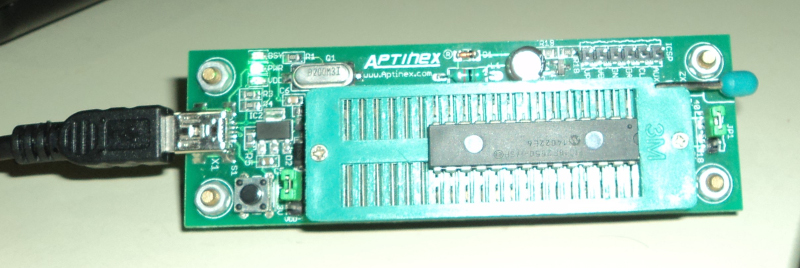}
    \end{center}
    \caption{\label{f:programmer}PIC programmer}
  \end{figure}

  First a low cost 8 bit P16F877A PIC microcontroller which had 368 bytes of RAM was selected 
  for the purpose. An Aptinex PIC programmer that can be connected to a computer via the 
  USB interface was used for programming the PIC and \autoref{f:programmer} is a photograph of that programmer. 
  To do coding Integrated Development Environment (IDE) called mikroC by MicroElektronica was used.
  First, small examples were tried while implementing the circuit on a breadboard. 
  After getting familiar with the process an AES implementation  done for 8 bit PIC was 
  obtained from \cite{aescode} and changes were done according to our need. But unfortunately the RAM on the   
  PIC P16F877A was not enough for running AES and therefore we had to go for a PIC with better 
  specifications.
  
\begin{figure}[htb]
    \begin{center}
      \includegraphics[width=12cm]{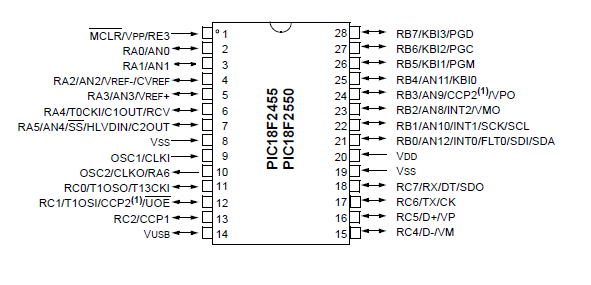}
    \end{center}
    \caption{\label{f:pic2550}Schematic diagram of PIC18F2550 from the datasheet}
  \end{figure}

\begin{figure}[htb]
    \begin{center}
      \includegraphics[width=5cm]{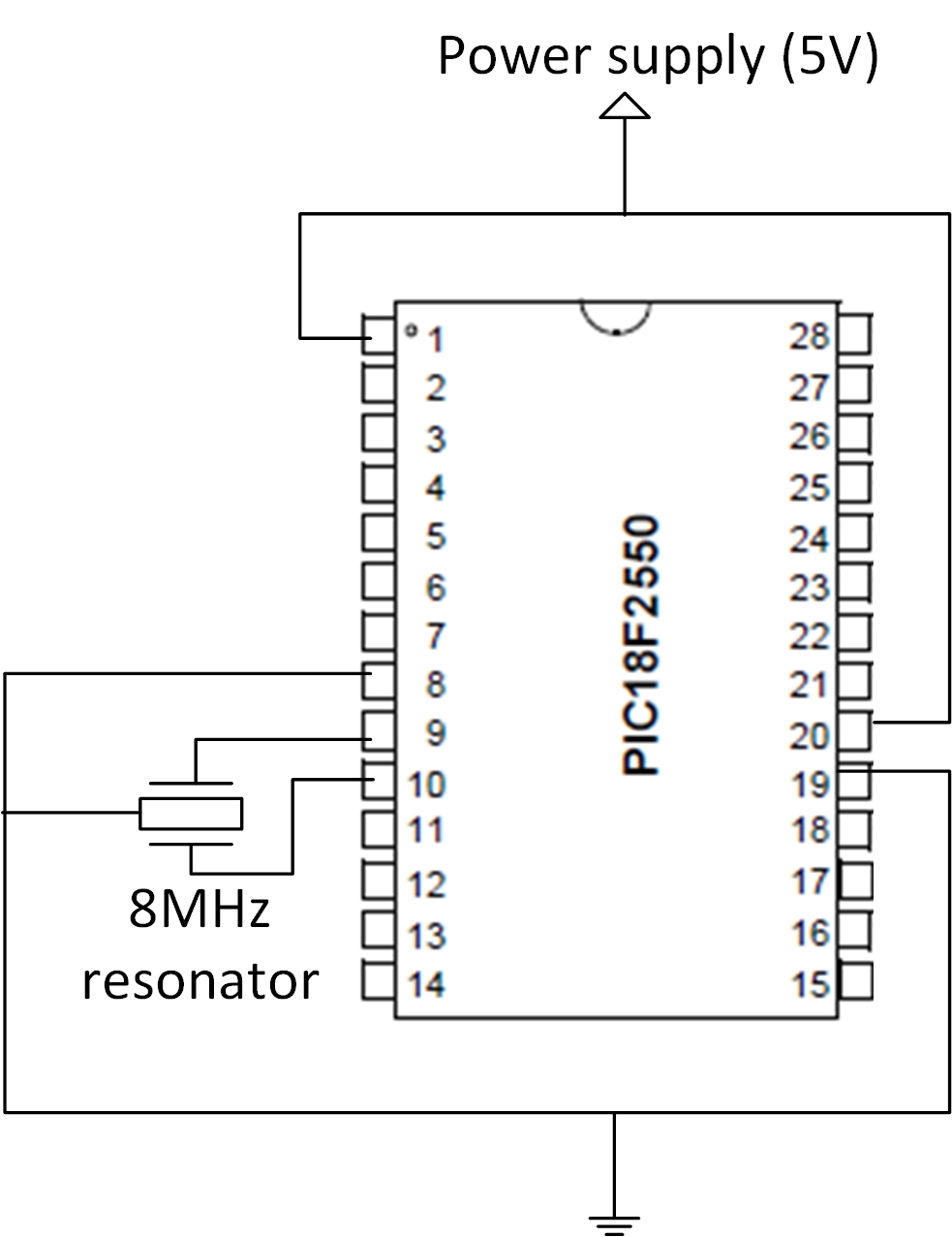}
    \end{center}
    \caption{\label{f:circuit1}Most basic circuit diagram for connecting the PIC}
  \end{figure}

\begin{figure}[htb]
    \begin{center}
      \includegraphics[width=1cm]{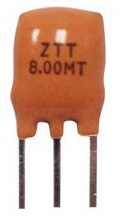}
    \end{center}
    \caption{\label{f:resonator}8MHZ resonator}
  \end{figure}

  PIC18F2550 was selected as it had 2048 bytes of RAM which is a quite enough value for our purpose. 
  Another reason that led to the selection of this PIC was the availability of an on chip USB peripheral
  module which would be helpful to interface with the computer to send plain text to the device. 
  Since this PIC is going to be used throughout the project, the pin diagram of the device is shown 
  in \autoref{f:pic2550}. The AES code was compiled and the hex file was burnt to the PIC. Here the program 
  was such that it encrypted a set of hard coded plain text using a hard coded key and then wrote the 
  result to the EEPROM(Electrically Erasable Programmable Read Only Memory) of the PIC. The sole purpose was to verify the accuracy of the code. 
  The circuit was setup up on a breadboard and the circuit diagram used is shown in \autoref{f:circuit1}. 
  Here the clock to the PIC has been provided via an 8MHz ceramic resonator 
  similar to the one shown in \autoref{f:resonator}.
  
  Then after turning on the power supply and waiting a while for the program to run, 
  the PIC was removed and put back to the PIC programmer to read the EEPROM data. The 
  cipher text in the EEPROM was compared with an online tool\cite{onlineaes}  that did AES encryption 
  to verify the accuracy of the AES code we used. 
  
  \subsection{\label{s:USB}USB interfacing of the PIC and encountered problems}

  A system that just encrypted a hard coded set of plain text is 
  not useful and the requirement is to build a system that takes plain text from a computer, 
  does the encryption on the device and then send back to the computer. Various possibilities to 
  do the interfacing were inspected including parallel port, serial port and USB. Parallel port 
  programming is easy but it is no longer found even on personal computers and hence cannot be used. 
  Serial port is also not available on computers such as laptops and because today laptops are so 
  widely used that option was also not a good one. In contrast USB is available on any computer and 
  therefore we selected to use that in spite of the difficulty to program.
  
  USB protocol is very complicated and it requires firmware from the microcontroller side and driver from
  the computer side to work. Therefore writing from the scratch would be a totally different project and 
  hence a library had to be used. MikroC has a USB library but unfortunately it was only for enumerating 
  the device as a Human Interface Device (HID). This would be suitable for a mouse or a joystick which are
  common human interface devices but for our device which needed to transfer text this would not suitable. 
  Even a software like Matlab doesn’t have a API (Application Program Interface) that directly writes or 
  reads text via such an HID interface and therefore an alternative solution had to be found.
  
\begin{figure}[htb]
    \begin{center}
      \includegraphics[width=7cm]{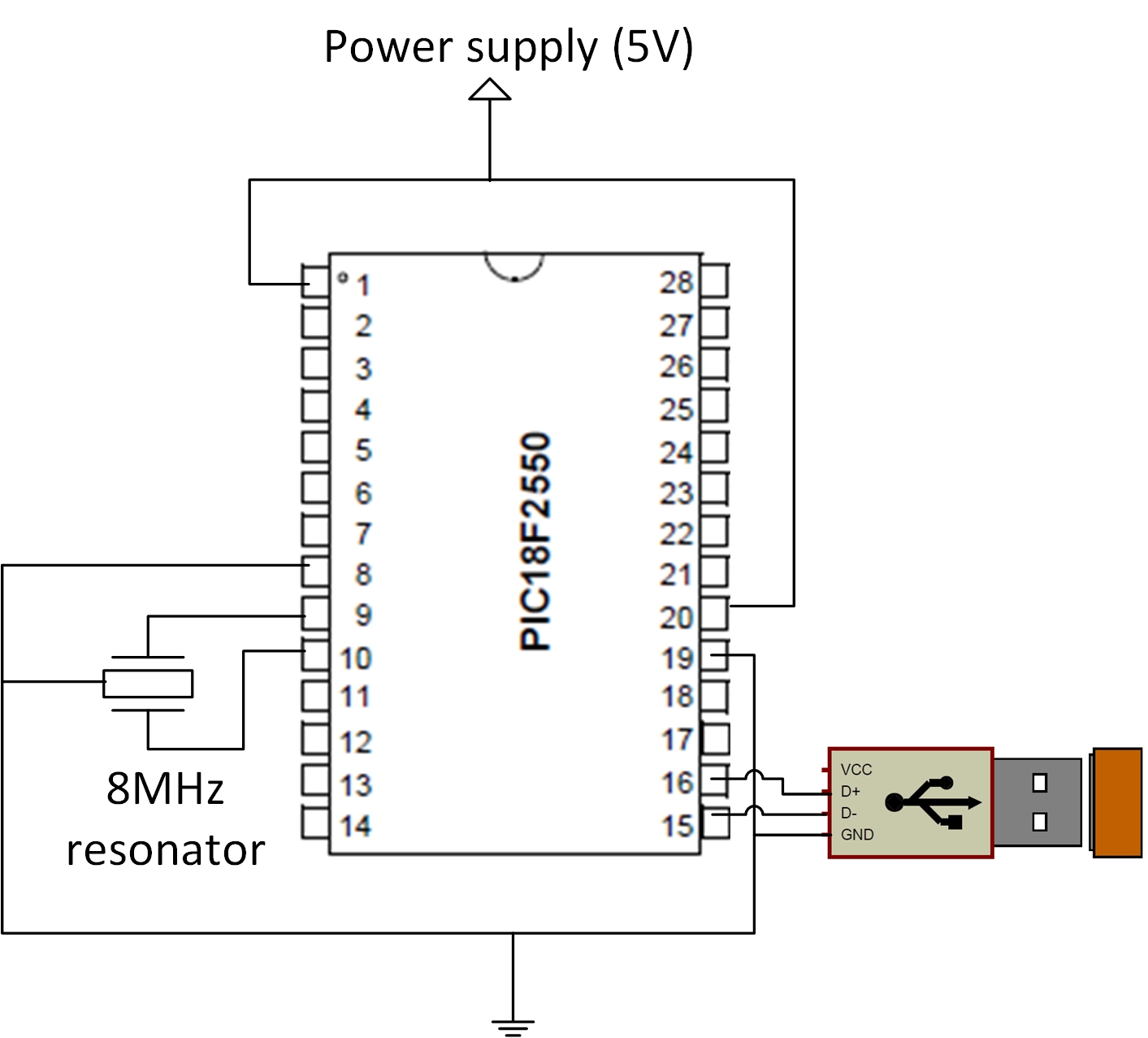}
    \end{center}
    \caption{\label{f:circuit2}Circuit after USB connection}
  \end{figure}

  After some investigation it was found out that the PIC C IDE by CCS had a library to 
  enumerate a PIC as a Communication Class Device (CDC) device. This type of devices are exposed 
  as a virtual serial port on the computer and hence programming is same as programming the serial port
  which is quite easy. Therefore we switched from Mikro C to PIC C. Getting familiar with PIC C which was
  bit different from MikroC took some time but after a while we did necessary modifications to the code 
  in Mikro C to make it compatible with PIC C. After verifying the accuracy of the code by an experiment
  similar to the one we did to Mikro C, we proceeded for USB interfacing. PIC C had easy to use library
  functions to handle USB and after getting familiar with them, the AES code was extended such that; it 
  accepts plain text via USB, do the encryption and send the cipher text back to the computer. 
  Finally it was compiled and burned to the PIC and the setup was done on a breadboard according 
  to the circuit diagram in \autoref{f:circuit2}. 
 
  Here D+ and D- in the USB port has been connected to respective pins in the PIC.
  Here we have designed the device as a self-powered USB device rather than a bus powered USB device.
  That is rather than providing power via the 5V VBUS pin in the USB port we give power using a 
  dedicated power supply. That is why the Vcc pin of the USB port in \autoref{f:circuit2} is not connected. 
  The reason is because USB power has lot of noise then the obtained power traces for the attack
  would consist of noise making the attack difficult\cite{noisepaper}.   
  
  The USB device was connected to the computer but it was not even detected by the computer. 
  By investigating a lot it was found out that USB full speed specification required a 
  48MHz clock while the one we used was 8MHz.  It was also found out that PIC18F2550 had a Phase 
  locked Loop (PLL) circuit that made it possible to generate a 48MHz clock signal internally using a 
  low frequency input clock signal. This internally generated 48MHz signal can be used to provide the 
  clock for USB. Several configuration bits had to be changed to achieve this. 
  
\begin{figure}[htb]
    \begin{center}
      \includegraphics[width=2cm]{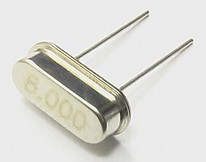}
    \end{center}
    \caption{\label{f:crystal}8MHz crystal}
  \end{figure}

\begin{figure}[htb]
    \begin{center}
      \includegraphics[width=8cm]{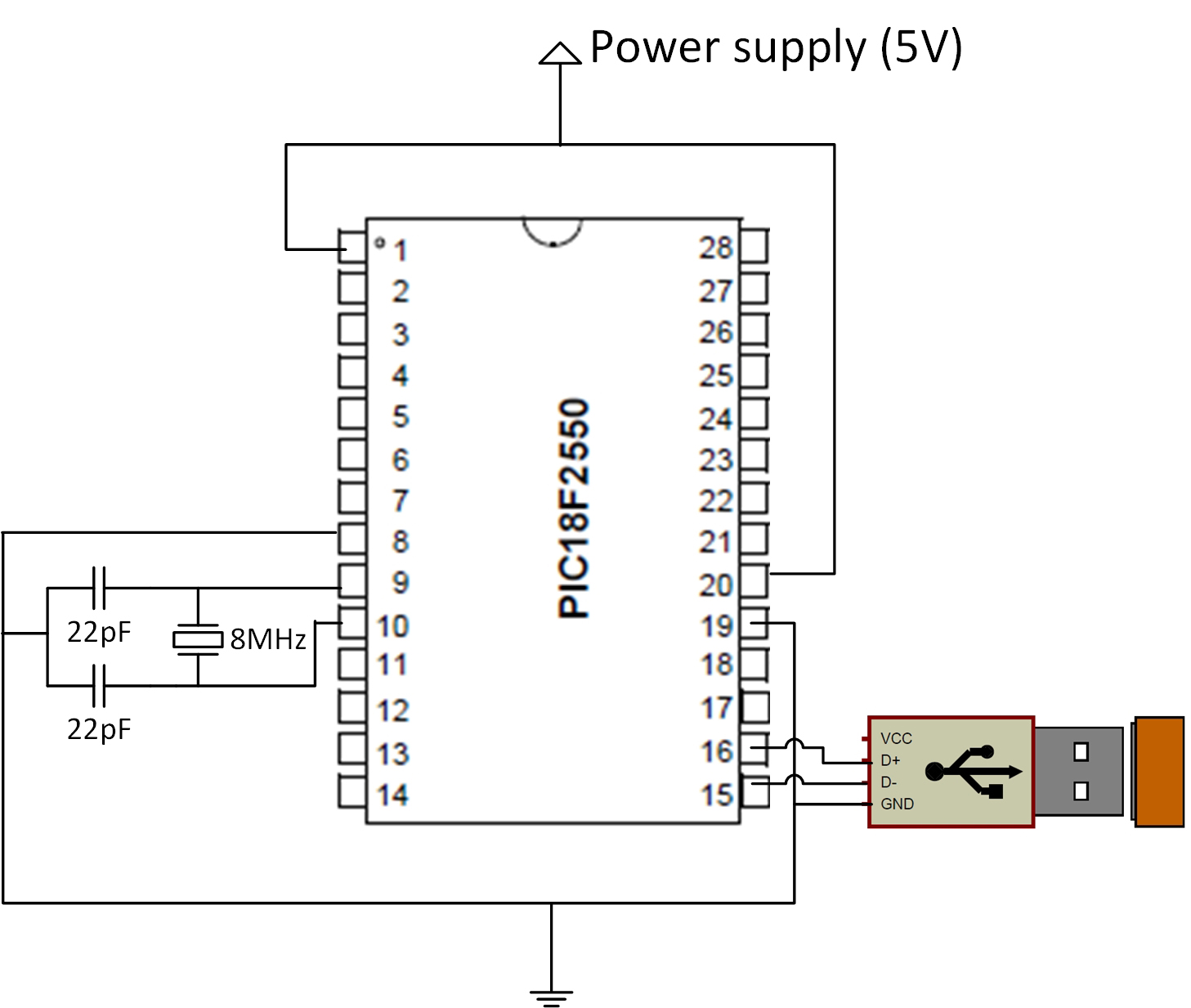}
    \end{center}
    \caption{\label{f:circuit3}Circuit after replacing the resonator with a crystal}
  \end{figure}

  After doing those changes, the device was detected by the computer but yet the ``device not recognized`` error
  came most of the times. Sometimes the device properly enumerated but still after a while, 
  the communication froze. Again after investigating and testing out various things it was found 
  out the culprit was the resonator we used. A resonator can have some deviation in the frequency it 
  generates from the stated value and this small deviation affects the 48MHz signal generated by the 
  PLL also to deviate. USB is too sensitive and such differences and such deviations cannot be tolerated. 
  As a solution we replaced the resonator with a crystal oscillator which has a better accuracy.
  \autoref{f:crystal} shows such a crystal oscillator. A crystal need two capacitors to work and hence the circuit now 
  looks as in \autoref{f:circuit3}. This circuit when connected via USB worked as expected. 

  But unfortunately more problems occurred with respect to USB when introducing the 
  power measurement resistor explained in \autoref{s:powermeasurementcircuit}. When the resistor is connected 
  serially to the device to compensate the voltage drop across the resistor, the voltage of the 
  power supply had to be increased and when doing this suddenly the USB connection disconnected. 
  After that it was no longer detected by the PC and after some investigation it was found out that 
  the internal voltage regulator in the PIC that provided power for USB communication had burnt. 
  As many users experiences in different forums said that the voltage regulator in the PIC is very 
  sensitive and burns unexpectedly, rather than buying a PIC we decided to supply a regulated 3.3V 
  to the pin 14 of the PIC from an external supply after setting the configuration options in PIC C 
  to turn off the internal voltage regulator. 
  
\begin{figure}[htb]
    \begin{center}
      \includegraphics[width=8cm]{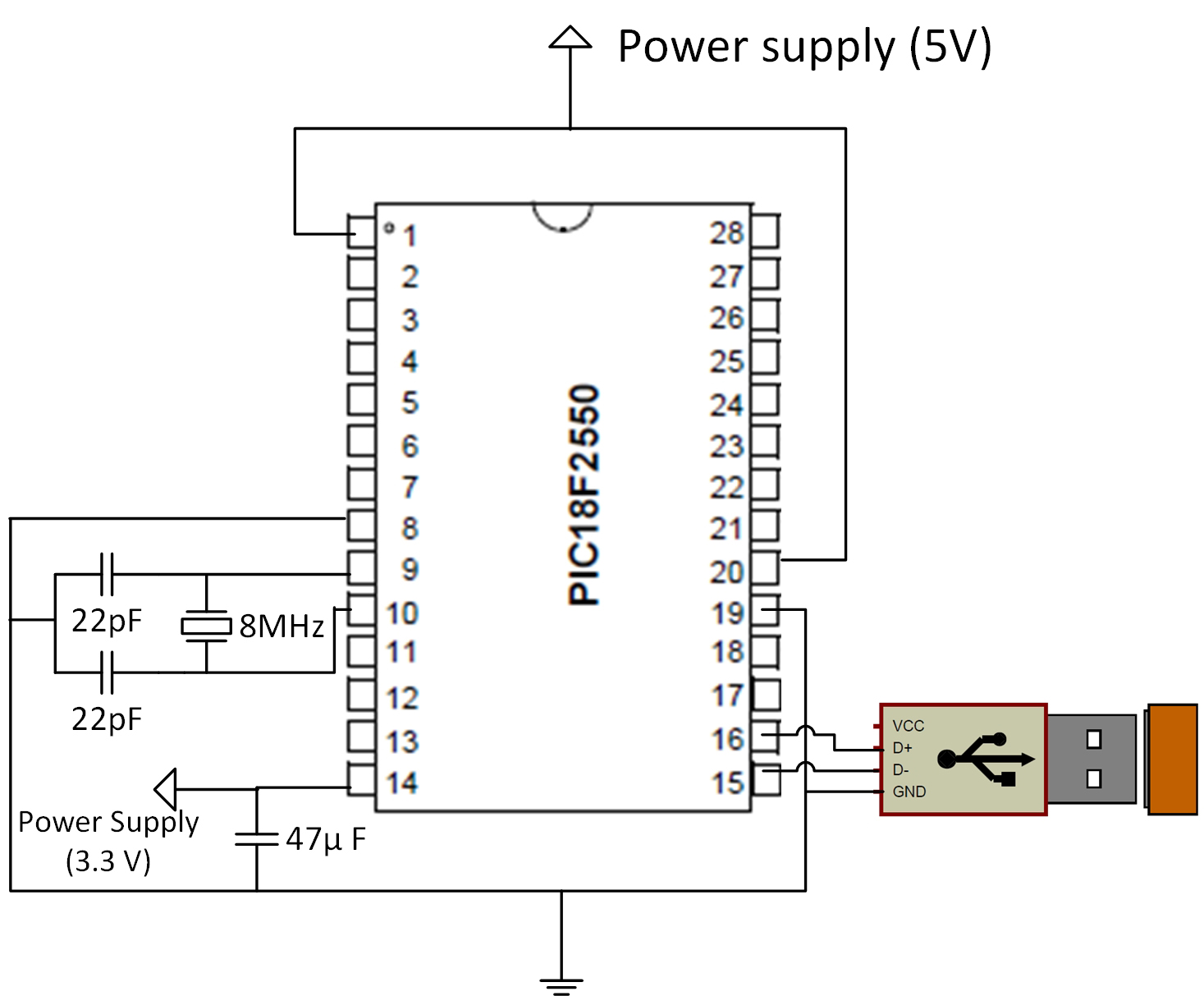}
    \end{center}
    \caption{\label{f:circuit4}Circuit after settling all USB errors}
  \end{figure}

  Problems did not stop at that point where errors occurred during transmission of data. 
  That is, data sent by the computer sometimes corrupted when being read by the PIC. 
  Also the communication froze unexpectedly sometimes. After some investigation it was found out 
  by putting a capacitor of value 0.47$\mu$F to pin 14, the problem can be solved. The final circuit 
  diagram is shown in \autoref{f:circuit4}.

  \section{\label{s:powermeasurementcircuit}Power measurement circuit}
  
  The attack requires to measure the power consumed during encryption.
  As devices to measure power consumption of a device directly are not available, 
  the necessity of an indirect method arise.  
  
\begin{figure}[htb]
    \begin{center}
      \includegraphics[width=5cm]{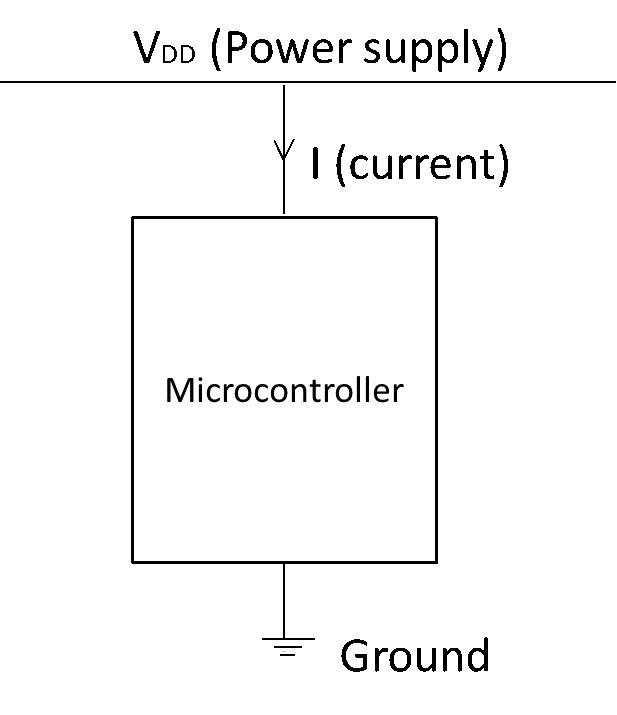}
    \end{center}
    \caption{\label{f:power1}Current drawn by a microcontroller}
  \end{figure}

  \autoref{f:power1} shows a simple setup of a microcontroller connected to a power supply. 
  The power consumption is of the microcontroller at an instance is given by the equation 
  $p=V_{DD}I$ where $I$ is the instantaneous current as explained \autoref{s:CMOS}. As $V_{DD}$ is constant, the current $I$ is 
  directly proportional to the power $p$. Therefore the power can be deduced by measuring the current. 
  But for measuring the current a device called a current probe is needed which is costly. 
  Therefore a method which an oscilloscope can be used is preferred but oscilloscopes 
  measure voltage rather than current. We elaborate two techniques which make it possible 
  for an oscilloscope to be used for measuring power.
  
  \subsection{\label{groundresistor}Ground resistor method to measure power}
\begin{figure}[htb]
    \begin{center}
      \includegraphics[width=7cm]{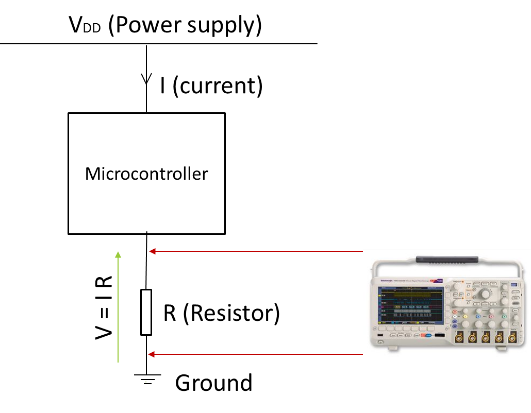}
    \end{center}
    \caption{\label{f:power2}Measuring power using ground resistor method}
  \end{figure}

  The first method is putting a resistor on the path in which the microcontroller 
  is grounded as in \autoref{f:power2}.  Now the voltage drop across the resistor is given by $V=IR$ 
  where $R$ is the resistance. As $R$ is a constant by measuring $V$ we can deduce $I$ which in 
  turn can be used to deduce the power. Therefore by connecting the oscilloscope across the 
  resistor power can be measured as in \autoref{f:power2}.   
  
  Power was captured using this method and the CPA algorithm explained in \autoref{s:cpa} was run to 
  check whether the correct key could be obtained but unfortunately it was not successful. It was 
  found out that two factors had affected the wrongs results.
  
  \begin{enumerate}
  \item Resistance of the resistor being too small
  \item Wrong grounding
  \end{enumerate}
  
  Mangard\cite[pp.~49-51]{mangard} in the example for his power measurement setup  has used a 1 ohm resistor. 
  But when we ran the CPA algorithm (explained in \autoref{s:aessteps}) on the traces collected using a 1 ohm
  resistor we could not derive the key. But later we found out that when we used a resistor which is 10 ohms or 
  higher we could derive the key successfully. 
  
  \begin{figure}[htb]
    \begin{center}
      \includegraphics[width=9cm]{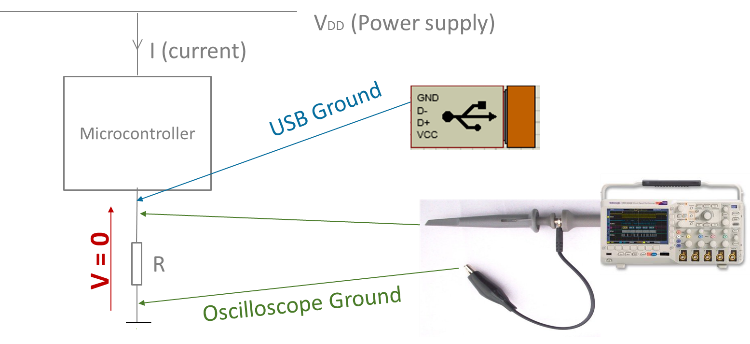}
    \end{center}
    \caption{\label{f:power3}Wrong way to ground the oscilloscope probe}
  \end{figure}      
  
  \begin{figure}[htb]
    \begin{center}
      \includegraphics[width=9cm]{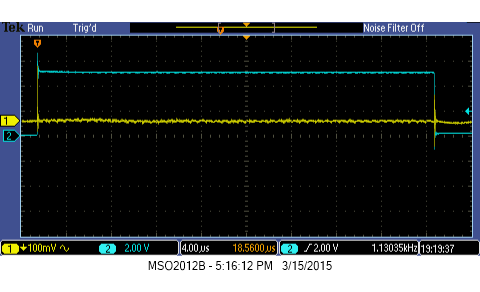}
    \end{center}
    \caption{\label{f:trace1}How the averaged power trace becomes 0 when oscilloscope ground is incorrectly connected}
  \end{figure}        
  
\begin{figure}[htb]
    \begin{center}
      \includegraphics[width=9cm]{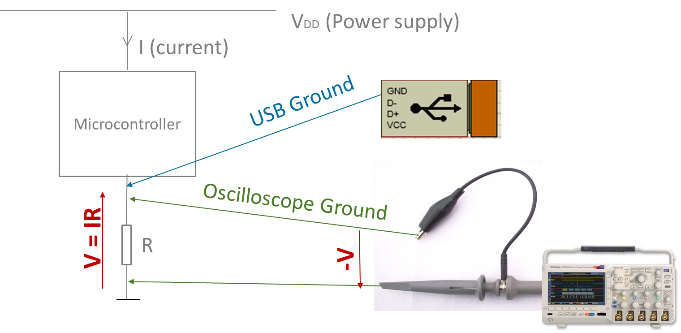}
    \end{center}
    \caption{\label{f:power4}Correct way to ground the oscilloscope}
  \end{figure}

\begin{figure}[htb]
    \begin{center}
      \includegraphics[width=9cm]{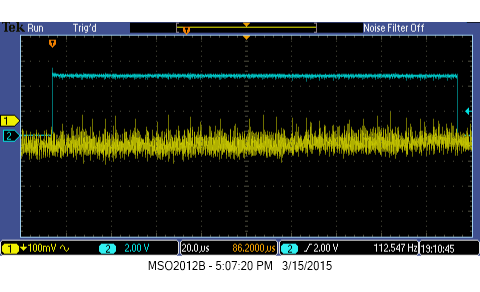}
    \end{center}
    \caption{\label{f:trace2}A good averaged power trace}
  \end{figure}

  Also connecting the ground wire of the oscilloscope it must be done correctly or else the attack would not work. 
  Initially we connected the oscilloscope probes as shown in \autoref{f:power3} and obtained the power traces. As the attack didn’t 
  work we took multiple power traces for the encryption of the same plain text and averaged it. This can be easily 
  done using the functions in the oscilloscope itself. When average of multiple traces are taken the most noise cancels 
  out and a trace that depicts an accurate power consumption can be taken\cite[p.~83]{mangard}.  The averaged trace looked like the one shown 
  in \autoref{f:trace1}. Here yellow wave which is almost a zero level 
  is the power trace (voltage across the resistor) and blue wave which is a square pulse is the trigger used which is 
  explained in \autoref{s:trigger}. As you can see, the power trace is almost 0 which means that something is wrong in the measurement.
  After investigation it was found out that wrong placing of ground wire of the oscilloscope was the issue. In \autoref{f:power3}  we have 
  connected the USB ground to the top of the resistor and the ground of the oscilloscope to the bottom of the resistor. 
  As both ground values are the same this puts the voltage across the resistor to 0 and this explains the power trace in \autoref{f:trace1}. 
  As a solution what we did was connecting the probe as shown in \autoref{f:power4}. Now both grounds are connected to the same place but the power
  trace would be an inverted one because we are measuring a negative voltage with respect to the oscilloscope ground. But in 
  oscilloscopes there are functions to easily invert a wave. After connecting in this correct fashion we got an average trace as 
  shown in \autoref{f:trace2}. Now the power trace shows some significant power consumption.

  \subsection{\label{vddresistormethod}$V_{DD}$ resistor method to measure power}
  
\begin{figure}[htb]
    \begin{center}
      \includegraphics[width=5cm]{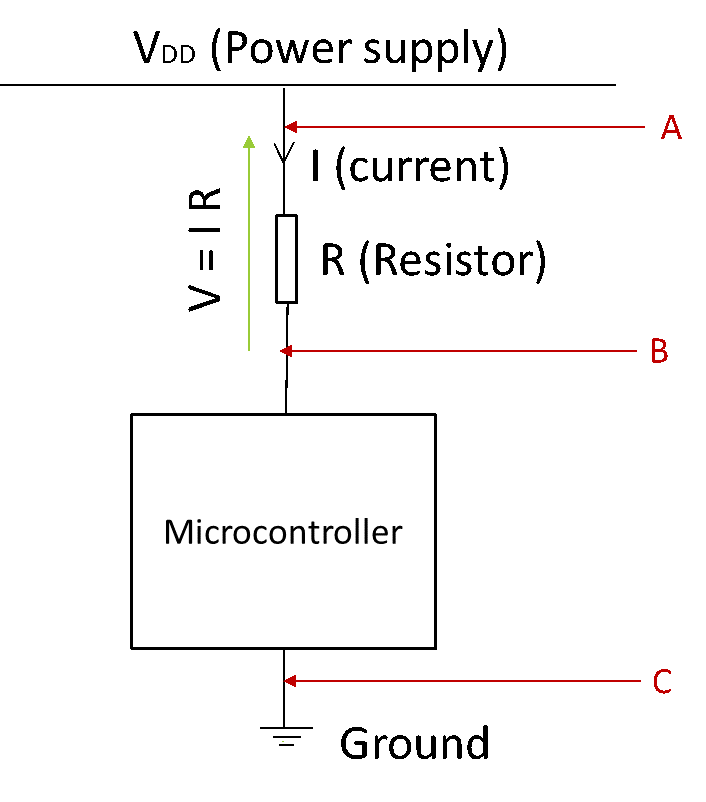}
    \end{center}
    \caption{\label{f:power5}Measuring power using $V_{DD}$ resister method}
  \end{figure}

  The second method is putting a resistor on the path to the power supply as shown in \autoref{f:power5} . 
  The concept is similar to the previous method but now if an ordinary oscilloscope probe is connected to 
  measure voltage across A and B in \autoref{f:power5} the resistor would go into flames in a few seconds just as happened to us. 
  The reason is that when you connect the ground wire of the oscilloscope probe to B, that point is grounded and if 
  the resistor value is small a large current will flow through it. Therefore a special probe called a differential 
  probe which is costly must be used to measure power across A and B. As an alternative two oscilloscope probes can be used. 
  One oscilloscope probe can be connected to measure voltage between A and C while another probe can be connected to B and C while 
  making sure the ground wires of both probes connect to C. Then by subtracting the voltage of the first probe by the voltage 
  of the second probe you can get the voltage between A and B.  One issue with this two probe method is the unnecessary
  overhead to do the subtraction of the wave forms. Another issue is most oscilloscopes usually comes with two probes 
  and when both probes are used like this we will not have another probe to use as the trigger that is explained in \autoref{s:trigger}  
  In spite of the difficulties $V_{DD}$ resistor method is still advantageous as elaborated in \autoref{s:resistordependence}
  because it requires less number of power traces 
  to do the attack which in turn will reduce the attack time.
  
  Later accidentally we figured out a new interesting method where just one probe can be used to 
  measure power using the $V_{DD}$ resistor method. That is just by connecting the oscilloscope probe between B and C.
  Therefore here we measure voltage across the microcontroller rather than across the resistor.
  Measuring using this method lets you do the attack using lesser number of traces than what is needed when 
  using ground resistor method. The comparison of number of traces needed is done in \autoref{s:resistordependence}.
  
  \subsection{\label{s:trigger}The trigger for the oscilloscope}
  
  Power measurement should be taken when the device is doing the encryption but not when it is 
  idling or communicating with the computer. Therefore the oscilloscope must be properly triggered to 
  measure the exact place we want to measure. As we made the device we have set the pin 21 of the PIC to set a 
  signal when is starts the encryption. This signal goes to 5V when it does the encryption and when it is not it goes to 0. 
  The blue wave form in \autoref{f:trace1} and \autoref{f:trace2} are the trigger signal set by pin 21. We have set the oscilloscope to trigger on 
  the occasion when the trigger signal goes from 0 to 5V. In a situation where the cryptographic device is not programmed 
  by the attacker yet there are methods to set the trigger but at this stage for convenience we ourselves set the trigger\cite[pp.~49-51]{mangard}.

  \section{\label{createpcb}Construction of the PCB}
  
\begin{figure}[htb]
    \begin{center}
      \includegraphics[width=7cm]{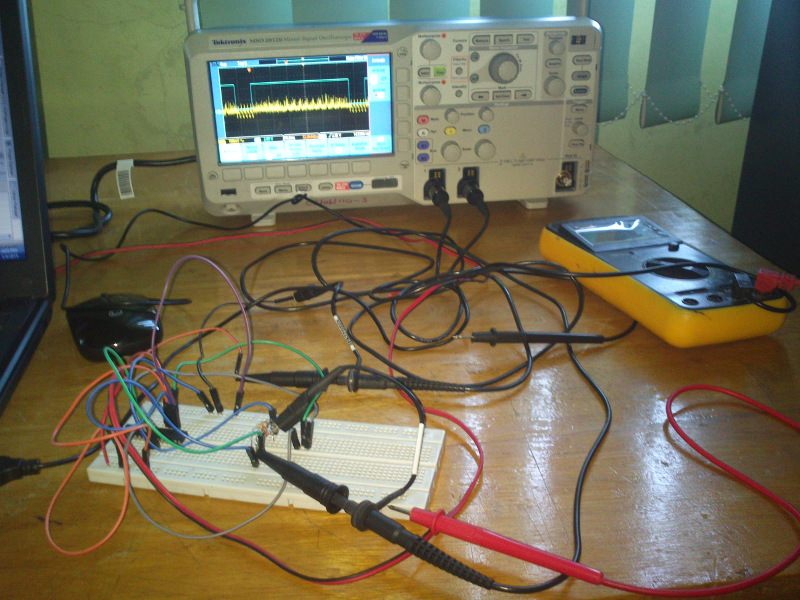}
    \end{center}
    \caption{\label{f:breadboard}The cryptographic device and the power measurement setup on a breadboard}
  \end{figure}

\begin{figure}[htb]
    \begin{center}
      \includegraphics[width=7cm]{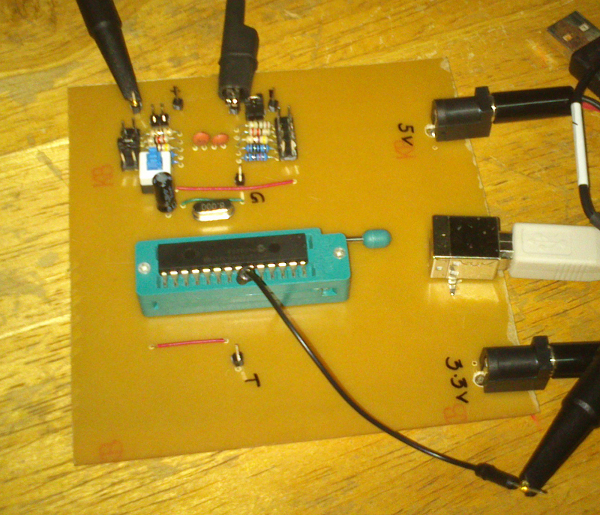}
    \end{center}
    \caption{\label{f:pcb}The cryptographic device and the power measurement setup on a PCB}
  \end{figure}

  As explained in the previous sections first we tested the cryptographic device and the 
  power measurement setup on a breadboard. The setup on the breadboard is shown in \autoref{f:breadboard}. 
  Then after fixing the issues encountered and came to a situation where the attack worked we made 
  the setup on a dot board. On dot board we tried several attacks and as it worked 
  we planned to create a PCB(Printed Circuit Board). The created PCB is shown in \autoref{f:pcb}. This PCB includes both power measurement methods 
  mentioned previously 
  where the needed one can be selected using a jumper. Also we soldered five resistors with different values
  for each power measurement method so that a required value can be selected for the power measurement. The advantage
  of having a PCB over a setup on a breadboard is that as the connections between components are much
  firm and as the wires going out here and there is less, the noise in the circuit is less. Therefore the number of traces
  needed to do an attack on the PCB is lesser than the number of traces needed to do the attack on the breadboard setup.
  
  \section{\label{s:automation}Power capturing automation}
  
\begin{figure}[htb]
    \begin{center}
      \includegraphics[width=8cm]{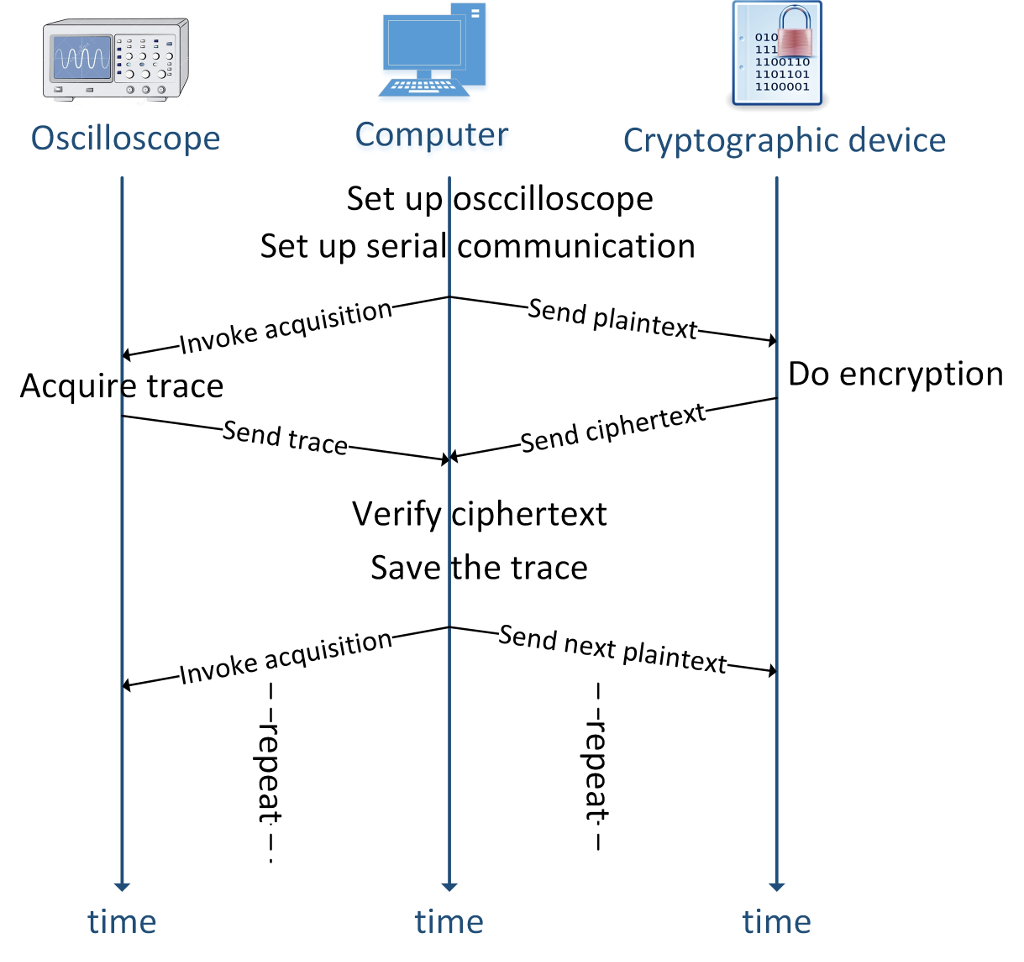}
    \end{center}
    \caption{\label{f:process}The overview of the basic steps in the automated power measurement setup}
  \end{figure}

  For the attack several hundred of power traces are necessary for encryption of different plain 
  texts and obviously the capturing of all these traces cannot be done manually. Therefore some automation is required. 
  The oscilloscope we used was Tektronix MSO2012B which had a USB interface to connect to a computer. 
  After installing drivers the communication could be established. For the purpose of automation we selected Matlab 
  as it had an easy to use API for handling instruments such as oscilloscopes. By doing the necessary configurations 
  in the Instrument Toolbox in Matlab, a Matlab script was written to automatically acquire traces from the oscilloscope.
  This Matlab script is also responsible for sending plain text to device. 
  As the cryptographic device has been enumerated as a USB CDC(Communication Device Class), it was also available as a virtual 
  serial port and therefore the serial port communication functions in Matlab could be used directly.
  In order to check whether the communication happened correctly and the cryptographic device 
  is working accurately after sending the plain text to the device and acquiring the power trace we acquire the 
  cipher text from the device and compare with the same encryption done on the computer.
  This step was much useful to detect errors during USB communication but after getting the system 
  working properly this step is no longer required. The overview of the basic steps in the automated power 
  measurement setup is shown in \autoref{f:process}. 

  \section{\label{s:summarytestbed}Summary}
  
  As explained above lot of difficulties arose when making the testbed and troubleshooting had to be done. 
  Also note that explained are the significant difficulties that we faced while many other little to
  moderate problems that were encountered have been omitted. Therefore it is very clear that creating a testbed is a 
  time consuming process and having an already built testbed would save a researcher interested in power analysis
  research from those difficulties and time consumed.
  
  \chapter{\label{c:aes}Attacking AES using CPA}
  
  As you have understood when reading \autoref{c:testbed}, the creation of testbed and doing a test attack
  are not two isolated processes. While creating the testbed we had to do several attacks to
  verify whether our testbed was working properly. By doing so only we figured out errors in our setup
  and rectified them. To test our testbed we used AES as the algorithm under attack. The reason to select
  AES was because it is the most widely used block cipher in the world and therefore lot of
  material on how to do a power analysis on AES are available. Since people had already attacked AES 
  we knew that if the attack does not work, the problem should be in our testbed. Otherwise if we directly
  started to attack an algorithm which is not yet tested we would not be able to determine whether the algorithm
  is not vulnerable for power analysis or whether our testbed is wrong. That is why we first
  tested on AES rather than directly attacking the Speck algorithm
  
  In this chapter we first show the steps followed when attacking AES. Then we elaborate some analysis we did about
  power analysis using AES as the algorithm under attack.
  
  \section{\label{s:aessteps}Steps of the attack}

  \subsection{\label{s:plaintext}Generating plain text}
  We perform the attack on AES as a known plain text attack. Therefore a set of plain text samples must be generated. 
  A C program was written to randomly generate a set of plain text. Then a set of 500 plain text samples were generated.
  
  \subsection{\label{s:selectfuncaes}Selection function for AES}
  
  Now  a selection function must be found. The concept of the selection function for a power analysis attack was 
  explained in \autoref{s:cpa}. 
  
   \begin{figure}[htb]
    \begin{center}
      \includegraphics[width=4cm]{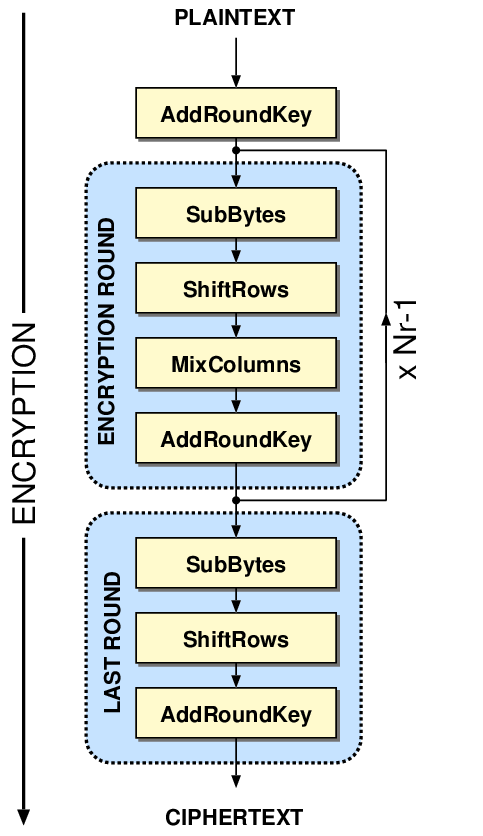}
    \end{center}
    \caption{\label{f:aessteps}Overview of AES\cite{aessteps}}
  \end{figure} 
  
  \autoref{f:aessteps} shows the overview of AES. First the operation called \textit{AddRoundKey} does a \textit{xor} operation between
  the plain text and the key. Then several encryption rounds are carried out where each of the round contains four operations
  called \textit{SubBytes}, \textit{ShiftRows}, \textit{MixColumns} and \textit{AddRoundKey}. Here \textit{SubBytes} operation does a 
  substitution box lookup. A substitution box lookup is called  a \textit{sbox} look up for short and in the rest of the chapter
  what is meant by a \textit{sbox} look up is the \textit{SubBytes} operation.
  
  What was selected as the intermediate value for the attack is the result coming after the \textit{SubBytes} operation
  in the first round. Therefore, the selection function is the combination of \textit{SubBytes} operation applied on the result
  of  \textit{AddRoundKey}\cite[pp.~138-141]{mangard}. 
  
 \subsection{\label{s:poweraes}Measuring power}
  
  The power capture should be done such that it includes a read or write of the intermediate value discussed in \autoref{s:selectfuncaes}.  
  Therefore we set the trigger to transit from 0 to 1 just before the \textit{AddRoundKey} at the beginning (just before the encryption starts)
  and then transit from 1 to 0 after the \textit{SubBytes} operation in the first round. Therefore as power is captured when the trigger is high, 
  it would include the power consumed when writing of the intermediate value. 
  
  The cryptographic device and the oscilloscope were connected appropriately with a $V_{DD}$ resistor of 100 ohm 
  selected as the power measurement circuit. Power was measured by connecting the oscilloscope probes across the PIC. 
  After connecting the devices to the computer, the generated plain text was given as the input to the Matlab script
  explained in \autoref{s:automation}. By running the Matlab script, 500 power traces were obtained such 
  that one trace for each plain text sample.
  
  \subsection{\label{s:cpaaes}Key derivation}
  
  The CPA code we obtained as mentioned in \autoref{c:relatedwork} was for attacks for AES on hardware and 
  therefore necessary changes were done. Here the code was for a known cipher text attack and therefore it was 
  first modified for doing a known plain text attack. Then the selection function was changed to the one that we used. 
  Then it was compiled and run on a computer having a NVIDIA GPU by giving plain text and the power traces as the input. 
  The PIC had been programmed such that it had \textquotedblleft67 76 89 79 88 98 A6 57 65 F7 65 77 5B 87 68 8C\textquotedblright \ as the key. 
  
  \begin{figure}[htb]
    \begin{center}
      \includegraphics[width=16cm]{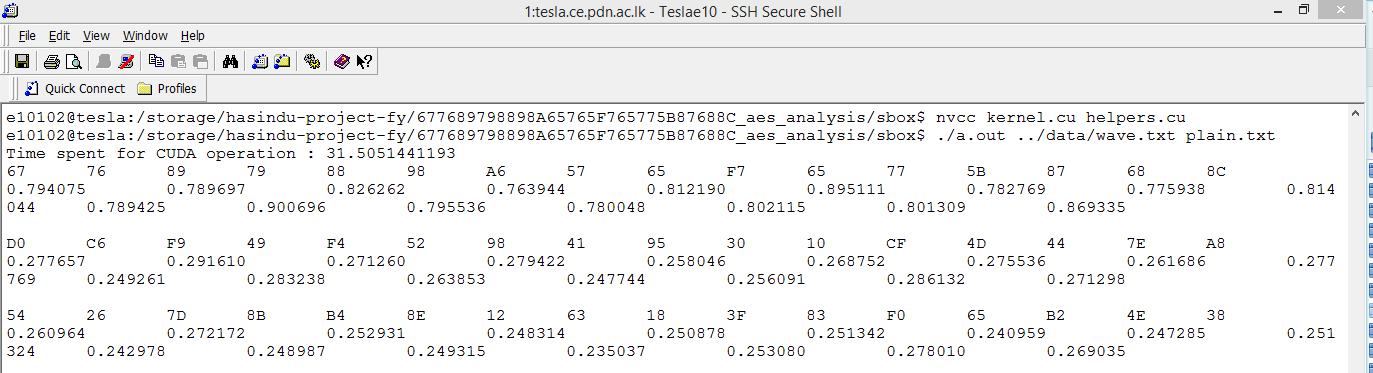}
    \end{center}
    \caption{\label{f:result}The results printed by the CPA program}
  \end{figure}

  As we discussed in \autoref{s:cpa}, the CPA algorithm is applied separately on subkeys (a subkey is defined as a part of the key)
  to recover each subkey separately. When attacking AES, a subkey is usually a part of the key which is one byte in size and
  therefore we can call it a keybyte.
  Since our key is 128 bits there are 16 keybytes.
  The result from the CPA code looks as shown in \autoref{f:result}. Here it prints the keybytes with the highest correlation
  coefficients in descending order. There are 16 columns as there are 16 keybytes.
  The first row there matches exactly to the key we have used.
  For convenience of comparing
  they are tabulated in \autoref{t:aes500}. Of course at the first 
  attack we could not derive the key. But doing lot of troubleshooting on the testbed as described in \autoref{c:testbed} and also 
  several debugging sessions on the CPA code only the results were obtained.
  In this  table the column named \textit{Key1} shows the set of keybytes which gave the highest correlation. The column \textit{Correlation1}
  is the correlation coefficients for each of the keybytes in \textit{Key1}. The \textit{key2} and \textit{Correlation2} are for the
  set of keybytes which has the next highest correlation and so on. As you can see, the column \textit{Key1} matches the exact key 
  we have used. The correlation coefficients for those keys are quite high where all are more than 0.7. But the correlation
  coefficients for the key with second highest correlation and the third highest correlation are much lesser 
  where both of them are less than 0.3.
  Therefore the fact that the correlation coefficients of \textit{Key1} and \textit{Key2} are having a large gap between them compared to the gap between the correlation
  coefficients of \textit{Key2} and \textit{Key3}, lets us confidently decides \textit{Key1} would give us the correct key.

  \begin{table}
  \newcommand{\mc}[3]{\multicolumn{#1}{#2}{#3}}
  \begin{center}
  \begin{tabular}{|l|l|l|l|l|l|l|}\hline
  × & \mc{1}{c|}{\textbf{Key1}} & \mc{1}{c|}{\textbf{Correlation1}} & \mc{1}{c|}{\textbf{Key2}} & \mc{1}{c|}{\textbf{Correlation2}} & \mc{1}{c|}{\textbf{Key3}} & \mc{1}{c|}{\textbf{Correlation3}}\\\hline
  \textbf{Keybyte 0} & \mc{1}{r|}{67} & \mc{1}{r|}{0.794} & \mc{1}{r|}{D0} & \mc{1}{r|}{0.278} & \mc{1}{r|}{54} & \mc{1}{r|}{0.261}\\\hline
  \textbf{Keybyte 1} & \mc{1}{r|}{76} & \mc{1}{r|}{0.79} & \mc{1}{r|}{C6} & \mc{1}{r|}{0.292} & \mc{1}{r|}{26} & \mc{1}{r|}{0.272}\\\hline
  \textbf{Keybyte 2} & \mc{1}{r|}{89} & \mc{1}{r|}{0.826} & \mc{1}{r|}{F9} & \mc{1}{r|}{0.271} & \mc{1}{r|}{7D} & \mc{1}{r|}{0.253}\\\hline
  \textbf{Keybyte 3} & \mc{1}{r|}{79} & \mc{1}{r|}{0.764} & \mc{1}{r|}{49} & \mc{1}{r|}{0.279} & \mc{1}{r|}{8B} & \mc{1}{r|}{0.248}\\\hline
  \textbf{Keybyte 4} & \mc{1}{r|}{88} & \mc{1}{r|}{0.812} & \mc{1}{r|}{F4} & \mc{1}{r|}{0.258} & \mc{1}{r|}{B4} & \mc{1}{r|}{0.251}\\\hline
  \textbf{Keybyte 5} & \mc{1}{r|}{98} & \mc{1}{r|}{0.895} & \mc{1}{r|}{52} & \mc{1}{r|}{0.269} & \mc{1}{r|}{8E} & \mc{1}{r|}{0.251}\\\hline
  \textbf{Keybyte 6} & \mc{1}{r|}{A6} & \mc{1}{r|}{0.783} & \mc{1}{r|}{98} & \mc{1}{r|}{0.276} & \mc{1}{r|}{12} & \mc{1}{r|}{0.241}\\\hline
  \textbf{Keybyte 7} & \mc{1}{r|}{57} & \mc{1}{r|}{0.776} & \mc{1}{r|}{41} & \mc{1}{r|}{0.262} & \mc{1}{r|}{63} & \mc{1}{r|}{0.247}\\\hline
  \textbf{Keybyte 8} & \mc{1}{r|}{65} & \mc{1}{r|}{0.814} & \mc{1}{r|}{95} & \mc{1}{r|}{0.278} & \mc{1}{r|}{18} & \mc{1}{r|}{0.251}\\\hline
  \textbf{Keybyte 9} & \mc{1}{r|}{F7} & \mc{1}{r|}{0.789} & \mc{1}{r|}{30} & \mc{1}{r|}{0.249} & \mc{1}{r|}{3F} & \mc{1}{r|}{0.243}\\\hline
  \textbf{Keybyte 10} & \mc{1}{r|}{65} & \mc{1}{r|}{0.901} & \mc{1}{r|}{10} & \mc{1}{r|}{0.283} & \mc{1}{r|}{83} & \mc{1}{r|}{0.249}\\\hline
  \textbf{Keybyte 11} & \mc{1}{r|}{77} & \mc{1}{r|}{0.796} & \mc{1}{r|}{CF} & \mc{1}{r|}{0.264} & \mc{1}{r|}{F0} & \mc{1}{r|}{0.249}\\\hline
  \textbf{Keybyte 12} & \mc{1}{r|}{5B} & \mc{1}{r|}{0.78} & \mc{1}{r|}{4D} & \mc{1}{r|}{0.248} & \mc{1}{r|}{65} & \mc{1}{r|}{0.235}\\\hline
  \textbf{Keybyte 13} & \mc{1}{r|}{87} & \mc{1}{r|}{0.802} & \mc{1}{r|}{44} & \mc{1}{r|}{0.256} & \mc{1}{r|}{B2} & \mc{1}{r|}{0.253}\\\hline
  \textbf{Keybyte 14} & \mc{1}{r|}{68} & \mc{1}{r|}{0.801} & \mc{1}{r|}{7E} & \mc{1}{r|}{0.286} & \mc{1}{r|}{4E} & \mc{1}{r|}{0.278}\\\hline
  \textbf{Keybyte 15} & \mc{1}{r|}{8C} & \mc{1}{r|}{0.869} & \mc{1}{r|}{A8} & \mc{1}{r|}{0.271} & \mc{1}{r|}{38} & \mc{1}{r|}{0.269}\\\hline
  \end{tabular}
     \caption{Results for 500 power traces} \label{t:aes500}
  \end{center}
  \end{table}

  For getting the correct key not even 500 traces were necessary. 
  Even using just 200 traces, the key could be derived correctly with a clear gap in the correlation coefficients
  for \textit{Key1} and \textit{Key2}. 
  Taking 200 traces usually takes at most 10 minutes and the CPA algorithm on CUDA for 200 traces takes just less than 1 minute. 
  Therefore AES on PIC could be broken in a time even lesser than 15 minutes.
  
  \section[Number of traces]{\label{s:numtraces}Effect of number of traces}
  
\begin{figure}[htb]
    \begin{center}
      \includegraphics[width=10cm]{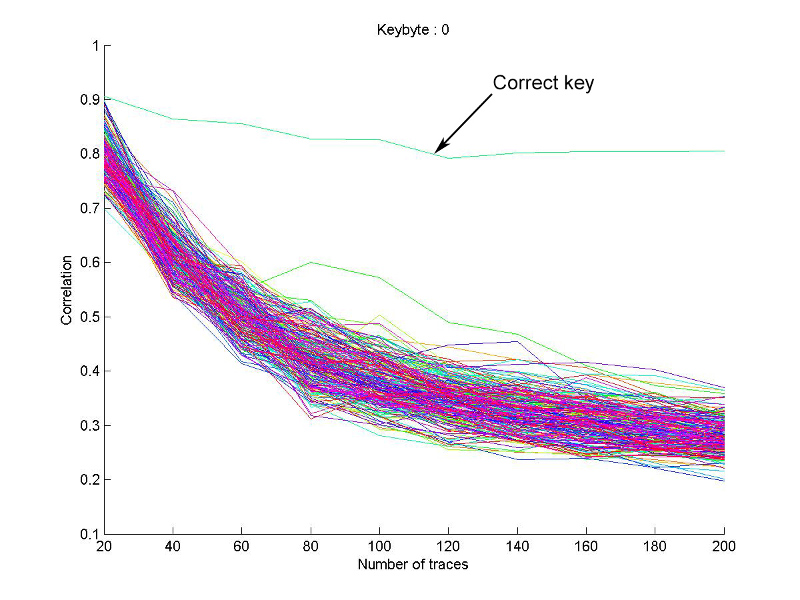}
    \end{center}
    \caption{\label{f:keybyte0}Plot for keybyte 0}
  \end{figure}

\begin{figure}[htb]
    \begin{center}
      \includegraphics[width=10cm]{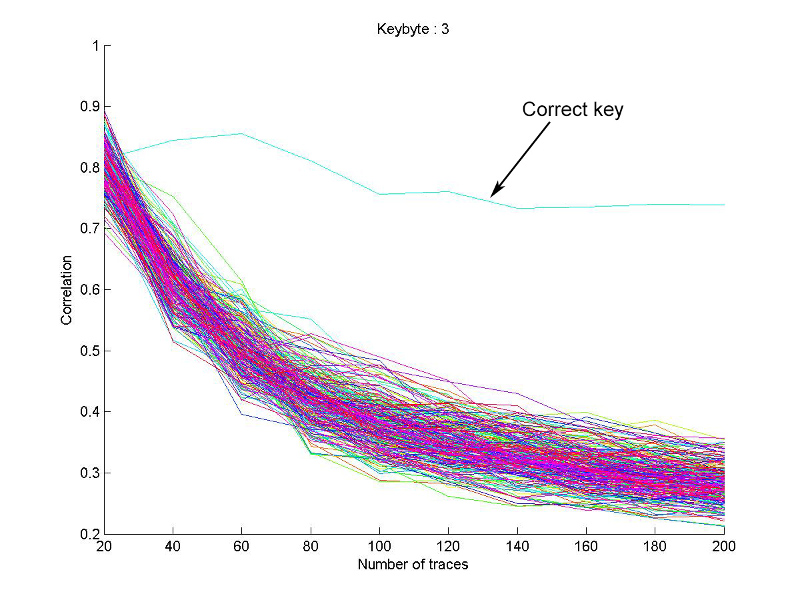}
    \end{center}
    \caption{\label{f:keybyte3}Plot for keybyte 3}
  \end{figure}

  When the number of traces used for the attack is too low the results get wrong. But when increasing the number of traces
  after a certain number of traces the answer gets correct. In order to check the minimum number of traces needed
  we plotted a set of graphs based on the power traces used for \autoref{s:aessteps}. The plot for the keybyte 0 is shown in 
  \autoref{f:keybyte0} and the plot for the keybyte 3 is shown in \autoref{f:keybyte3}. The plots for the rest of the bytes
  are not shown here as two keybytes are enough for this elaboration. In these plots, the x axis represents the number of traces
  used and the y axis represents the value of the correlation coefficient. As a keybyte can have values from 0 to 255,
  in each plot there are 256 lines that represent each of these possibilities for that keybyte. As they have fallen on top of each
  other obviously you will not be able to count them. In both plots a unique line that maintain a high correlation 
  when compared to other lines is visible. This corresponds to the correct key. This correct key still maintains a higher
  correlation value while the other keys go down as the number of traces increase. Also it is notable that
  the gap between the correlation values between this correct key and the other keys are increasing with the number of traces.
  When the number of traces are 20 there is no significant gap between them. In the plot for keybyte 3, the correct key 
  is not even at the top at this number of traces. Therefore by using a few number of traces like 20 the correct key cannot be derived.
  But when the number of traces is like 100 clearly you can say what is the correct key.
  
  For plotting these graphs first the CPA code was modified such that it printed all correlation coefficients for all keys
  rather than  the top most ones. Then using a shell script the CPA program was called for different number of traces.
  The processing of the results and plotting them was automated by writing a Matlab script.

  \section[Measuring circuit effect]{\label{s:resistordependence}Effect of the power measuring circuit}
  
  In \autoref{s:powermeasurementcircuit} we discussed that we implemented two power measuring methods 
  in our testbed. Though the $V_{DD}$ method traditionally required a differential probe which we did not have, 
  using the method we invented where power is measured across the microcontroller rather than across the resistor
  we could  still measure power using this method.  In this section we compare how the number of power traces needed 
  to successfully do the attack changes depending on the power measuring method used.
  
  First we collected 200 traces using the ground resistor method and ran the CPA code.
  Then we collected another 200 traces now using the $V_{DD}$ resistor method. The key of the cryptographic 
  device when doing these measurements had been set to all zeros. The value of the resistor used in both situations was 100 ohms.
  \autoref{t:ground} shows the summary of the results for the ground resistor method and 
  \autoref{t:vdd} shows the summary of the results for the $V_{DD}$ resistor method. 
  
  \begin{table}
  \begin{center}
  \begin{tabular}{|l|r|r|r|r|r|r|r|r|}\hline
  \textbf{Keybyte number} & \textbf{0} & \textbf{1} & \textbf{2} & \textbf{3} & \textbf{4} & \textbf{5} & \textbf{6} & \textbf{7}\\\hline
  \textbf{Best key match} & 0 & 0 & 0 & 0 & 8B & 0 & 45 & 5B\\\hline
  \textbf{Best correlation} & 0.413 & 0.384 & 0.479 & 0.501 & 0.406 & 0.621 & 0.378 & 0.382\\\hline
  \textbf{Next correlation} & 0.399 & 0.362 & 0.362 & 0.332 & 0.360 & 0.360 & 0.363 & 0.359\\\hline
  \textbf{Difference} & 0.013 & 0.022 & 0.117 & 0.169 & 0.046 & 0.260 & 0.014 & 0.024\\\hline
  \end{tabular}
  \caption{\label{t:ground}Results for ground resistor method}
  \end{center}
  \end{table}

  \begin{table}
  \begin{center}
  \begin{tabular}{|l|r|r|r|r|r|r|r|r|}\hline
  \textbf{Keybyte number} & \textbf{0} & \textbf{1} & \textbf{2} & \textbf{3} & \textbf{4} & \textbf{5} & \textbf{6} & \textbf{7}\\\hline
  \textbf{Best key match} & 0 & 0 & 0 & 0 & 0 & 0 & 0 & 0\\\hline
  \textbf{Best correlation} & 0.720 & 0.754 & 0.736 & 0.777 & 0.708 & 0.850 & 0.745 & 0.742\\\hline
  \textbf{Next correlation} & 0.387 & 0.375 & 0.400 & 0.385 & 0.380 & 0.402 & 0.361 & 0.386\\\hline
  \textbf{Difference} & 0.334 & 0.379 & 0.336 & 0.391 & 0.328 & 0.449 & 0.384 & 0.356\\\hline
  \end{tabular}
  \caption{\label{t:vdd}Results for $V_{DD}$ resistor method}
  \end{center}
  \end{table}
  
  In both tables
  we only show the first 8 keybytes of the key rather than showing all 16 keybytes to save space.
  There the first row shows the index of the keybytes. Since we are only considering first 8 keybytes this is from 0 to 7.
  The second row has the key that had highest correlation while the third row shows the value of  the correlation coefficient for that key.
  The fourth row has the value of the second highest correlation coefficient. In this table we don't show
  the key with respect to that second highest correlation as it is not needed for our comparison.
  The last row has the difference between the correlation coefficients in row 3 and 4. 
  As you can see, though the same number of traces were used the results in the two cases are different.
  In $V_{DD}$ resistor method all derived keybytes are 0 which is correct. Here the correlation values for those key
  bytes are quite high which is greater than 0.7 and the difference is  also quite high. But in the ground resistor
  method several keybytes have gone wrong. Even the keybytes which are 0 does not have high correlation
  values as in the $V_{DD}$ case. Also the difference between the correlation values here are much smaller.
  
  Therefore we can say that the $V_{DD}$ method for measuring power is better than the ground resistor method
  when the number of traces needed is compared, that in turn saves the time taken for the attack.
  
  \section[Noise filter effect]{\label{s:noisefilter}The effect of using a noise filter}

    \begin{table}
   \begin{center}
  \begin{tabular}{|l|l|r|r|r|r|r|r|r|r|}\hline
  × & \textbf{Keybyte number} & \textbf{0} & \textbf{1} & \textbf{2} & \textbf{3} & \textbf{4} & \textbf{5} & \textbf{6} & \textbf{7}\\\hline \hline
  \textbf{100 MHz} & \textbf{Best key match} & 0 & 0 & 0 & 0 & 0 & 0 & 0 & 24\\\hline
  \textbf{×} & \textbf{Best correlation} & 0.417 & 0.384 & 0.444 & 0.393 & 0.374 & 0.522 & 0.388 & 0.384\\\hline
  \textbf{×} & \textbf{Next correlation} & 0.379 & 0.376 & 0.385 & 0.386 & 0.372 & 0.374 & 0.376 & 0.379\\\hline
  \textbf{×} & \textbf{Difference} & 0.038 & 0.008 & 0.059 & 0.007 & 0.002 & 0.147 & 0.012 & 0.005\\\hline \hline
  \textbf{75MHz} & \textbf{Best key match} & 0 & 0 & 0 & 0 & D & 0 & 0 & 0\\\hline
  \textbf{×} & \textbf{Best correlation} & 0.437 & 0.471 & 0.492 & 0.490 & 0.379 & 0.598 & 0.440 & 0.405\\\hline
  \textbf{×} & \textbf{Next correlation} & 0.375 & 0.397 & 0.403 & 0.381 & 0.368 & 0.369 & 0.367 & 0.399\\\hline
  \textbf{×} & \textbf{Difference} & 0.061 & 0.074 & 0.088 & 0.110 & 0.011 & 0.229 & 0.073 & 0.006\\\hline \hline
  \textbf{42MHz} & \textbf{Best key match} & 0 & 0 & 0 & 0 & 0 & 0 & 0 & 0\\\hline
  \textbf{×} & \textbf{Best correlation} & 0.52 & 0.55 & 0.50 & 0.58 & 0.45 & 0.64 & 0.46 & 0.52\\\hline
  \textbf{×} & \textbf{Next correlation} & 0.36 & 0.38 & 0.36 & 0.37 & 0.37 & 0.38 & 0.36 & 0.39\\\hline
  \textbf{×} & \textbf{Difference} & 0.15 & 0.17 & 0.13 & 0.21 & 0.07 & 0.26 & 0.10 & 0.13\\\hline \hline
  \textbf{21MHz} & \textbf{Best key match} & 0 & 0 & 0 & 0 & 61 & 0 & 0 & 0\\\hline
  \textbf{×} & \textbf{Best correlation} & 0.437 & 0.472 & 0.444 & 0.511 & 0.398 & 0.568 & 0.427 & 0.394\\\hline
  \textbf{×} & \textbf{Next correlation} & 0.394 & 0.390 & 0.381 & 0.402 & 0.364 & 0.399 & 0.406 & 0.378\\\hline
  \textbf{×} & \textbf{Difference} & 0.043 & 0.082 & 0.063 & 0.109 & 0.034 & 0.169 & 0.021 & 0.015\\\hline \hline
  \textbf{11MHz} & \textbf{Best key match} & 0 & 0 & 0 & 0 & 0 & 0 & 0 & 0\\\hline
  \textbf{×} & \textbf{Best correlation} & 0.470 & 0.390 & 0.402 & 0.432 & 0.418 & 0.608 & 0.454 & 0.441\\\hline
  \textbf{×} & \textbf{Next correlation} & 0.391 & 0.378 & 0.396 & 0.365 & 0.374 & 0.382 & 0.386 & 0.370\\\hline
  \textbf{×} & \textbf{Difference} & 0.079 & 0.012 & 0.006 & 0.067 & 0.044 & 0.226 & 0.068 & 0.071\\\hline
    \end{tabular}
    \caption{\label{t:filterdependence}The dependence of results on the cut-off frequency of the high pass filter}
    \end{center}
    \end{table} 
  
   According to \cite{microchip} one instruction cycle of PIC18F2550 consists of four oscillator periods.
   Therefore as we used an 8MHz crystal with PLL enabled to derive a 48MHz clock, the real frequency of instructions would be 12MHz
   (48MHz/4). Nyquist sampling theorem states that when sampling a continuous signal to obtain a discrete signal,
   the sampling rate should be at least twice the maximum frequency component in the continuous signal in order to retain
   all the useful information about the original signal.
   Therefore in this case a sample frequency of 24MHz (12MHz X 2) or more is necessary.
   Using a lesser value than that would reduce the accuracy of the power consumption measurements of each instruction,
   but on the other hand using a very high sampling frequency also would be disadvantageous as then it would only increase
   the amount of noise in the measurements rather than improving the accuracy.
   The oscilloscope we used (Tektronix MSO2012B) has a frequency range of 100MHz and this whole 
   bandwidth would not be therefore useful in our case. As the oscilloscope had a function to activate a high pass 
   filter known as the noise filter we applied various cut off frequencies it supported and obtained power traces. Here we
   used the ground resistor method to measure power and took 200 power traces for each cut-off frequency
   100MHZ(no filter applied), 75MHz, 42MHz, 21MHz, and 11 MHz which were the cut-off frequencies
   supported by our oscilloscope. In all these cases the key of the cryptographic device
   had been set to all zeros.

   In \autoref{t:filterdependence} the results have been tabulated. As similar to the results in
   \autoref{s:resistordependence} only the first 8 keybytes have been shown and what are depicted by each row in each sub table
   for different cut-off frequencies are also same as \autoref{s:resistordependence}. Here if you observe \textit{Difference} row
   for each cut-off frequency for correct keybytes it is visible that the largest difference between the highest and the second highest 
   correlation coefficients exist for 42MHz. This means that at 42MHz the number of traces required for an attack is low. 
   The reason why 42MHz is better than 100MHz or 75MHz would be because using 42MHz means the noise at frequencies
   above 42MHz would be cut-off making the traces more clear. The reason why the lower frequencies such as
   21MHz or 11MHz provide worse results would be because those frequencies are lesser than the Nyquist frequency
   and hence useful information are removed from traces.

 \chapter{\label{c:speck}Speck implementation}
 
 \section{\label{s:speckalgo}Speck encryption algorithm}

 Speck is a very recent cryptographic algorithm that was released by NSA(National Security Agency) in June 2013\cite{speck}.
 This algorithm falls into the class of ciphers called block ciphers just as the famous AES algorithm. The specialty of this new algorithm is
 that it is light weight and it is optimized for performance in software implementations. Therefore the performance of this
 algorithm on a microcontroller would be impressive.
 
 \subsection{\label{s:speckop}Operations in Speck}

 Speck involves only three simple operations. Namely \textit{add}, \textit{rotate} and \textit{xor}. Therefore this cipher is known as a 
 ARX (add-rotate-xor) cipher. Simple arithmetic addition between two integers is what is referred by \textit{add}. 
 Speck needs both left rotation and right rotation where the rotation is done bitwise. If you consider a rotate to right by $n$ number
 of bits that means the number has to shifted $n$ bits to the right while the overflowing bits must be wrapped
 around to the beginning. The third operation is the famous \textit{xor} operation. All these three operations
 are simple calculations rather than memory references and that is the major reason that makes the algorithm
 light weight.

 \subsection{\label{s:speckvariants}Variants of Speck}
 
  \begin{table}
  \begin{center}
  \begin{tabular}{|l|l|l|}\hline
  \textbf{Block size (bits)} & \textbf{Key size (bits)} & \textbf{No. of Rounds}\\\hline
  32 & 64 & 22\\\hline
  48 & 72 & 22\\\cline{2-3}
  × & 96 & 23\\\hline
  64 & 96 & 26\\\cline{2-3}
  × & 128 & 27\\\hline
  96 & 96 & 28\\\cline{2-3}
  × & 144 & 29\\\hline
  128 & 128 & 32\\\cline{2-3}
  × & 192 & 33\\\cline{2-3}
  × & 256 & 34\\\hline
  \end{tabular}
  \end{center}
  \caption{\label{t:speckvar}Ten variants of Speck}
  \end{table}
  
 The algorithm has ten variants with different block sizes and key sizes\cite{speck}. The number of rounds
 depend on the block size and the key size. \autoref{t:speckvar} shows the parameters for the different variants in speck.
 As you can see the smallest version just requires a 32 bit block size and a 64 bit key size with 22 rounds of encryption.
 The largest version requires a 128 bit block size with a 256 bit key and 34 encryption rounds.

  \begin{figure}[htb]
    \begin{center}
      \includegraphics[width=6cm]{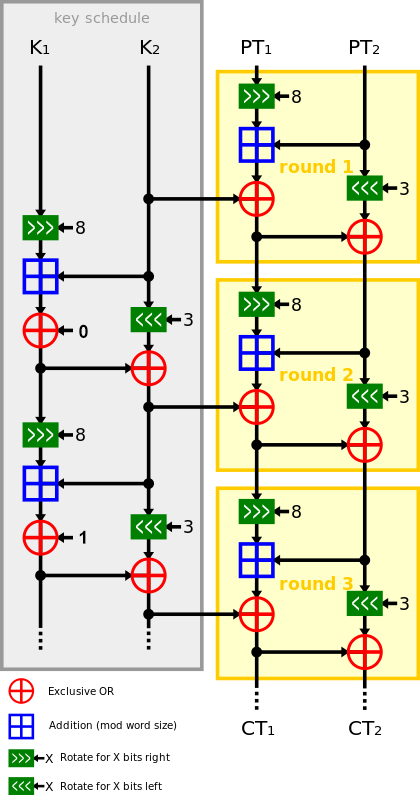}
    \end{center}
    \caption{\label{f:speck_overview}Overview of Speck algorithm\cite{speckfig}}
  \end{figure}  
  
  \subsection{\label{s:specksteps}Steps in the algorithm} 
  
 The overview of Speck algorithm is shown in \autoref{f:speck_overview}. Here only three encryption rounds are shown because anyway
 the rest looks the same. Plain text is broken into two equal parts as PT1 and PT2. Key is broken into two equal parts
 as K1 and K2. On the left side, the key generation happens. On the right side, the transformations to the plain text and the mixing with the key happens. 
 First PT1 is rotated to the right by 8 bits and then it is added with PT2. Then the result (lets call this $r$) is subjected to \textit{xor} with 
 the right half of the key K2. PT2 is first rotated left by 3 bits and is subjected to \textit{xor} with the result $r$. This is what happens in the first round.
 This sequence of actions are repeated through the rest of the rounds. For the first round, K2 was directly used
 but then for the next rounds, a modified version of the key called the round key is used. These round keys are formed
 using \textit{add}, \textit{rotate} and \textit{xor} operations carried step by step for K1 and K2 as shown in \autoref{f:speck_overview}. 

 \section{\label{s:differencewithaes}Difference with AES}
 
 Both AES and Speck are symmetric block ciphers but several notable differences exist among them. The steps in the encryption algorithm
 as well as the key generation algorithms are obviously different but we focus on the following differences as well.

 The block size, key size and the number of rounds
 in different variants of each algorithm are different. In AES the block size is always 128 bits. But as you could see in \autoref{t:speckvar}
 Speck supported several block sizes. AES only support three different key sizes 128, 192 or 256 bits which the number of
 encryption rounds are 10,12 and 14 respectively. But on the other hand as listed in \autoref{t:speckvar}, Speck gives flexibility
 to select the key size from a wide range. Also the number of rounds in Speck is somewhat larger than the number of rounds in AES.
 
 AES has operations such as \textit{sbox} lookups and therefore such operations need a considerable amount of memory.
 Moreover AES has heavy computations. For example \textit{MixColumns} step in AES performs considerable number of
 matrix multiplications. 
 Also the implementation of AES cipher is a comparatively long code and therefore the binary file
 would be larger. Due to these reasons
 AES is not suitable to be run on a low end microcontroller. Speck on the other hand does not have any \textit{sbox} lookups or any
 other complicated operations.Instead it has only three simple operations namely \textit{add}, \textit{rotate} and \textit{xor}
 as discussed in \autoref{s:speckvariants}
 Also the implementation is very small and therefore it is possible  to be stored even on a very low cost microcontroller with
 very less memory.
 
 Another great difference is how the key is mixed with plain text. As there are several variants in each algorithm and
 as the key mixing in each have some differences, for comparison lets take 128 bit variants of both.
 If the 128 bit key version of AES is considered, the first step of encryption is the mixing of the 128 bit key with 
 the 128 bit plain text block using an \textit{xor} operation. \autoref{f:aessteps} showed the basic overview of AES and
 we discussed in \autoref{s:selectfuncaes} that the function shown as \textit{AddRoundKey} is the one that is responsible
 for the mixing of the key with plain text. If the 128 bit key and 128 bit block version of Speck is considered, the way in which the key and plain text are
 mixed is completely different when compared with the process in AES. \autoref{f:speck_overview} showed the overview of Speck algorithm
 and there as you can see, the whole key is not mixed at once with the plain text. The key is first split into two equal parts
 as K1 and K2 and similarly plain text is split into two parts as P1 and P2. P1 which is the left half of plain text is
 subjected to round and add operations and the result is mixed with K2 which is the right half of the key, by using 
 \textit{xor} operation. K1 is used in the next rounds to generate the round keys.

 \section[Speck for 8 bit]{\label{s:speck8bit}Implementation of Speck for 8 bit PIC}
 
 A reference implementation done in C for the Speck variant with a 128 bit block size and 128 bit key 
 is available in \cite{speck}. The implementation there is done using 64 bit unsigned integers.
 But unfortunately we are using PIC18F2550 microcontroller in our testbed which is an 8 bit
 microcontroller and the CCS PIC compiler for 8 bit PIC that we use for
 generating binaries for the microcontroller does not support 64 bit integers. Since we could not
 find a Speck implementation done for 8 bit microcontrollers we decided to implement it.
 As the reference source code was for 128 bit key and 128 bit plain text our implementation for 8 bit was also 
 for the 128 bit key and 128 bit plain text variant of Speck.
 
 The approach was to use a unsigned char array(byte array) to represent an unsigned 64 bit integer.
 A 64 bit unsigned integer is made up of 8 bytes and therefore a byte array of size 8 can represent such
 64 bit unsigned integer. For example take a 64 bit unsigned integer which is equal to 0706050403020100 in hexadecimal 
 (0x0706050403020100). \autoref{t:bytearray} shows how this number can be shown using a byte array.
 There the first row indicate the index of the array and the second row shows the value stored in that slot
 in hexadecimal.
 
  \begin{table} 
  \begin{center}
  \begin{tabular}{|l|l|l|l|l|l|l|l|l|}\hline
  \textbf{Index} & 0 & 1 & 2 & 3 & 4 & 5 & 6 & 7\\\hline
  \textbf{Value} & 0x07 & 0x06 & 0x05 & 0x04 & 0x03 & 0x02 & 0x01 & 0x00\\\hline
  \end{tabular}
  \end{center}
  \caption{\label{t:bytearray}Representation of a 64 bit unsigned integer using a byte array}
  \end{table}

 Though the representation is straight forward, the challenge was to implement the basic
 operations needed for carrying out the encryption. Following subsections discuss how those operations were implemented based on the
 unsigned char array.
 
 \subsection{\label{s:speck_add}ADD operation} 
 
\begin{figure}[htb]
    \begin{center}
      \includegraphics[width=16cm]{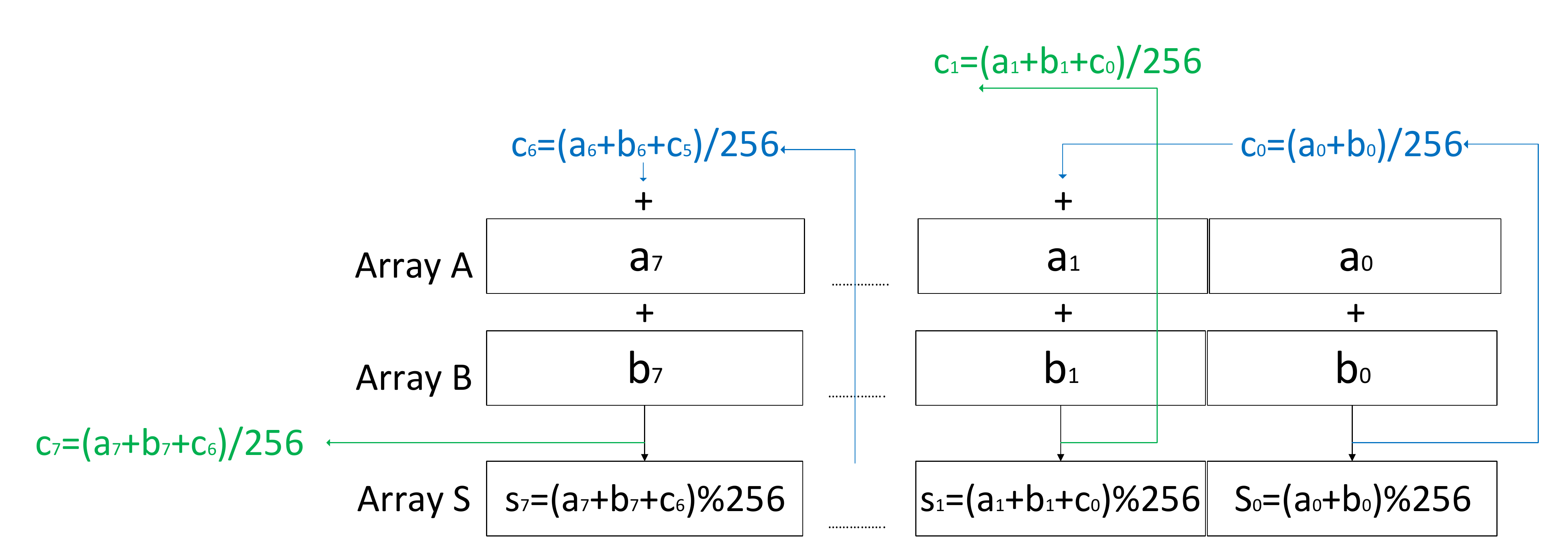}
    \end{center}
    \caption{\label{f:add}Addition of 64 bit unsigned integers using a byte array}
  \end{figure}

 Addition of two byte arrays can be implemented by adding corresponding bytes in a loop starting from
 the most significant byte but the overflow should be treated properly. \autoref{f:add} shows the addition of two byte arrays namely \textit{A} 
 and \textit{B} to get the answer byte array \textit{S}. First, the least significant bytes must be added ($a_{0}$ and $b_{0}$)  and if the addition does not exceed
 255 which is the maximum value supported by unsigned char, then the result of the addition can be directly put there.
 But if the result of the addition is a value greater than or equal to 256, the carry must be taken and added to the left adjacent byte
 as depicted in \autoref{f:add}.
 
 Fortunately PIC C compiler supported 16 bit integers despite the PIC is 8 bit. Therefore the addition of two bytes, for example the addition
 of ($a_{0}$ and $b_{0}$) was taken to a 16 bit unsigned variable. Then simply by taking the modulo operation of this value with 256 
 the value for $s_{0}$ can be found ($s_{0} = (a_{0}+b_{0})\ \% \ 256$). The carry can be found by doing an integer division
 with 256 ($c_{0} = (a_{0}+b_{0})/256$. $s_{0}$ can be put to the respective slot in array \textit{S} and $c_{0}$ should be 
 added  to the addition of $a_{1}$ and $b_{1}$ when finding $s_{1}$ and $c_{1}$. This process continues till the most significant byte of 
 the array is reached. The carry obtained for the most significant byte $c_{7}$ is dropped as anyway even in 64 bit addition this part is dropped.

 \subsection{\label{s:speck_or}ROTATE operation} 

 Rotate operation was the most difficult operation to implement. Speck needs both left rotation and right rotation but as
 left rotation is like the inverse of right rotation here we discuss only about right rotation.
 
 There are no direct rotate functions in C programming language but such a function can be easily implemented using \textit{shift} and \textit{or} operations.
 For example say we want to rotate a 64 bit unsigned number called $n$ to the right by 8 bits. The result can be easily obtained
 by first shifting the number  $n$ to the right by 8 bits ($n>>8$), then shifting the number $n$ to the left by
 56 bits ($n<<8$) and finally doing an \textit{or} operation on the two shifted intermediate numbers ( $ (n>>8)|(n<<56) $). 
 But since we use a byte array, the necessity of implementing \textit{shift right}, \textit{shift left} and \textit{or} operations for the byte array arise.
 
\begin{figure}[htb]
    \begin{center}
      \includegraphics[width=15cm]{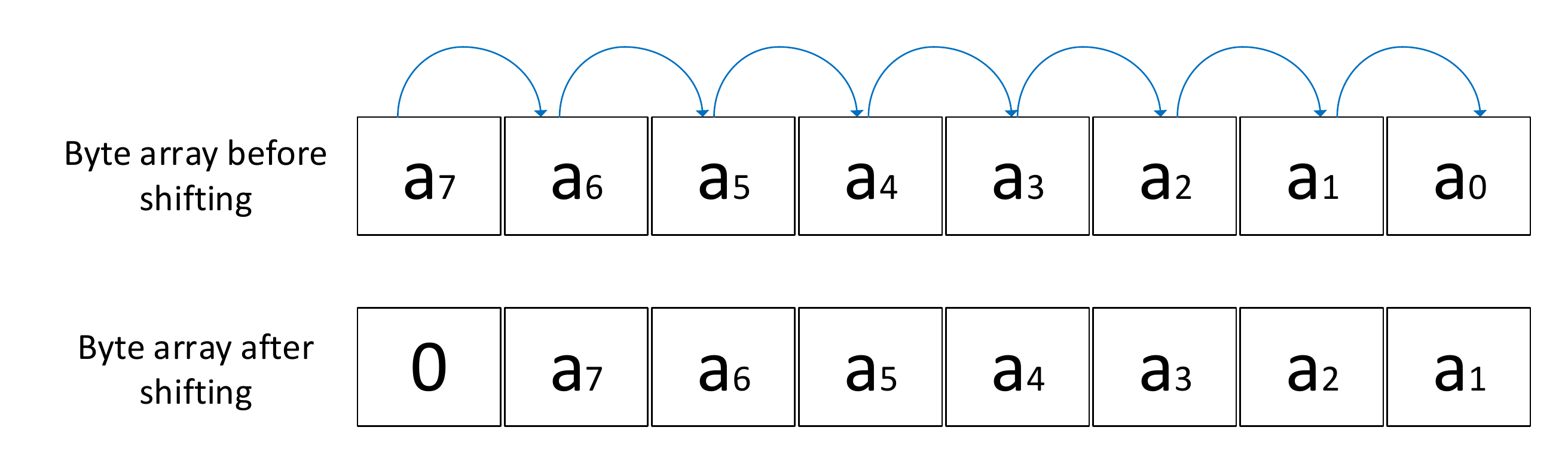}
    \end{center}
    \caption{\label{f:shiftr}An 8 bit shift right operation of a 64 bit unsigned integer using a byte array}
  \end{figure}

 First let us consider the implementation of \textit{shift right} operation. Imagine an example where the number (byte array) must be
 shifted by 8 bits to the right. Because 8 bits means one byte we can easily shift the arrays elements to the right one by one
 as shown in \autoref{f:shiftr}. Therefore similarly for a shift by $k$ number of bits in a situation where $k$ is a multiple of 8
 we can shift the elements in the byte array by $k/8$ times. 
 
\begin{figure}[htb]
    \begin{center}
      \includegraphics[width=16cm]{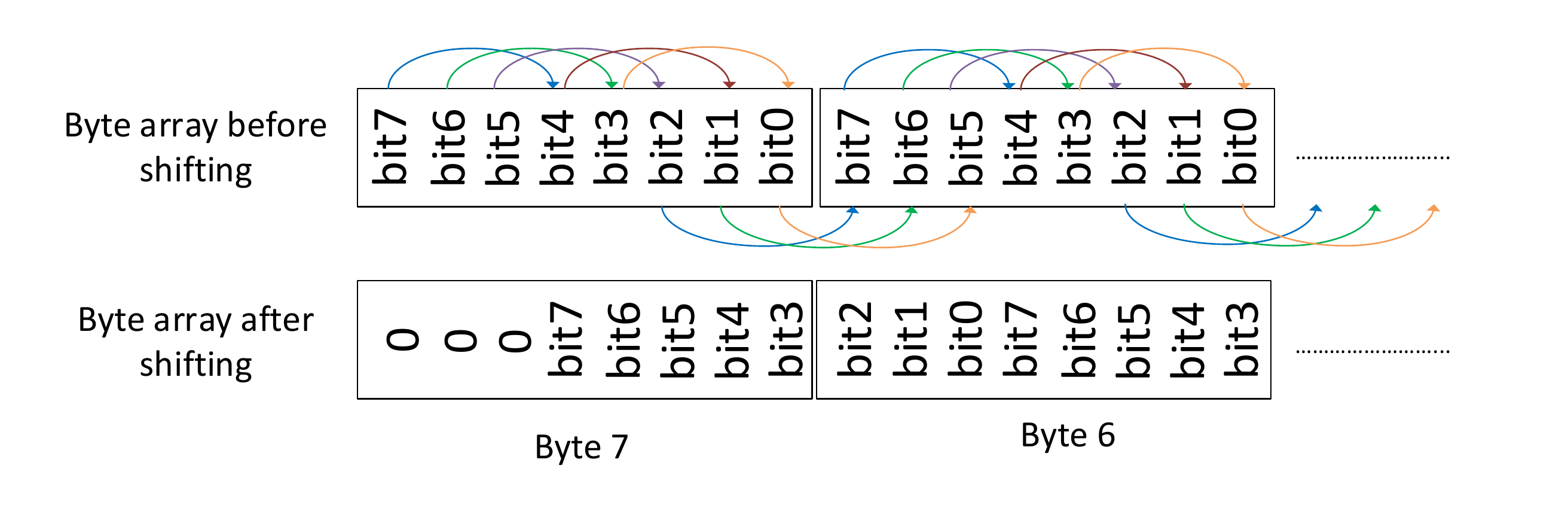}
    \end{center}
    \caption{\label{f:shiftright3}bitwise right shift of a byte array by 3 bits}
  \end{figure}

\begin{figure}[htb]
    \begin{center}
      \includegraphics[width=14cm]{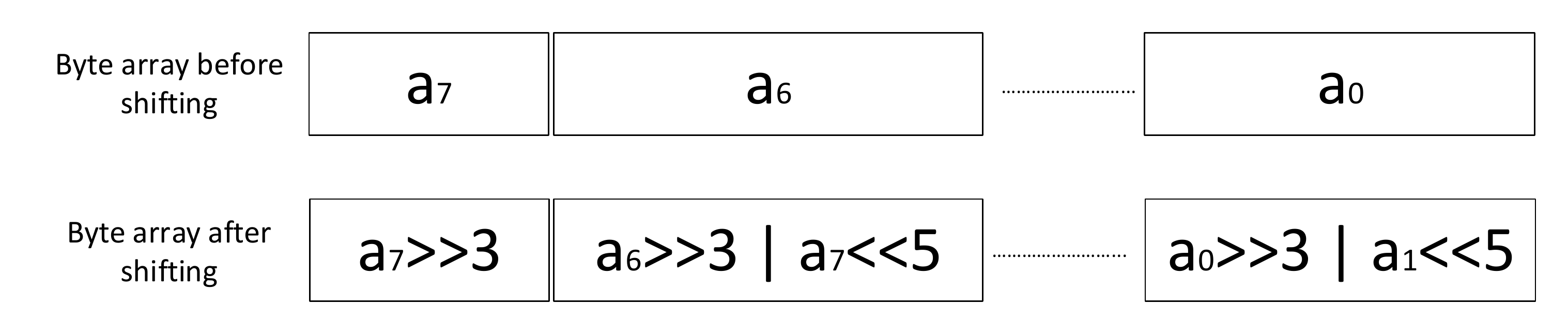}
    \end{center}
    \caption{\label{f:shiftright3how}Implementation of 3 bit right shift}
  \end{figure}

 Though the above case is very straight forward, the difficulty comes when $k$ is not a multiple of 8. For example say $k$ is equal to 3 bits. Now the array cannot be shifted
 by whole elements but instead each elements must be internally shifted while taking the overflown bits to the the beginning of the 
 next consecutive byte. Consider \autoref{f:shiftright3} where the byte array is right shifted by 3 bits. 
 As you can see, the most significant byte (byte 7) can be just shifted by 3 bits as the  leftmost 3 bits anyway becomes 0.
 But the other bytes cannot be just shifted like that. For example if the byte 6 is considered, just shifting the value to the right by
 3 bits is not enough but further it is necessary to bring the last 3 bits of the most significant byte as shown in \autoref{f:shiftright3}.
 How the implementation was done is shown in \autoref{f:shiftright3how}. Here we right shift the value in the 6\textsuperscript{th} slot by 3 bits
 and apply \textit{or} operation with the value in 7\textsuperscript{th} slot shifted left by 5 bits. This trick brings the overflown bits of the 
 7\textsuperscript{th} slot to the beginning of the 6\textsuperscript{th} slot.
 
 Now consider a more difficult situation where a value has to be shifted by a number of bits $k$, where $k$ is larger than 8 and $k$ is not a multiple of 8.
 For example consider a shifting by 19 number of bits. Now the method in the previous paragraph will not work
 as shifting a byte by 19 bits will make it all zero as the byte it self is just 8 bits. Therefore in such situations, the solution is to first shift by 16 bits
 which means shifting the whole array elements by two and then to do a shift by 3 as in the previous paragraph.
 Therefore generally if the shifting is to be done by $k$ bits $k/8$ number of bytes must be shifted and then $k\%8$ number
 of bits must be shifted according to the method described in the previous paragraph. Here the sign `$\%$` refers to modulo operation.
 
 The left shift is the inverse of the right shift which was discussed up to now and hence it is not elaborated here. Finally the \textit{or} operation must be
 implemented but luckily it does not have any problems like overflowing and therefore \textit{or} operations of corresponding bytes
 can be done independently using a simple loop. The implementation of \textit{or} is exactly similar to the implementation 
 of \textit{xor} operation discussed in \autoref{s:speck_xor}.

 \subsection{\label{s:speck_xor}XOR operation}
 
 The \textit{xor}  operation of a byte array is straightforward to implement. Here there is no dependence
 between different bytes and therefore each pair of bytes can be just subjected to  \textit{xor} to get the result.

 \section{\label{s:specktest}Testbed for Speck}
 
 The implementation described in \autoref{s:speck8bit} was first done on the computer. Then the accuracy of
 the implementation was tested using the test vectors given in \cite{speck}. 
 
 Now the requirement to be fulfilled to carry out an attack on Speck is to create a testbed that does encryption using Speck algorithm.
 But fortunately microcontrollers are reprogrammable. Therefore the same testbed we created for attacking AES which is described 
 in \autoref{c:aes} can be changed to a Speck cryptosystem by reprogramming the PIC with a program that carries out
 Speck encryption. For this, the 8 bit implementation done on the computer was imported to PIC C and it was modified such that it became compatible with the PIC C compiler. Then functions such as accepting input 
 from a computer and sending result to the computer via the USB interface were added in a similar fashion
 described in \autoref{s:USB}. Then it was compiled and then PIC microcontroller was programmed to get a working Speck cryptosystem.
 
 Then finally the Speck cryptosystem was tested for the accuracy and proper functionality.
 
 \section[Speck for 16 bit]{\label{s:speck16bit}Implementation of Speck for 16 bit PIC and testbed} 
  
 Due to the reasons described in \autoref{s:speck16attack} it was decided to implement the Speck algorithm for 16 bit PIC microcontrollers as well.
 Fortunately CCS PIC C had a compiler for 16 bit PIC. For the most convenience, this compiler supported 64 bit integers
 and therefore an implementation based on arrays such as the one described \autoref{s:speck8bit} was not necessary.
 Instead we could directly use the reference code we found for Speck which is described in the beginning of \autoref{s:speck8bit}, by doing 
 a some modifications to make it comply with the PIC C compiler.
 
 The testbed we discussed so far was based on the 8 bit microcontroller PIC18F2550 and for running the 16 bit implementation
 a 16 bit microcontroller had to be used. Since several PIC24FJ64GA002 were already available with us, it was decided
 to be used. But unfortunately the pin layout of this microcontroller was completely different with the layout of the 
 microcontroller that was previously used and therefore the earlier testbed and the PIC programmer
  was not compatible. Further this microcontroller did not have any USB controller and therefore
 some other type of communication was needed. 
 
\begin{figure}[htb]
    \begin{center}
      \includegraphics[width=6cm]{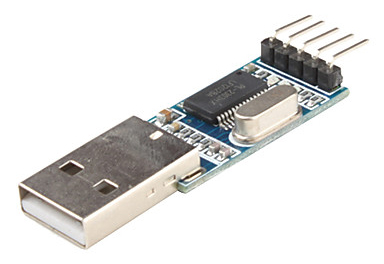}
    \end{center}
    \caption{\label{f:ttl}USB to RS232 TTL converter}
  \end{figure}

 It was figured out that using the facility called In-Circuit Serial Programming (ICSP) that is available in PIC microcontrollers, 
 the same PIC programmer that was used for programming the earlier 8 bit PIC can be still used by deriving several pins from the programmer through wires
 and then connecting to the 16 bit microcontroller. Then on a breadboard we implemented separate circuits;
 one for programming the PIC and the other for the Speck testbed. For communication we decided to use an off-the-shelf
 USB to RS232 TTL (USB to serial 
 Transistor-Transistor Logic) converter. The converter we used is shown in \autoref{f:ttl}. By connecting the converter to appropriate pins in the 
 microcontroller and then by modifying the CCS PIC C code to support traditional serial communication the objective was achieved.
 
 \chapter{\label{c:speckattack}Attack on Speck}
 
 Speck which is a very recent light weight algorithm for microcontrollers as described in \autoref{c:speck} has a good possibility
 of becoming popular for embedded device industry in the future. Therefore it would be a good idea to
 check whether this algorithm is also vulnerable for power analysis attacks so that any countermeasures can be 
 implemented. According to best of our knowledge there is no work on power analysis on Speck and therefore this chapter discusses
 new experiences. First in \autoref{s:speck_challenge}, the differences in Speck with AES that makes it impossible
 to directly use the known attack method on AES is discussed. Then in \autoref{s:spechhow} the step by step methodology
 that should be followed by someone to reproduce an attack on Speck is described. Next \autoref{s:firstattempt} describes
 the experiences we did undergo when formulating the attack including the challenges that were faced and how they were successfully
 overcome. After that in \autoref{s:speckvul} the effort taken for a power analysis attack on Speck is compared with a similar attack on AES.
 Finally in  \autoref{s:speck16attack} we show that the attack is successful not only
 on 8 bit microcontrollers but also on 16 bit microcontroller and hence how the concept can be generalized for any bit microcontroller.
 
 \section{\label{s:speck_challenge}Challenge in attacking Speck}
 
 In \autoref{s:differencewithaes} the differences that Speck has when compared with AES were discussed. Some of these changes
 make it impossible to apply the approach followed in breaking AES using power analysis directly to Speck algorithm. 
 Here we discuss the differences that
 made the attack on Speck more challenging than attacking AES and the approaches found by us to treat the issues.
 
 \subsection{\label{s:keygeneration}Difference in mixing the key}
 
 In \autoref{s:differencewithaes}  the difference in Speck and AES in the key mixing procedure was
 thoroughly discussed. In AES because the whole 128 bit key is mixed at once with the plain text, if one set of
 power traces that include
 the power consumption for writing or reading of an appropriate intermediate value such as the result of the \textit{sbox} 
 is taken, the whole key is obtainable as soon as you run the CPA algorithm once.
 
 But in Speck the key is split into two halves and only the right half of the key K2 was the one that was directly
 mixed with plain text (most precisely not plain text but plain text that is subjected to \textit{rotate} and \textit{add}).
 Therefore if a proper selection function is selected appropriately the right half key 
 can be derived. 
 
 Deriving the left half of the key K1 is not straight forward. K1 is not anywhere directly mixed with the plain text or
 even modified plain text. It is in the second round 
 that K1 is used but even then instead of mixing K1 directly, K1 is first subjected to various operations to
 derive a round key and finally it is this round key that is mixed. This round key is not just formed
 using operations on K1 alone, but also K1 is subjected to a mixing with a modified version of K2.
 Therefore we cannot directly derive K1 but instead the round key must be first derived and reverse operations
 must be done to this round key with the help of already found K2 to derive original K1.
 
 Therefore process of attacking the key in Speck must be done in two phases. First a set of power traces must be collected
 that involve power consumption for right half of the key and CPA algorithm must be run on  those traces
 to derive that key. After that, another set of power traces has to be taken that involves power consumption for
 left half of the key and CPA algorithm in collaborated with the already found key must be run to derive this part of the key.
 Due to the need of carrying out in two phases, the overall time to do an attack is definitely high for
 Speck than when compared with AES.
 
 \subsection{\label{s:nosbox}Lack of sbox operations}
 
 As discussed in \autoref{s:differencewithaes} AES has substitution box lookups while that is not in Speck.
 For attacking AES as we discussed earlier, the power consumption
 for the result of this \textit{sbox} look up is the one that is used for attacking. The lack of \textit{sbox} look up in Speck
 make it necessary to find a new selection function. As a selection function, a function that gives a intermediate result
 based on plain text and the key must be found and therefore the \textit{xor} operations that happen in Speck to mix the keys with plain text
 can be used. Of course using \textit{xor} as the selection function can give troubles such as getting
 wrong key as the result and the requirement of large number traces, where such issues are discussed in \autoref{s:firstattempt}.

 \section{\label{s:spechhow}Steps to attack Speck}
 
 In this section we describe the step by step methodology that the reader should follow to attack a Speck cryptosystem
 while explaining the reasons behind the steps. Since the creation of the testbed was discussed previously at this point we assume that
 the attacker already has a testbed and elaborate from that point onwards.
 
  \begin{figure}[htb]
    \begin{center}
      \includegraphics[width=12cm]{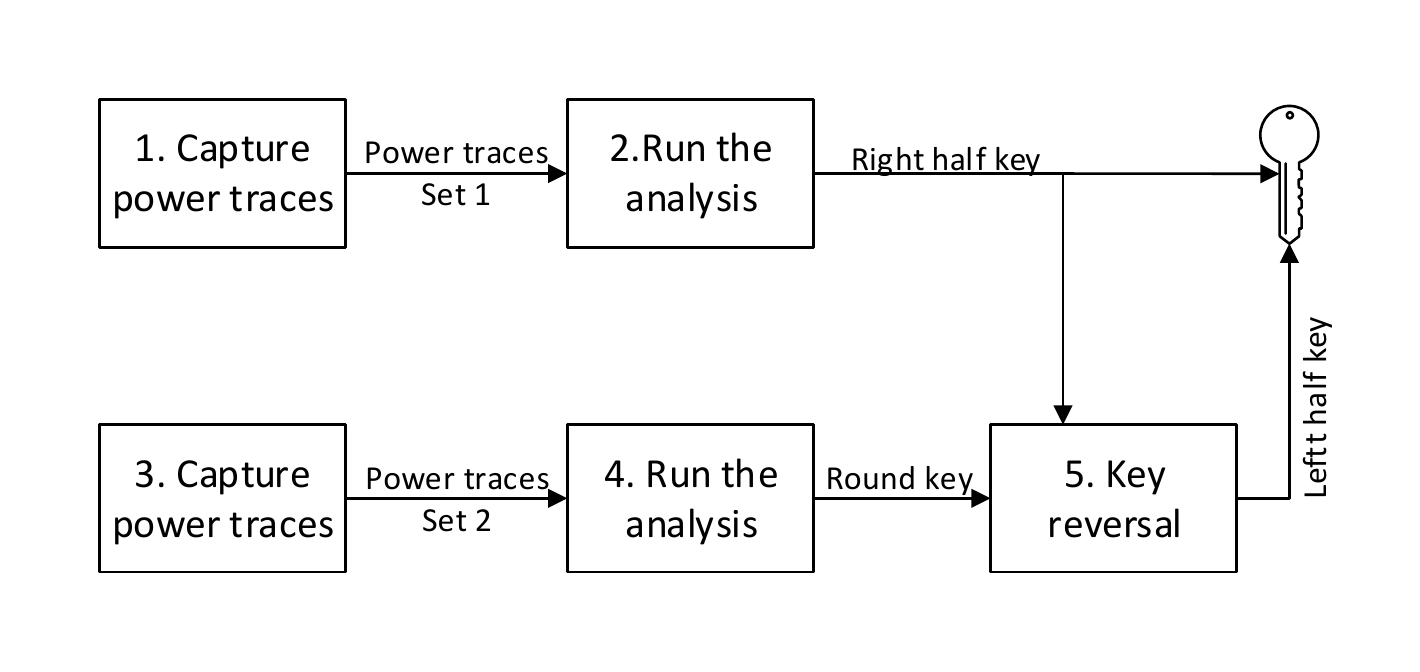}
    \end{center}
    \caption{\label{f:specksteps}Basic steps to attack Speck}
  \end{figure}  
 
 \autoref{f:specksteps} shows the most basic steps that should be followed to attack Speck. Each block represents
 a task that should be followed while the arrows represents the data flow. The order of how tasks should be executed
 is shown using a number inside the blocks. 

  \begin{figure}[htb]
    \begin{center}
      \includegraphics[width=8cm]{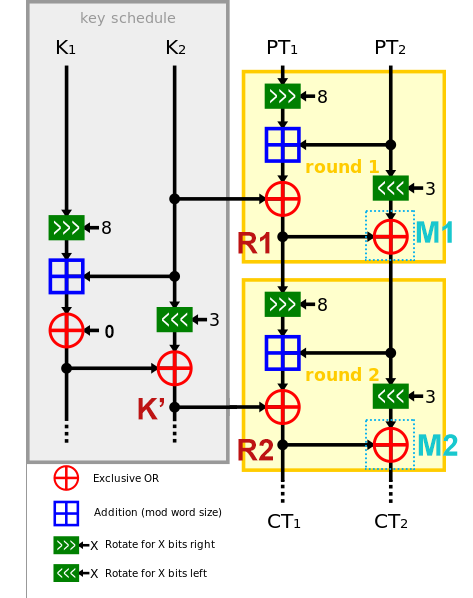}
    \end{center}
    \caption{\label{f:powerpoints}Power capturing points for Speck Speck}
  \end{figure}  
 
 The first step as shown in \autoref{f:specksteps} is collecting of power traces. This set of power traces must
 include power consumption of reading or writing of an intermediate value that is based on the right half of the key K2. 
 The point marked in \autoref{f:powerpoints} as R1 which is the result that comes after mixing K2 with 
 transformed plain text is such a good point. We suggest that you measure power during the \textit{xor} operation M1 as marked 
 in \autoref{f:powerpoints}. When power is measured across this operation by setting the trigger explained in \autoref{s:trigger}
 to go high at the beginning of this M1 operation and then to go down at the end of M1 operation it will be possible to 
 capture the power consumption that would happen when loading the value at R1 from memory. 
 
 After the power traces are collected, as the  second step, the CPA algorithm should be run by giving those power traces as the input.
 Of course the plain text used for collecting power also should be given as input though it is not shown in \autoref{f:specksteps}.
 The CPA implementation used for AES cannot be used here. Instead some modifications must be done such that it comply with 
 Speck. For example changes must be done to the selection function
 such that now it calculates the intermediate value at R1. Also as we are attacking half of a 128 bit
 key, now the algorithm no longer has to do calculations for 16 bytes but just 8 bytes. Once the CPA algorithm is run it will output 
 you the right half of the key K2.
 
 Then the third step as shown in \autoref{f:specksteps} is the capturing of other set of power traces. This set of traces
 should include power consumption of reading or writing of a value that is based on the left half of the key K1. The point 
 marked as R2 in \autoref{f:powerpoints} is such a good point. This is the first place in the algorithm where
 K1 is mixed with plain text but note that K1 is already mixed with K2. We suggest that you measure power 
 across the M2 operation shown in \autoref{f:powerpoints} similar to a way that was described for M1.
 
 As the fourth step, the CPA algorithm should be again run by giving the new set of power traces as the input.
 The CPA implementation used at step 2 cannot be used here directly as now the intermediate value is now R2 but not R1. 
 Therefore that implementation should be changed appropriately and run to get
  the round key at the position marked as K' in \autoref{f:powerpoints} as the result.
 
\begin{figure}[htb]
    \begin{center}
      \includegraphics[width=4cm]{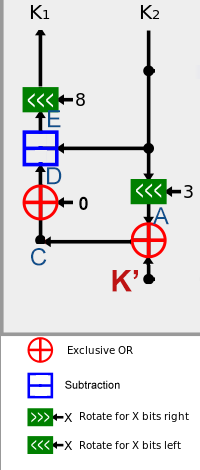}
    \end{center}
    \caption{\label{f:reverse}Deriving left half key from the round key}
  \end{figure}

 The final step is to derive original left half of the key K1 using the round key. For this, the round key at K' must be applied 
 with reverse operation until K1 is obtained. \autoref{f:reverse} denotes the order of reverse operations. It is well 
 known that the reverse operation of \textit{xor} is \textit{xor} it self. Therefore by doing an \textit{xor} between K' and the
 left rotated K2 at A, value at C can be found out.  The \textit{xor} of C with 0 is C itself and therefore value C is equal to D.
 Then instead of addition subtraction must be done. By subtracting K2 from D value E would be found out.
 Here care must be taken when D is less than K2 as such value for D is a result of a overflow that happened when finding 
 D using E and K2 at addition during the key generation. Therefore in such situation we should subtract K2 from D plus removed carry during addition.
 The reverse operation of rotate right is rotate left and by doing that K1 can be derived successfully.
 
 Now as shown in \autoref{f:specksteps} we have both K1 and K2 where by concatenating them the correct key can be found out.
 
 \section[Experiences and difficulties]{\label{s:firstattempt}Experiences and difficulties faced when attacking Speck} 
 
 This section describes the steps we sequentially followed when formulating the attack on
 Speck together with the various experiences we got. Further, various problems and failures 
 we came across are elaborated together with the solutions that
 were found.
 
 \subsection{\label{s:First}First attempt of the attack}
 
 As discussed in \autoref{c:speck} we already had created a working Speck Cryptosystem based on an 8 bit PIC18F2550 microcontroller
 and therefore an attack could be attempted. Just as in attacking AES, first a set of plain text were generated. Now the challenge
 was to identify an appropriate selection function. For AES this challenge was not there as the selection function for AES was
 well known. But Speck being a new algorithm did not have any previous research on power analysis and therefore the 
 selection function had to be identified ourselves. Due to the differences in Speck algorithm
 when compared to AES which were discussed in \autoref{s:keygeneration}, a new
 approach had to be followed where two sets of power traces were separately taken
 and each half of the key had to be attacked separately as discussed in \autoref{s:spechhow} 
 
 Since the intermediate value R1 in \autoref{f:powerpoints} was the one that was identified by us as a good place to attack the right half of the key K2,
 power traces were taken such that it included the \textit{xor} operation that resulted R1 (not the M1 \textit{xor}
 operation shown in \autoref{f:powerpoints} but the \textit{xor} operation before that). How the power trace capturing is
 automated using Matlab was discussed well in \autoref{s:automation} and therefore it is not repeated here.
 The next task was to modify the CPA algorithm written CUDA C such that now it can derive the key of Speck from a set of power traces rather 
 than AES. Because now we are attacking one half of the key which is just 64 bits it has only 8 bytes. The CPA algorithm for 
 AES was to attack a full 128 bit (16 byte) key and therefore necessary changes were done such that now it only did calculations
 for 64 bits(8 bytes). Further due to the fact that the selection function in Speck is very different from AES, major changes were done 
 to the CPA code such that the intermediate value R1 is calculated to be used for finding the hypothetical power values.
 
 Then the CPA code was run on the power traces expecting to get the correct key, but unexpectedly an all zero key came as the result
 while the key we used was not that. This issue, the reason behind and how we solved it is described in 
 \autoref{s:zerokey}.

  \subsection{\label{s:zerokey}Zero key issue}

  Getting all zeros as the result of the CPA algorithm for any key used for encryption was surprising. 
  We did various attempts such as increasing the number of power traces but the result was still
  the same. Therefore we got a suspicion whether this issue is due to some special characteristic of Speck algorithm which is not in AES.
  The power capturing in Speck was done over a single \textit{xor} operation while in AES this was done over a combination
  of a \textit{xor} and a \textit{sbox} lookup and it was guessed
  whether this difference was the cause of the issue. Fortunately AES also has \textit{xor} operations and therefore we 
  decided to run attacks on AES while using only \textit{xor} operations.
  
  As the \textit{AddRoundKey} which does \textit{xor} at the beginning of 
  the encryption in AES is also an operation which is a function of the plain text and the key, it was selected as 
  the selection function. As the power traces we initially took for \autoref{s:aessteps} also included writing and 
  reading of the value obtained after the \textit{AddRoundKey}, the same traces could be still used for the attack.
  Let's call this set of power traces as \textit{first set of power traces} during the rest of the subsection.
  Therefore in the CPA code, modifications
  were done to change the selection function and after running the algorithm, 
  for the most surprise the correct key could be still successfully obtained for AES despite the selection function used. 
  
  Even though the selection function was changed, the power measurements still included the \textit{sbox} as well.
  We though this might be the reason why the attack on AES was successful despite the selection function. 
  A test was carried out where now the power capturing was taken just only for the \textit{AddRoundKey} operation which is a \textit{xor} operation. 
  That is, the trigger was set to go high at the beginning of \textit{AddRoundKey} and it was set to low after the operation.
  The difference with the previous traces is that now the traces do not include power consumption during \textit{sbox} lookup.
  Let's call this the \textit{second set of power traces}.
  Now the CPA algorithm with \textit{AddRoundKey} as the selection function is expected to return the correct key
  because the traces still include the moment when the result after \textit{AddRoundKey} operation is written to the memory.
  But after running the algorithm we figured out that we always get all zeros as the result despite of what is set as the key in the 
  cryptosystem.
  
  With that result we could decide that when measuring power only for the \textit{xor} operation some problem is caused.
  As this was an unexplainable result it was decided to investigate further by plotting how the correlation
  coefficient changed with the time (sampled time in power traces) for each possibility of each keybyte for each selection function
  . We guessed that by comparing the plots for each selection function it would be possible to get a hint regarding the reason.
 
  As there are 256 possibilities per each keybyte where there are 16 separate keybytes, 4096 correlation versus time datasets had to
  be generated while running the CPA algorithm. For this, the CPA code was modified such that it printed
  these necessary information to separate files. But as this raised a necessity of retaining large amount of data
  in the memory we had to do some more changes to the code such that now the analysis was done part by part while
  transferring calculated portions of data from memory to disk. First the modified CPA code was run while using
  the the combination of \textit{xor} and \textit{sbox} as the selection function. Then the modified CPA code was
  run while using just \textit{xor}  as the selection function. For both situations, the \textit{first set of power traces} was used.
    
  After generating the two data sets which each was several gigabytes in size, a Matlab script 
  was written to to draw the plots. The script was run separately on the two datasets to get
  two sets of 4096 graphs. From these graphs necessary graphs were selected to do the comparison.
 
    \begin{figure}[htb]
    \begin{center}
      \includegraphics[width=12cm]{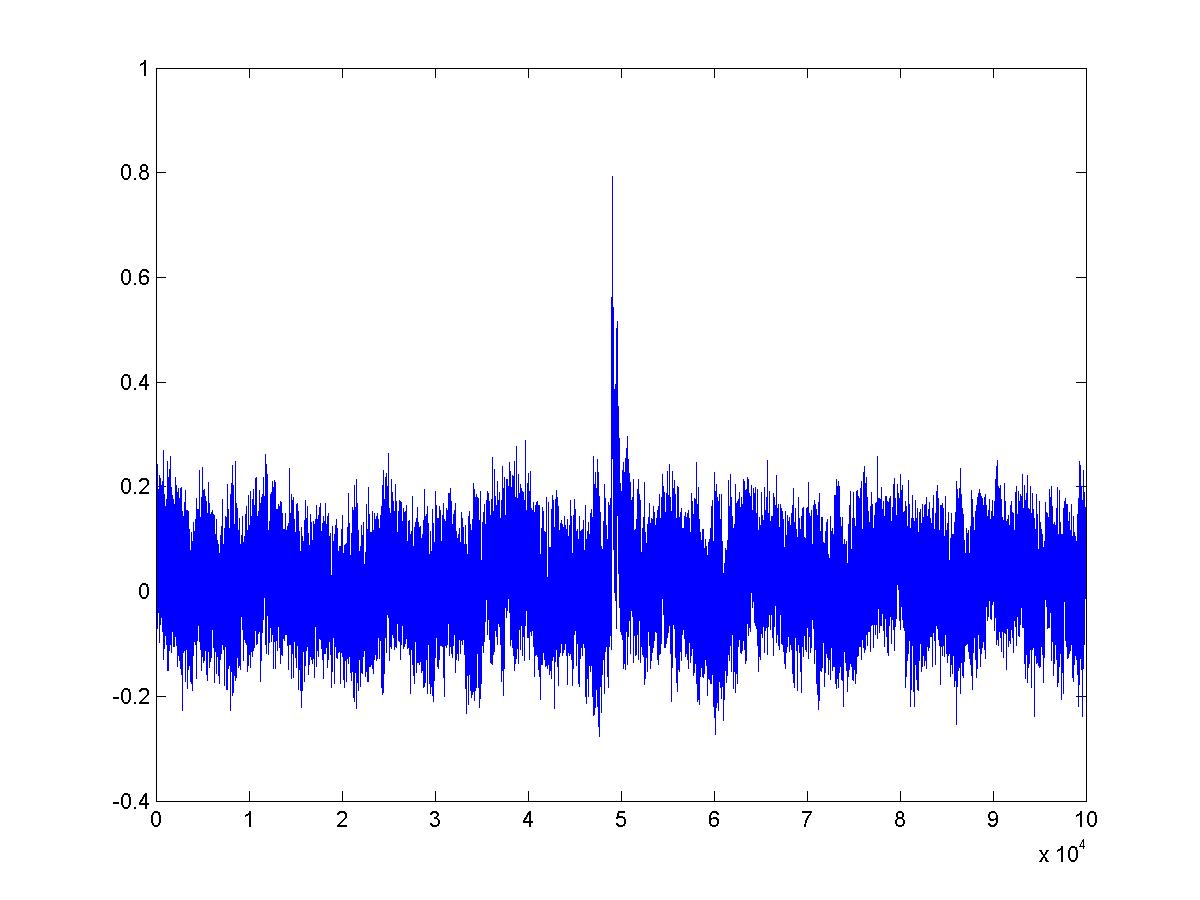}
    \end{center}
    \caption{\label{f:sboxcorrect}The correlation vs time graph for a correct key when using \textit{sbox} lookup as the selection function}
  \end{figure} 
  
     \begin{figure}[htb]
    \begin{center}
      \includegraphics[width=12cm]{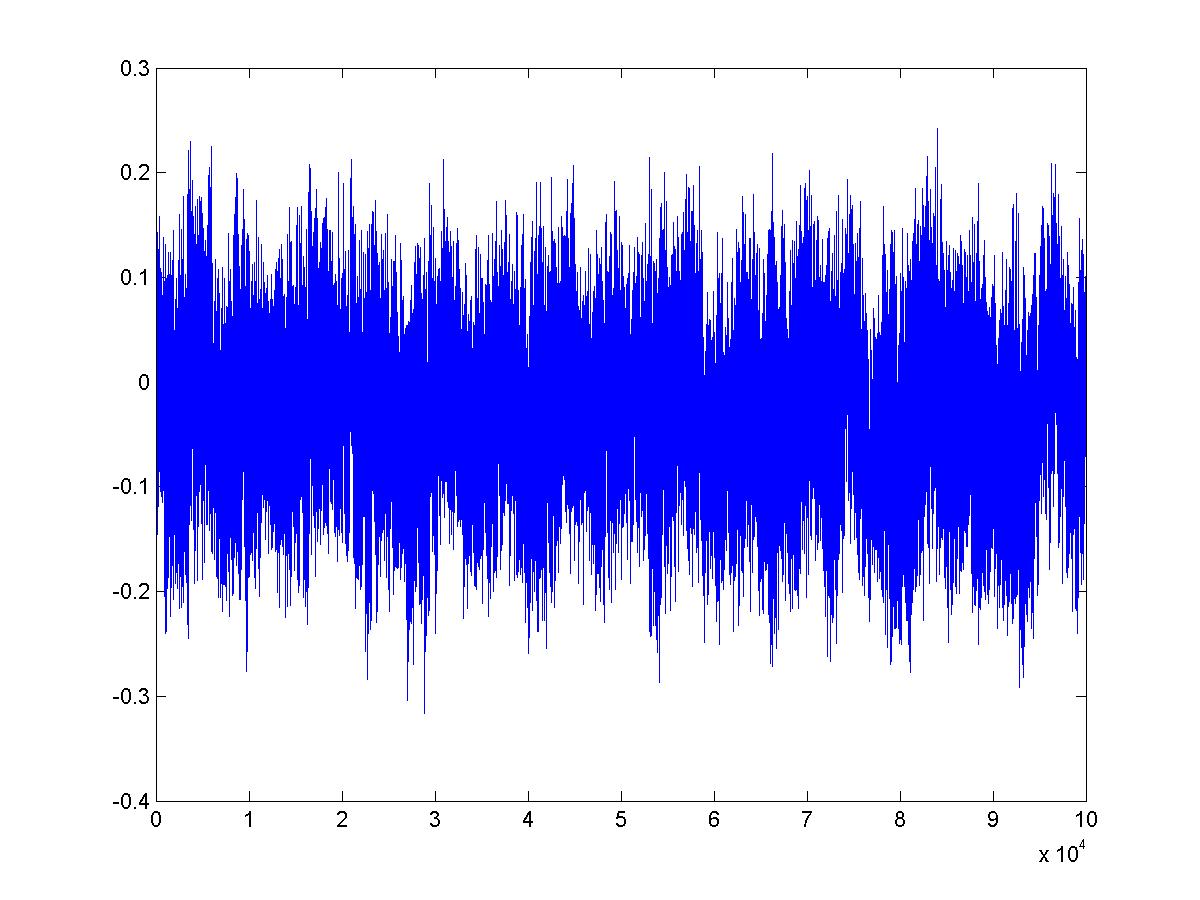}
    \end{center}
    \caption{\label{f:sboxwrong}The correlation vs time graph for a wrong key when using \textit{sbox} as the selection function}
  \end{figure}

  \autoref{f:sboxcorrect} is the correlation vs time graph for the correct key guess for a certain keybyte when \textit{sbox} lookup is used as
  the selection function (most precisely combination of \textit{xor} and \textit{sbox} look up is the selection function).
  The x axis represents the time. Actually this is the sample number of the power traces. As our power traces had 100 000
  sample points x axis range from 1 to $10 \times 10^{4}$. The y axis is the correlation coefficient for the considered key guess.
  As you can see there is a sudden peak in the middle. This is the place where the result from the \textit{sbox} lookup is written to the memory.
  Correlation coefficient goes high because the real power consumption at this moment matches the hypothetical
  power consumption values that we calculated. \autoref{f:sboxwrong} shows  a similar graph but for a wrong key guess.
  There we do not see any sudden peak in the correlation values because there is no such correlation
  between the real and hypothetical power values.

    \begin{figure}[!htbp]
    \begin{center}
      \includegraphics[width=12cm]{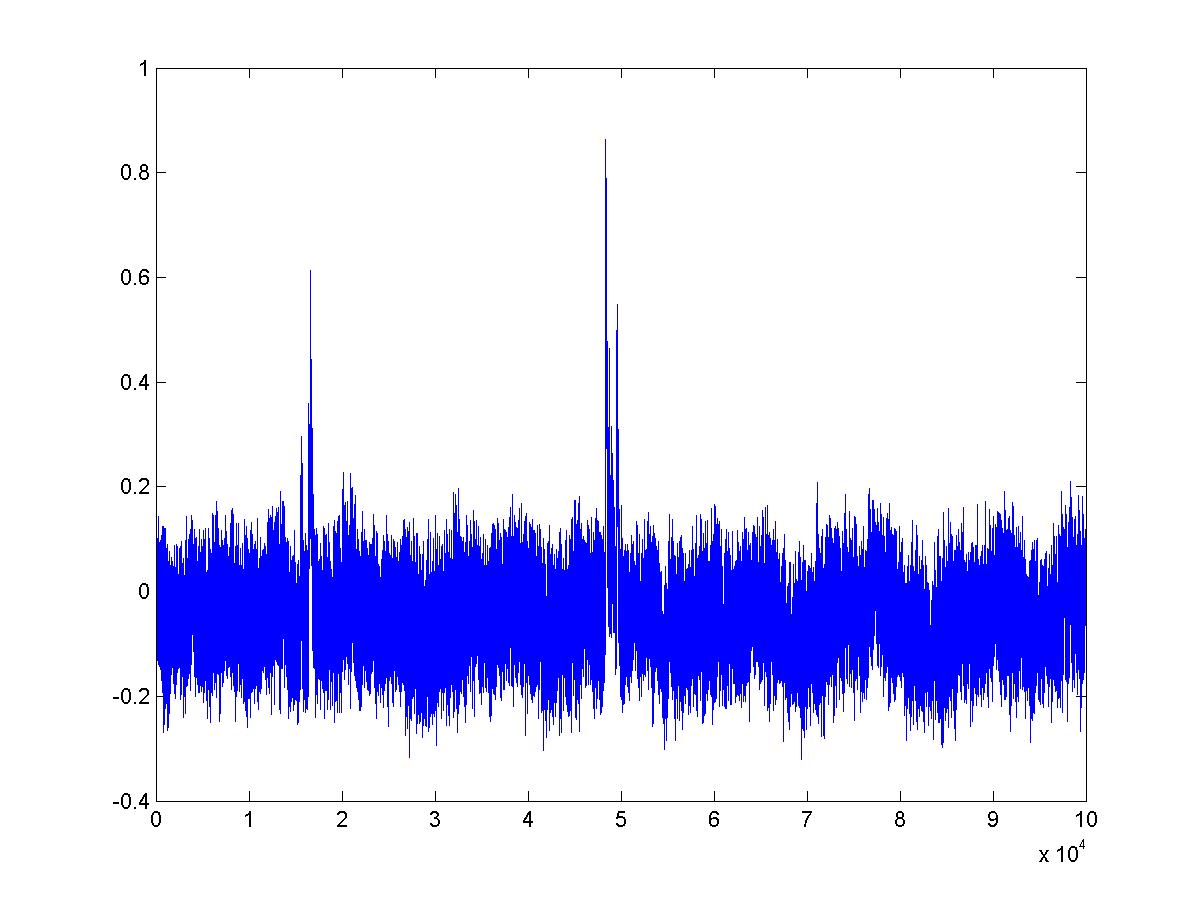}
    \end{center}
    \caption{\label{f:xorcorrect}The correlation vs time graph for a correct key when using \textit{xor} as the selection function}
  \end{figure}   
  
   \begin{figure}[!htbp]
    \begin{center}
      \includegraphics[width=12cm]{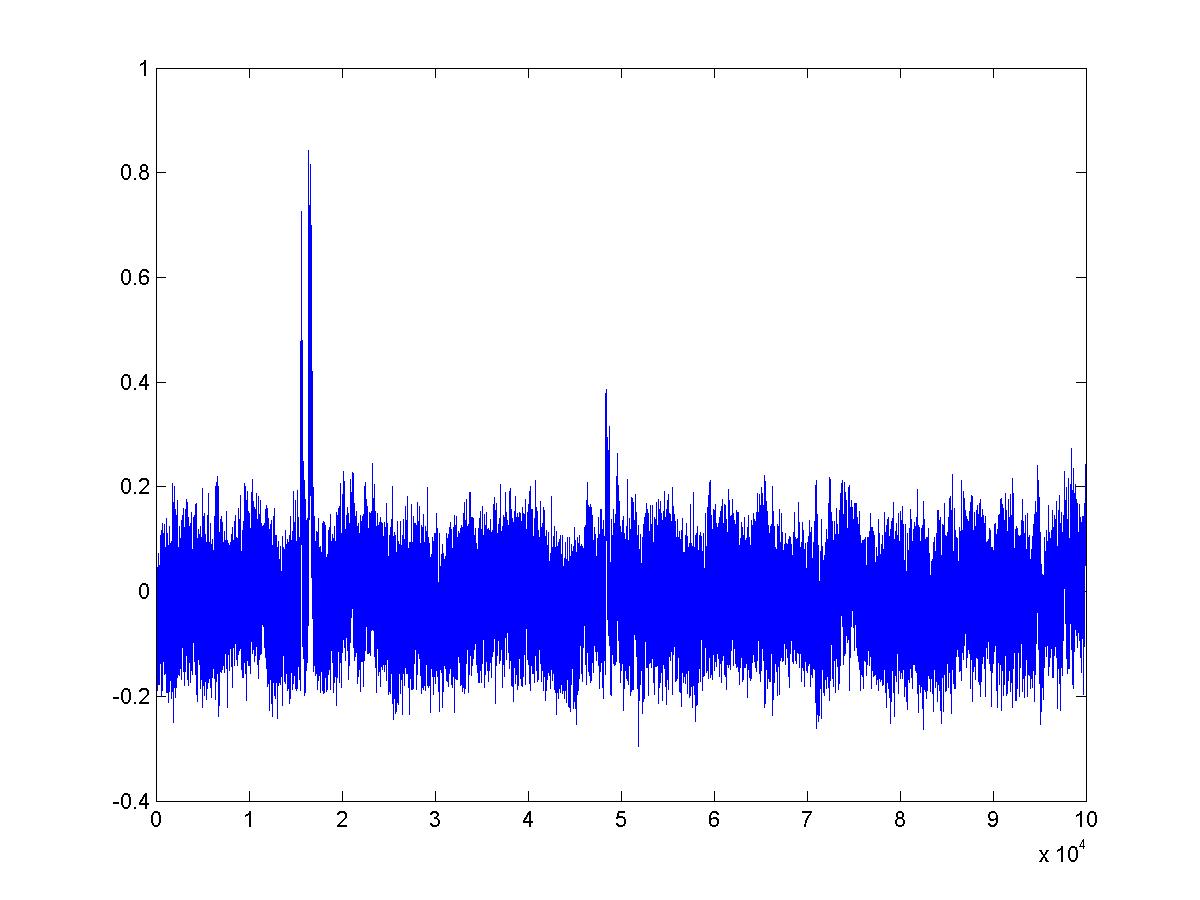}
    \end{center}
    \caption{\label{f:xor0issue}The correlation vs time graph for the zero key guess when using \textit{xor} as the selection function}
  \end{figure}

  \autoref{f:xorcorrect} is the graph drawn for the correct key guess for the same keybyte as the previous case but now \textit{xor} is the selection
  function used. There two peaks are visible. The left peak is when the result of the \textit{xor} operation (let's call this value $R$) is written to the memory.
  When the \textit{xor} operations of all keybytes finish, then starts the \textit{sbox} lookup and $R$ has 
  to be again loaded from memory. The peak in the right side is due to this loading of $R$, to be used for the \textit{sbox} look up.
  If you consider the location of this right peak with the single peak in \autoref{f:sboxcorrect}, it can be noted that they
  lie very near. The reason is that because the loading of $R$ for \textit{sbox} lookup from memory and then writing
  the result of \textit{sbox} lookup to the memory happen just after the other.
  
  The graphs for a wrong key guess when \textit{xor} is used as the selection function still looks very same to \autoref{f:sboxwrong} where there are no peaks.
  But for key guess 0 though it is a wrong key, still it has some peaks as shown in \autoref{f:xor0issue}. This high correlation
  should be the reason that falsely returned zero as the result by the CPA algorithm for the \textit{second set of power traces} we used. In this second
  set of power traces we only captured power for the \textit{xor} operation and therefore the right peak in \autoref{f:xorcorrect}
  would be no longer there now as those samples for the \textit{sbox} lookup is no longer there. Therefore if you consider the 
  left peak in \autoref{f:xorcorrect} (correlation when writing $R$ to the memory)
  is less than the significant peak that is occurring in \autoref{f:xor0issue}. Therefore 0 is returned as the 
  most possible key. If you observe the location of the significant peak in \autoref{f:xor0issue} carefully
  you can see that this lies just before the left peak in \autoref{f:xorcorrect}. As the peak in \autoref{f:xorcorrect} 
  was corresponding to the writing of result after the \textit{xor}, we can deduce that the significant peak in \autoref{f:xor0issue} is the moment when
  plain text is loaded from memory in order to be subjected to \textit{xor}. 
  
  When a number is subjected to \textit{xor} with 0 the result is
  the number itself. Therefore if plain text is subjected to \textit{xor} with 0 the result is the plain text itself.
  When we use \textit{xor} as the selection function, for key guess 0 the CPA algorithms correlates the power consumption during
  loading of plain text memory to the hypothetical power consumption calculated for the \textit{xor} of plain text with key guess 0.
  Finally this was the reason that makes a peak for 0 key guess.

  Since now the cause for the zero key issue was known we decided to do another attempt which is going to discussed in 
  \autoref{s:second}.
   
  \subsection{\label{s:nummoretraces}The need of more number of traces} 
 
  While on the path to solve the zero key issue as explained in \autoref{s:zerokey}
  it could be noted that the difference between the correlation coefficient of the best key match and the correlation
  coefficient of the second key match
  is less when using \textit{xor} as the selection function. As previously discussed in \autoref{s:cpaaes} having a larger difference
  means that the possibility of the best key match becoming the correct key is high. \autoref{t:selectf} shows this
  difference value for each 
  selection function. We have not included the key guesses or the correlation coefficient of each key guess
  as for our comparison just the difference values are adequate. Also out of the 16 keybytes only the first 8 is shown here. The row named \textit{xor} is for the selection function 
  \textit{xor} while the row named \textit{sbox} is for the selection function consisting of combination of 
  both \textit{xor} and \textit{sbox}. For both selection functions the 
  same set of power traces have been used.
  Here as you can see, when \textit{sbox} is used the difference value is around 0.5.
  This is a quite significant difference to deduce that the result returned by the CPA algorithm is the correct one with much confidence.
  But the difference value when only \textit{xor} is used as the selection function is even lesser than 0.1 most of the time. 
  Therefore here we cannot say with much confidence that the best key match is the correct key. Therefore in order to 
  get a more significant difference value it would be necessary to take more power traces. 
  Therefore when \textit{xor} is used as the selection function more number of power traces has to be taken than when using \textit{sbox}.
  
  Therefore at this point we learnt that for breaking Speck the number of power traces that should be taken is
  higher than what is needed for AES because in Speck we do not have \textit{sbox} lookups.
  
  \begin{table}
  \begin{center}
  \begin{tabular}{|l|l|l|l|l|l|l|l|l|}\hline
  \textbf{Keybyte number} & \textbf{0} & \textbf{1} & \textbf{2} & \textbf{3} & \textbf{4} & \textbf{5} & \textbf{6} & \textbf{7}\\\hline
  \textbf{xor} & 0.122 & 0.051 & 0.043 & 0.049 & 0.029 & 0.035 & 0.030 & 0.073\\\hline
  \textbf{sbox} & 0.516 & 0.498 & 0.555 & 0.485 & 0.554 & 0.626 & 0.507 & 0.514\\\hline
  \end{tabular}
  \end{center}
  \caption{\label{t:selectf}The effect of the selection function on the difference between the the highest and the second highest correlation coefficient}
  \end{table} 
  
  One of the reasons for this necessity of larger number of power traces when using \textit{xor} for the attack rather 
  than \textit{sbox}, can be explained by comparing the confusion property \textit{sbox}.
  In cryptography the idea of confusion is used to hide the relationship between the key and the cipher text\cite{forouzan}.
  High confusion means that if a single bit in the key is changed, many bits in the cipher text are changed.
  The technique used in block ciphers to increase the confusion is the \textit{sbox} lookup. 
  In power analysis the key is found by comparing the real power consumption values to the hypothetical power consumption
  values calculated by the power model. When an \textit{sbox} is there a change in one bit of the key would change lot of bits
  in the resultant value and hence when calculating hypothetical power consumption values a keyguess with even one bit
  change to the correct key would give a very different hypothetical power value than the real power consumption value. Therefore correlation
  for the correct key would be significantly higher than for the key guesses that are even closer to the correct key, making it possible
  to derive the correct key with less number of power traces. In contrast, in an operation like \textit{xor} a change of one bit in the key 
  would maximally change only one bit in the resultant value and therefore it makes it harder to distinguish hypothetical 
  power consumption values for the keys closer to the correct key. Because of that larger number of power traces are required.
  
  Though the original purpose of \textit{sbox} and the concept of confusion itself was to make a cipher more secure it seems  
  that for power analysis having a \textit{sbox} becomes an advantage to  the attacker.
 
  \subsection{\label{s:second}Second attempt of the attack}
  
  Based on that experience got by the analysis done in the previous subsections we did a second attempt
  to attack Speck where we became successful at the end.
  As explained in \autoref{s:zerokey} in order to get the correct key, power should
  be measured such that it does not include the loading of the value to be subjected to  \textit{xor}, but only include the moment
  when writing the result of \textit{xor} after the operation. One solution may be to preprocess the power
  traces such that the loading part is trimmed but as \textit{xor} operation happen eight number of times
  per each keybyte finding the exact locations is difficult.
  
  Then we came up with a solution which used a different power measurement point that still made it possible 
  to use the same old selection function. The result  at R1 in \autoref{f:powerpoints} which is the result after the \textit{xor} operation is
  of course loaded back from memory later when the algorithm starts the next encryption round (2\textsuperscript{nd} round). Therefore if
  power was measured when this loading happen, that is during M1 operation in \autoref{f:powerpoints}, then the problem would be solved. 
  The reason is that now we do not have the loading of the 
  value before \textit{xor} which resulted in the troublesome correlation. After new power traces were taken
  based on this approach the correct key could be successfully derived after running the CPA algorithm.
  
  Since the right half key of was found out successfully now the next step was to find the left half key.
    As it was discussed in \autoref{s:keygeneration}, a round key is generated by doing
  various modifications to the left half key including a mixing with the right half key. It is this round key that is
  subjected to \textit{xor} with modified plain text to get an intermediate value at R2 in \autoref{f:powerpoints}.
  Therefore a new set of power traces must 
  be collected that includes writing of this value R2. Based on the experience with zero key issue
  power was measured across the operation M2 in \autoref{f:powerpoints}
  
    On the power traces obtained, CPA algorithm was run with right half key as an input
  to derive the round key. But of course the final objective is to find the original key but not a round key. Therefore the round key 
  must be reversed by applying the reverse operations backwards. By applying those reverse operations as elaborated in \autoref{s:spechhow}, 
  the original left half key could be found out.
  This key reversal functionality was added into the CPA code so that it directly printed the original key.
 
  \section[Speck vulnerability]{\label{s:speckvul}Vulnerability of Speck for power analysis}
  
  Speck, even though it is a recent encryption algorithm targeted mainly for embedded systems it is still vulnerable to
  power analysis attacks. Embedded systems are the target of power analysis attacks and therefore countermeasures
  must be implemented in any such system that is going to use Speck as the encryption algorithm.
  
  Although Speck is vulnerable to power analysis, when compared with the power analysis on AES, the effort needed for the attack on
  Speck is high. \autoref{t:aessuccesstime} shows the time consumption for each of the steps in attacking AES 
  while \autoref{t:specksuccestime} shows the time consumption for each step in attacking Speck (8 bit implementation). Data in both tables are
  for experiments done in same conditions except the algorithm. As you can see attacking AES involves two steps
  while attacking Speck involves four steps. The reason is that when attacking Speck each half of the key must be recovered
  in two separate phases and therefore
  capturing power  and running the algorithm is done twice shown as round 1 and round 2 in \autoref{t:specksuccestime}.
  The need for carrying out two rounds instead of one obviously increases the effort necessary as well as the time taken.
  Further if the values in the tables are observed carefully it is notable that time taken for power collection for
  AES took about 610 seconds while it took around 1570 seconds for the same task for Speck. Not only power
  collection but also the time for running CPA algorithm is similarly higher for Speck than AES.
  This is because for Speck more power traces had to be taken than for AES. For example for the attacks shown here
  200 traces only were taken for AES while 500 traces had to be taken for Speck (per each round) for successfully
  deriving the key. The necessity of more number of traces for Speck is due to the fact that it does not have
  substitution box look ups like in AES.
  
  Discussed results show that though Speck is somewhat less vulnerable than AES with respect to power analysis still
  good countermeasures are needed as it is possible to break Speck in a time less than 1 hour.

  {%
  \begin{table}
  \newcommand{\mc}[3]{\multicolumn{#1}{#2}{#3}}
  \begin{center}
  \begin{tabular}{|l|l|}\hline
  \textbf{Step} & \textbf{Time taken / s}\\\hline
  Collecting power traces & \mc{1}{r|}{613.22}\\\hline
  Running CPA & \mc{1}{r|}{12.29}\\\hline
  \textbf{Sum} & \mc{1}{r|}{\textbf{625.51}}\\\hline
  \end{tabular}
  \end{center}
  \caption{\label{t:aessuccesstime}Time taken for a successful attack on AES}
  \end{table}     
  }%
 
  {%
  \begin{table}
  \newcommand{\mc}[3]{\multicolumn{#1}{#2}{#3}}
  \begin{center}
  \begin{tabular}{|l|l|}\hline
  \textbf{Step} & \textbf{Time taken / s}\\\hline
  Round 1: Collecting power traces & \mc{1}{r|}{1566.25}\\\hline
  Round 1: Running CPA  & \mc{1}{r|}{28.97}\\\hline
  Round 2: Collecting power traces  & \mc{1}{r|}{1572.05}\\\hline
  Round 2: Running CPA & \mc{1}{r|}{28.63}\\\hline  
  \textbf{Sum} & \mc{1}{r|}{\textbf{3195.90}}\\\hline
  \end{tabular}
  \end{center}
  \caption{\label{t:specksuccestime}Time taken for a successful attack on Speck 8 bit implementation}
  \end{table}     
  }%

 \section[Attack for 16 bit]{\label{s:speck16attack}Attacking Speck on 16 bit microcontroller}
 
  As it was mentioned in \autoref{s:cpa} when we are applying the CPA algorithm it is done separately on subkeys which are parts
  of the whole key. During all the situations discussed before including attacks on both AES and Speck
  the subkeys were always 8 bits (1 byte) in size. When performing AES whether the microcontroller is 8 bit, 16 bit, 32 bit or whatever,
  the substitution box lookups are always happening byte by byte. Therefore for any microcontroller, AES can be attacked using CPA 
  by attacking on each byte of the key.
  
  But Speck on the other hand which comprises of only \textit{add}, \textit{rotate} and \textit{xor} operations does not have any byte
  wise operation. If you consider an 8 bit microcontroller these operations would of course happen byte wise
  because registers are anyway 8 bits. In \autoref{s:speck8bit} we described how we did the implementation
  of Speck for an 8 bit microcontroller and there you would have noticed that all operations were done byte wise using loops.
  Even if a compiler that supports larger integers is used for an 8 bit microcontroller, the compiler would
  still use the same approach in assembly level. Because of this it can be thought that CPA attack on Speck which attacked each byte 
  separately would have only worked only because the implementation was 8 bit. The result of the attack is therefore 
  to be checked in 
  a situation when 16 bit or higher microcontrollers are used.
  
  For a 16 bit  microcontroller a straight forward method to attack would be to attack using
  CPA on subkeys which are 16 bits in size. Since registers are 16 bits all operations would happen in terms of
  values which are 16 bit in size and therefore the attack will definitely work. We tried this approach by using the 
  16 bit implementation of Speck done for PIC described in \autoref{s:speck16bit}. Here the CPA code had to be modified
  considerably such that it now attacked 2 bytes at a time than 1 byte at a time. But this analysis took
  considerably larger time when compared to a byte wise attack as now a 16 bit number which has 65536 possibilities
  increases the number of computations necessary. Therefore if the microcontroller was 32 bit, the number of possibilities
  is much more and the time for a CPA attack also would be high. Therefore the approach of
  increasing the size of a subkey is not going to work for microcontrollers with large registers.
  
  The next approach was to check the feasibility of a byte wise CPA attack on a 16 bit microcontroller.
  Though operations happen 16 bits at a time, still there should be some correlation between the total power 
  consumption for the 16 bits and the power consumption for 8 specific bits that we try to attack at a time. 
  When more and more power traces
  are taken for different plain text samples, the power consumption for the rest of the bits except the specific
  8 bits that are attacked at a time would be treated as 
  noise\cite[pp.~138-141]{mangard} automatically by the CPA algorithm. Taking this to consideration we did a byte wise CPA analysis for a set of power
  traces captured when encrypting using the 16 bit microcontroller. It was figured out that it was still possible to
  recover the key successfully. But the number of power traces needed was much higher than the number necessary for
  the attack on 8 bit microcontroller. But still the complete attack was possible in several hours.
  
  When a 32 bit microcontroller is considered it should be still possible to do a byte wise CPA attack
  because now 24 other bits apart from the 8 bits we are targeting at a time for the attack can be considered noise.
  But more power traces
  would be definitely needed for such. Therefore it can be deduced that the more number of bits the microcontroller is, the more number of power traces
  are needed. Capturing more number of power traces takes time while the execution time for CPA algorithm
  also takes more time when more traces have to be analysed. But still a time period like 1 day or even 1 week 
  is still a feasible time for an attacker and hence
  the vulnerability is there even for microcontrollers with large registers.
  Therefore countermeasures are still needed despite the register size.
   
  \chapter[Circuit Countermeasures]{\label{c:hardware}Circuit based Countermeasures} 
  
  Circuit based countermeasures as introduced in \autoref{s:countermeasures_intro} is a type of hardware countermeasures
  that introduce new circuit components or modify existing circuit at macro level opposed to micro architecture level changes. In this chapter we test the effectiveness of
  adding power line filters to the circuit \cite{mangard} and later test the effectiveness of our own ideas.
  
    We implemented different countermeasure circuits, connected to our PIC2550 based testbed that runs AES described in \autoref{c:testbed} and attacked them to evaluate their effectiveness. We selected AES algorithm rather than Speck since AES is the most vulnerable out of the two and for evaluating countermeasures weak one is the better. From the two power measurement methods discussed in
    \autoref{s:powermeasurementcircuit} we used the Vdd resistor method as it was the method that required less number of power traces. 
    The used resistor value throughout the setups in this chapter is 100 ohms if not stated otherwise.
  
  \section[filters]{Filter based countermeasures}
  
  Electronic filters \cite{filter} take an input signal and output the signal with certain frequency components removed. A filter that removes high frequency components of a signal and passes only the low frequency components is called a low pass filter. In practical implementations unwanted frequency components cannot be completely removed, instead they are attenuated. Hence, in a low pass filter frequency components beyond a certain frequency called the cut off frequency are attenuated such that they are lesser than -3 dB. Filters that are made using passive components such resistors, capacitors and inductors are called passive filters, while ones which are made using active devices such as operational amplifiers are called active filters. A passive filter made using resistor and capacitors are called RC filters. A passive filter made using inductors and capacitors are called LC filters. A special branch of filters called power line filters are used to eliminate noise coming from the power supply to provide smooth voltage across the appliance. Line filters are usually 
  passive filters because they can filter the input power signal itself without need of an additional power supply. Therefore we
  implement some passive filters to test effectiveness of power filters introduced in \autoref{relatedcounter}.
  
  \autoref{f:trace2} showed how a power trace looked like during encryption. The power trace consists of large number of peaks and these peaks are the points that leak most of the secrets out \cite{mangard}. A waveform with lot of peaks and sudden variations has a large bandwidth. When such a signal is passed through a low pass filter to get high frequency components removed, then the peaks in the signal are flattened making the signal look smoother. \autoref{f:1000uFpic} shows a power trace that is filtered by such low pass filter. Here all the peaks found in \autoref{f:trace2} have flattened and the signal looks completely different. The power line filter based countermeasures uses this principle to minimize the leakage. Though this idea has been proposed some time back \cite{mangard}, still any work that presents the effectiveness of them is not to be found. Therefore we decided to practically implement them and test them. 
  
	The rest of this section discusses each filter we implemented, their results and finally the comparison of the effectiveness of them. 
  
  \subsection{Single capacitor}
  
\begin{figure}[htb]
    	\begin{center}
    		\includegraphics[width=8cm]{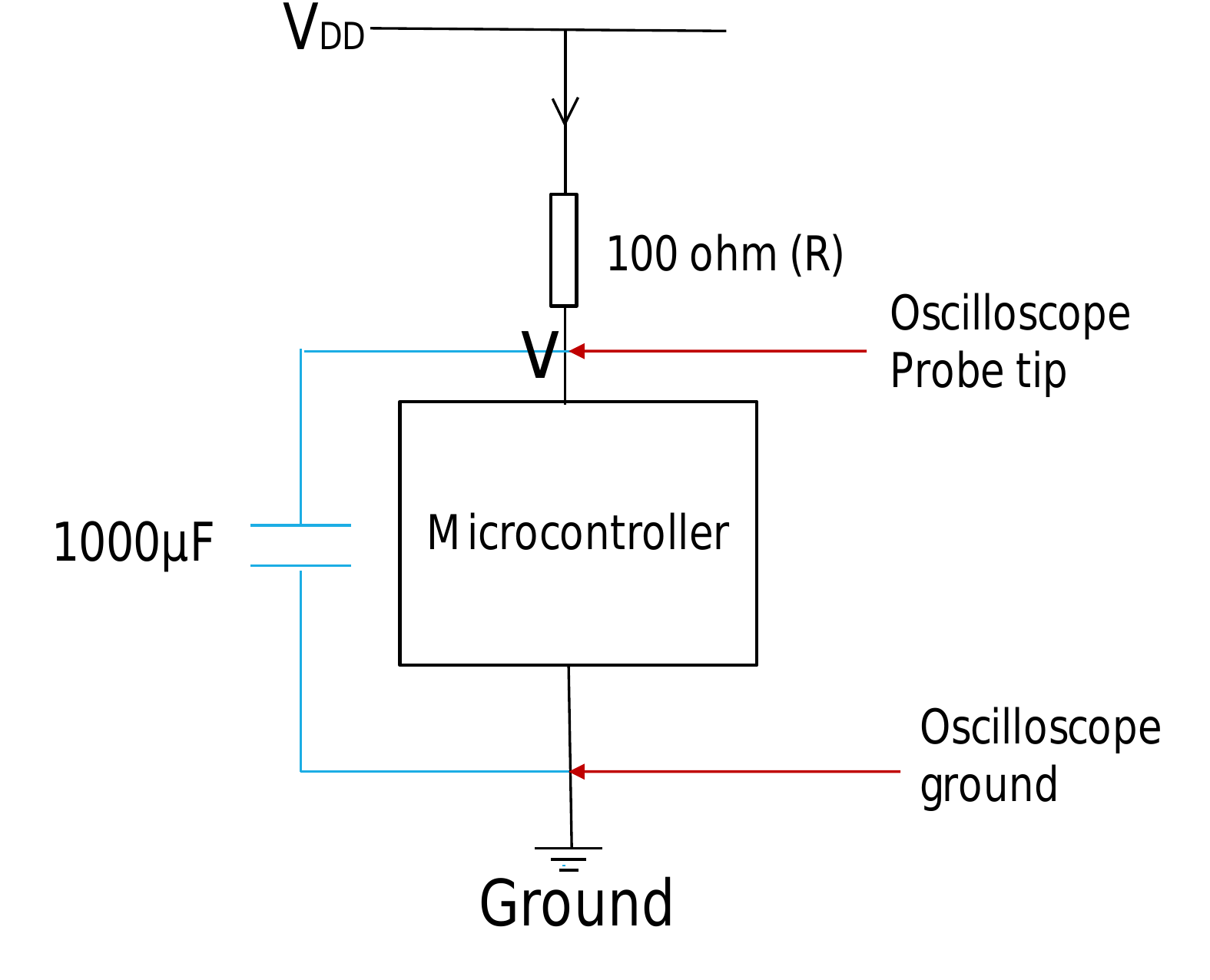}
    	\end{center}
    	\caption{\label{f:1000uF}Single capacitor connected in parallel to the microcontroller}
    \end{figure}

\begin{figure}[htb]
    	\begin{center}
    		\includegraphics[width=9cm]{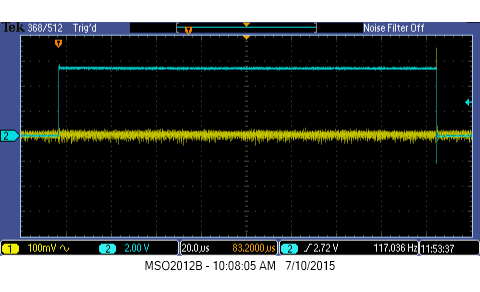}
    	\end{center}
    	\caption{\label{f:1000uFpic}Power trace when a capacitor filter is used}
    \end{figure}

  A single capacitor can be used to implement a power line filter \cite{capfilter}. Capacitor was connected in parallel to the microcontroller as shown in \autoref{f:1000uF}. Here a $1000 \mu F$ capacitor has been used.
  A capacitor when connected in parallel tries to keep the voltage across it constant by repeatedly charging and discharging. 
  In \autoref{f:1000uF}, 100 ohms resistor is the Vdd resistor for measuring power. Oscilloscope probe is connected as shown in \autoref{f:1000uF}.
  The power traces taken with the capacitor is shown in \autoref{f:1000uFpic}. When comparing with the power trace with no capacitor in \autoref{f:trace2}, it is seen that the variation in the signal is quite minimized. Because the capacitors are not ideal 
  and  therefore takes a time to charge and discharge, a perfect constant
  voltage is not observed.
  
  The capacitor connected as shown in \autoref{f:1000uF} is a low pass filter. The impedance of a capacitor is given by the following equation where f is the frequency and C is the capacitance.
  
    \[
    X = \frac{1}{2 \pi j f C}
    \]
  
  When the frequency is smaller, the impedance is larger and hence signal components with smaller frequencies are reluctant to pass through a capacitor.
  But when  the frequency is larger, impedance gets smaller and those signals pass through the capacitor. When the microcontroller in
  \autoref{f:1000uF} draws varying currents, it causes noise in the power supply line. The capacitor conducts and shorts out all the
  high pass signals, filtering out only the low pass ones. This smooth out the large variations of the signal.
  
\begin{figure}[htb]
    	\begin{center}
    		\includegraphics[width=9cm]{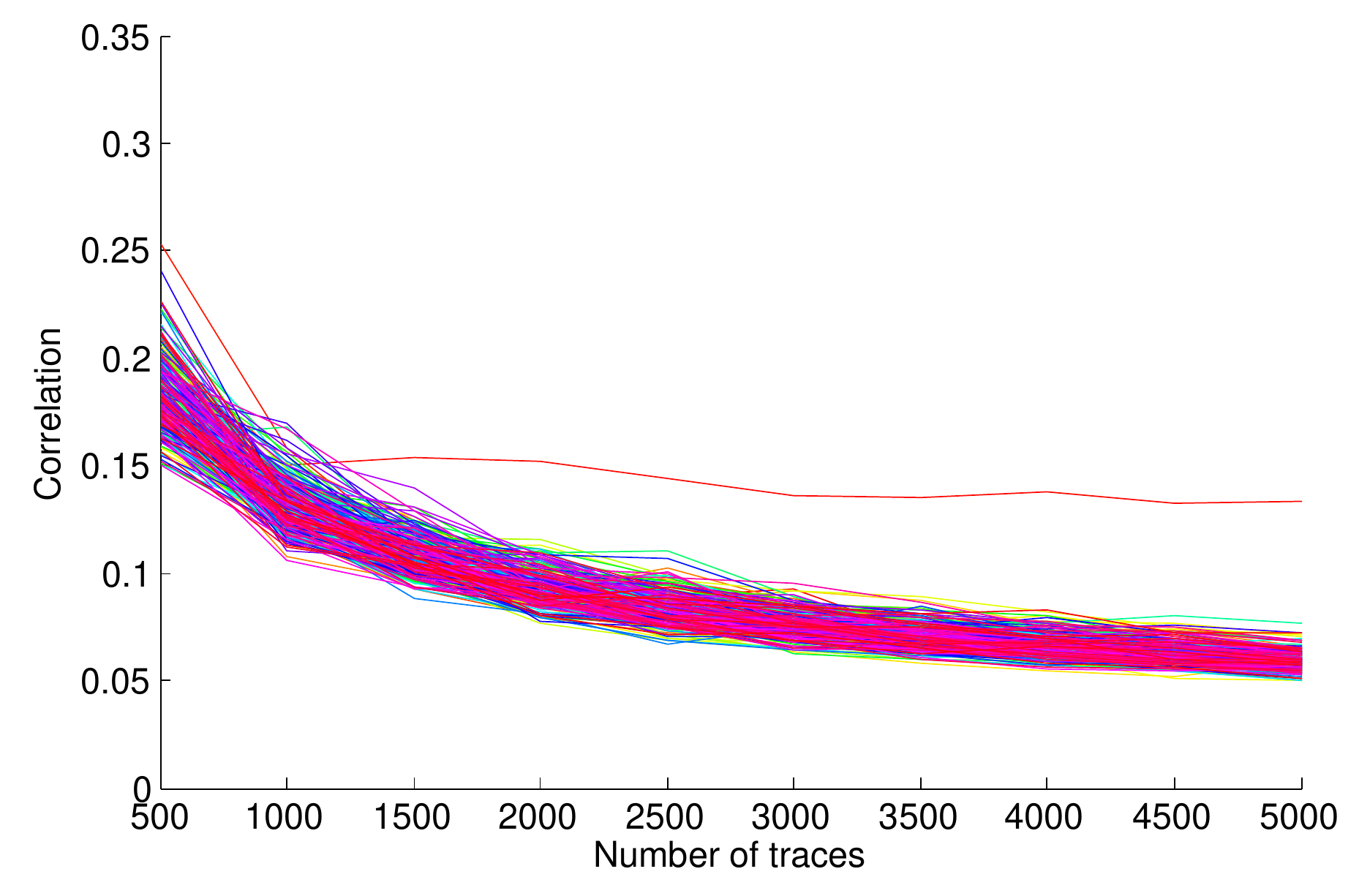}
    	\end{center}
    	\caption{\label{f:1000uFgraph}Number of power traces needed when a capacitor filter is used}
    \end{figure}

 Though the signal is affected, still there is a visible variation and this is enough to derive the key of the system. But this requires more number of power traces. \autoref{f:1000uFgraph} shows how the correlation coefficient changes with the number of power traces during an attack. We saw in \autoref{f:keybyte0} that when no capacitor is used even 50 power traces were enough to figure out the correct key.
  But now in \autoref{f:1000uFgraph} at least 1500 power traces are required to identify the correct key out of the others. But the time required to collect 1500 power traces and analyse them was only about 1.5 hours. This is still a very feasible time for an attacker.

  \subsection{Single inductor}
  
\begin{figure}[htb]
    	\begin{center}
    		\includegraphics[width=7cm]{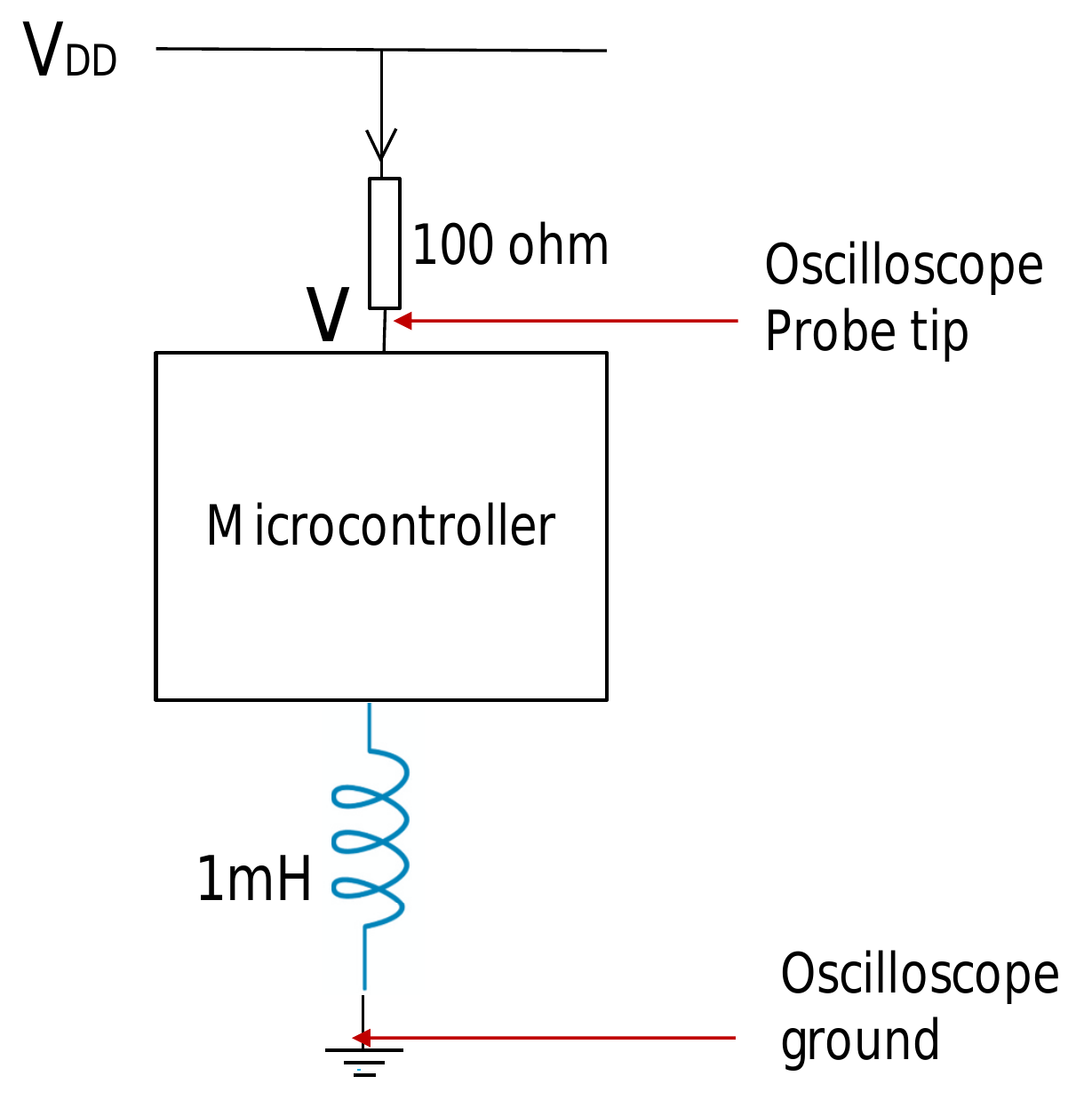}
    	\end{center}
    	\caption{\label{f:inductor}Single inductor connected serially to the microcontroller}
    \end{figure}

\begin{figure}[htb]
    	\begin{center}
    		\includegraphics[width=9cm]{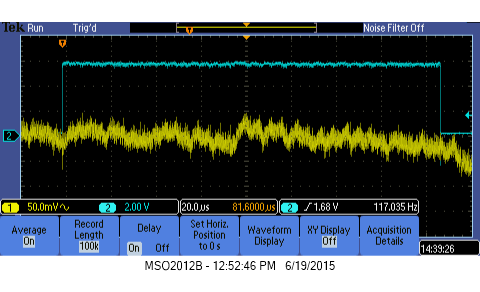}
    	\end{center}
    	\caption{\label{f:inductorpic}Power trace when a inductor filter is used}
    \end{figure}

  A one Milli Henry inductor was connected serially to the microcontroller as shown in \autoref{f:inductor}. An inductor tries to keep the current constant on the branch it is connected. This phenomenon changes the power traces to the one shown in \autoref{f:inductorpic}.
  
  The impedance of an inductor can be written as follows.
      \[
      X = 2 \pi j f L
      \]  
  
  Here, f is the frequency of the signal and L is the inductance. 
  According to this equation when the frequency increases, the impedance increases. Hence an inductor is reluctant to pass the high frequency signals through it. As the inductor is connected serially in \autoref{f:inductor}, the high frequency signals in the whole branch is
  attenuated. That is why the variation of the signal in \autoref{f:inductorpic} is not as high as when no inductor was used.
  
\begin{figure}[htb]
   	\begin{center}
   		\includegraphics[width=9cm]{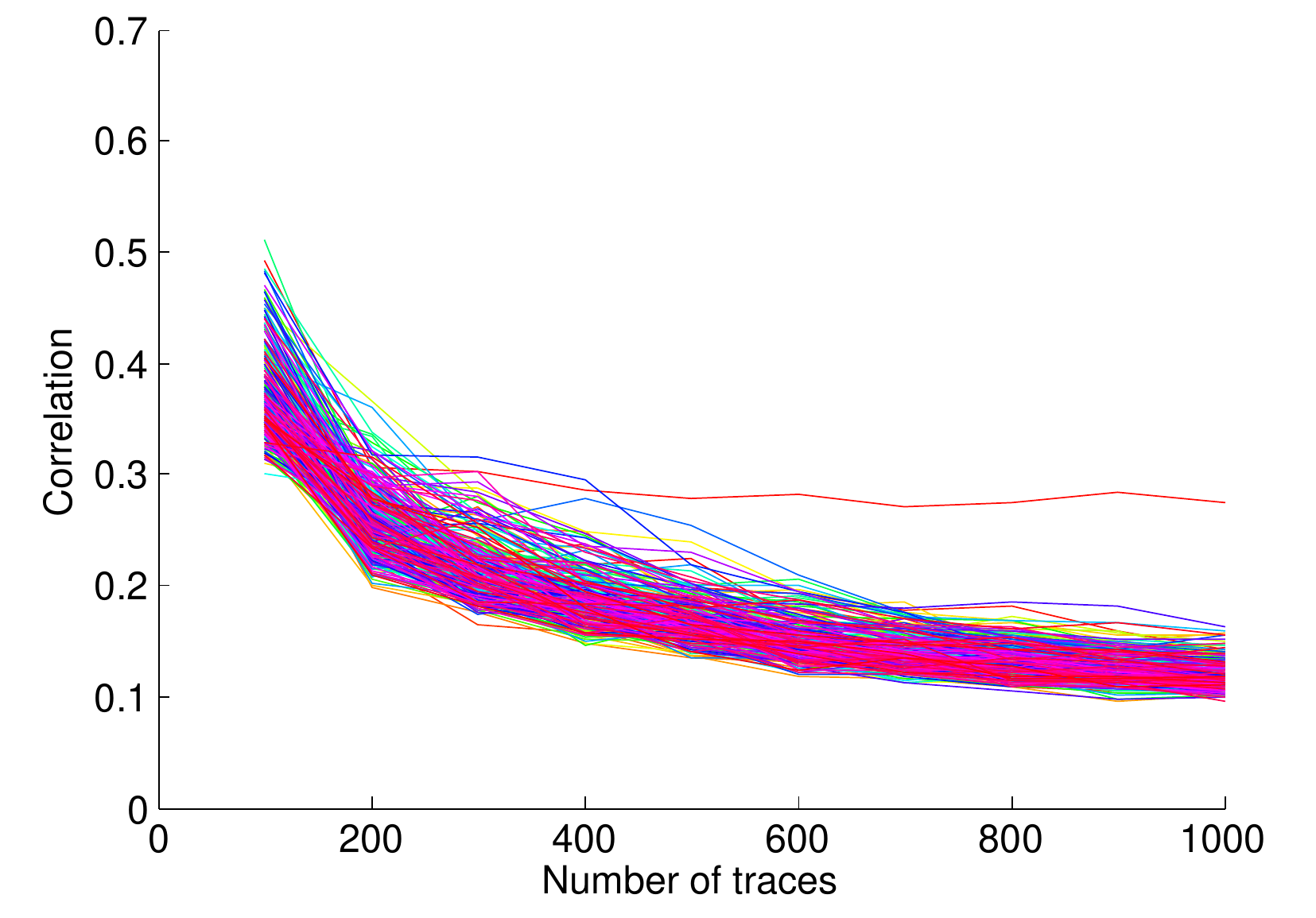}
   	\end{center}
   	\caption{\label{f:inductorgraph}Number of power traces needed when an inductor filter is used}
   \end{figure}

  The number of traces required to attack the system with the inductor connected can be found from \autoref{f:inductorgraph}. As you an see now at least 500 traces are required to distinguish the key correctly. This is not as good as the capacitor. But yet better than nothing was used.

  \subsection{\label{s:filter}LC filter}
  
\begin{figure}[htb]
    	\begin{center}
    		\includegraphics[width=9cm]{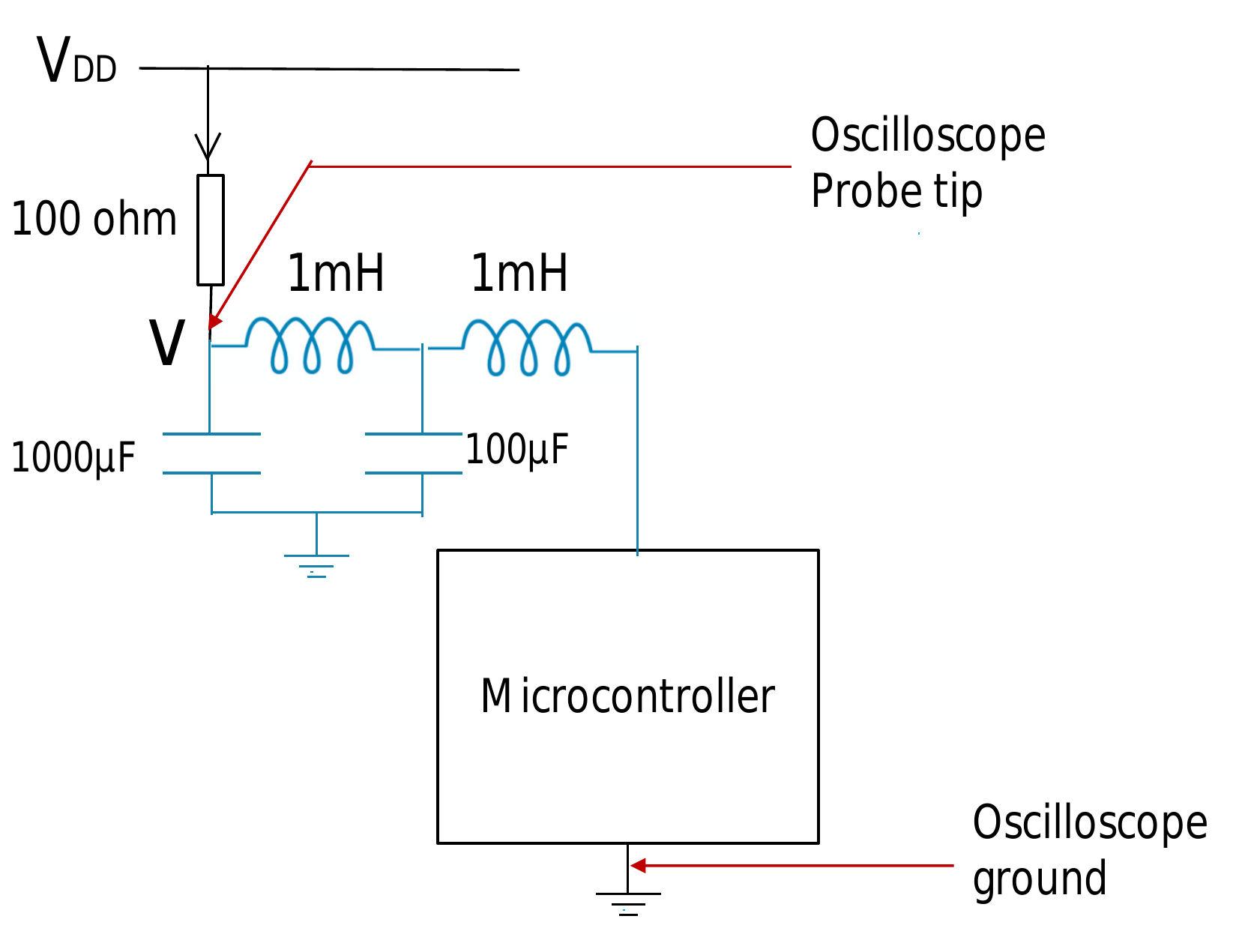}
    	\end{center}
    	\caption{\label{f:lc}Providing power to the microcontroller through an LC filter}
    \end{figure}

\begin{figure}[htbp]
    	\begin{center}
    		\includegraphics[width=9cm]{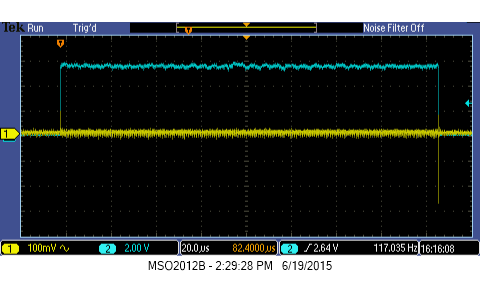}
    	\end{center}
    	\caption{\label{f:lcpic}Power trace affected by a second order LC filter}
    \end{figure}

  A low pass second order LC filter was implemented as shown in \autoref{f:lc}, using two $1000 \mu F$ capacitors and two 1 mH inductors. 
  As this is a second order filter it has a sharp cut off slope. Therefore the frequencies greater than the cut off frequency are 
  much more attenuated when compared to low oder filters considered before.  
  The power trace in this case in shown in \autoref{f:lcpic}. As you can see now the 
  power trace is very much smoothened than all previous cases. 
  
    \begin{figure}[htbp]
    	\begin{center}
    		\includegraphics[width=9cm]{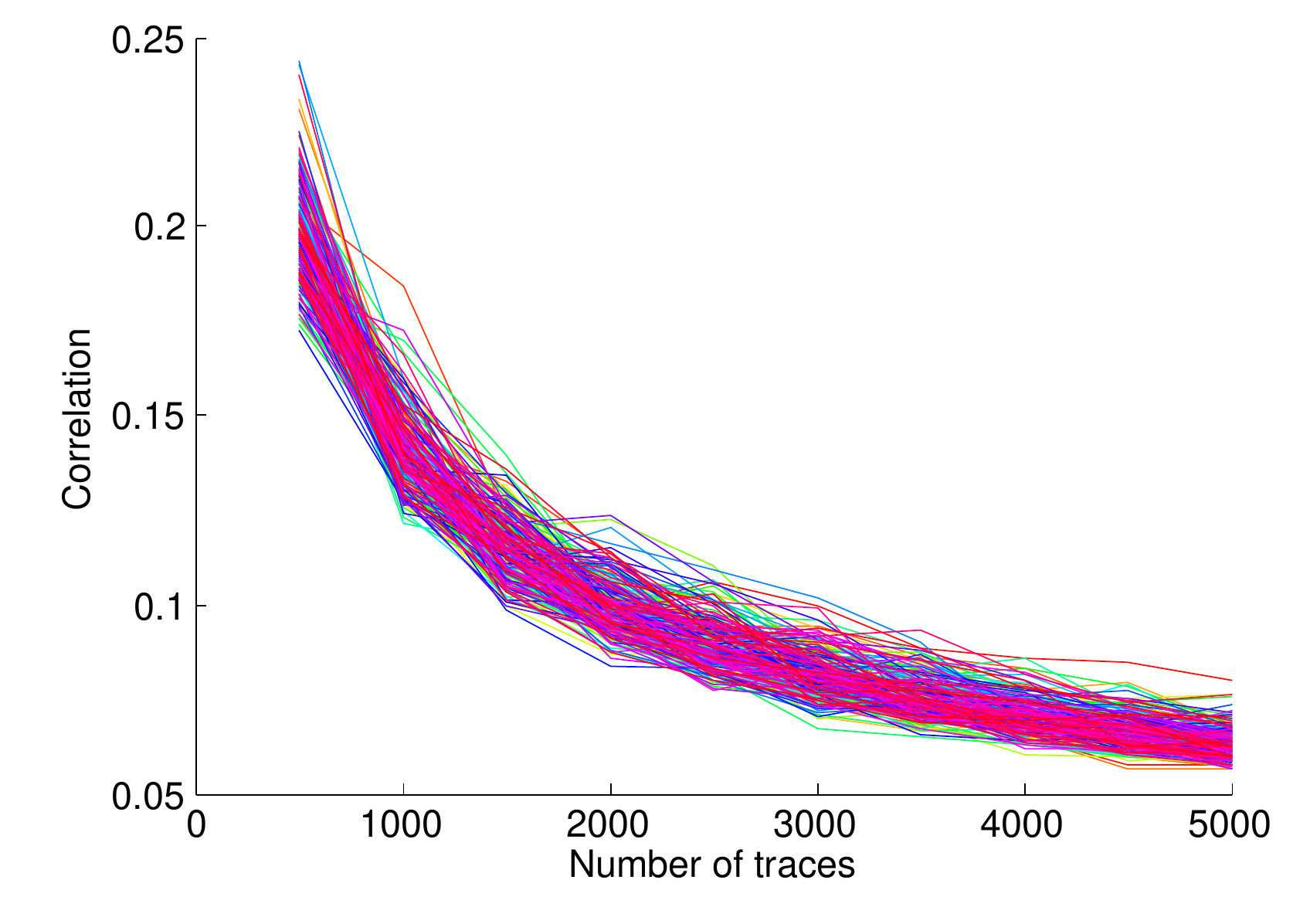}
    	\end{center}
    	\caption{\label{f:LCgraph}Number of power traces needed when an LC filter is used}
    \end{figure}         
    
   As expected, the number of power traces required for an successful attack increased by a considerable amount compared to the others methods.
   As seen from \autoref{f:LCgraph} it is only at about 5000 power traces, the correct key starts to become significant from other keys. 5000 is a comparatively large number of traces but yet the time taken for an attack is about 4.5 hours, which is feasible for an attacker.

  \subsection{Effectiveness of tested filters}

  \autoref{t:filters} shows a comparative summary of the effectiveness of different filters we have implemented. It can be seen that, whatever the filter
  implemented will increase the number of power traces than when nothing is implemented. The most effective method
  out of them is the second order LC filter which required more than 5000 traces with about 4.5 hours to complete the attack. But yet
  non of them are good enough to make an attack infeasible. If higher order filters are used, the effectiveness would increase.
  But yet the number of components increase with the order of the filter which compromise the size of the circuit and the power dissipation as well.

  \begin{table}
  	\begin{center}
  	\begin{tabular}{|l|l|l|}\hline
  		\textbf{Method} & \textbf{Approximate minimum } & \textbf{Approximate minimum}\\
  		& \textbf{number of traces} & \textbf{time}\\\hline
  		Without filters & 50 & 5 minutes\\\hline
  		Capacitor (1mF) connected in parallel & 1500 & 1.5 hours\\\hline
  		Inductor (1mH) connected serially & 500 & 30 minutes\\\hline
  		LC (Inductor-capacitor) second order filter & 5000 & 4.5 hours\\\hline
  	\end{tabular}
  \end{center}
    	\caption{Results for different filters}
    	\label{t:filters}
  \end{table}

 \section{Methods we tried}
 
 Then we tested some ideas of our own. First we tested how regulation of the voltage across the microcontroller affects power analysis. Then we tested the effect of providing power via a non linear device such as an operation amplifier. Next we tried implementing a constant current source with the intention of making the current drawn constant so that the leakage is minimized.
 Moreover, we also tested how connecting multiple chips in parallel to the same power supply affects power analysis. The rest of this section discusses those approaches tried by us, their results and finally a comparison of their effectiveness. 
 
 \subsection{Voltage regulator}
 
\begin{figure}[htb]
    	\begin{center}
    		\includegraphics[width=9cm]{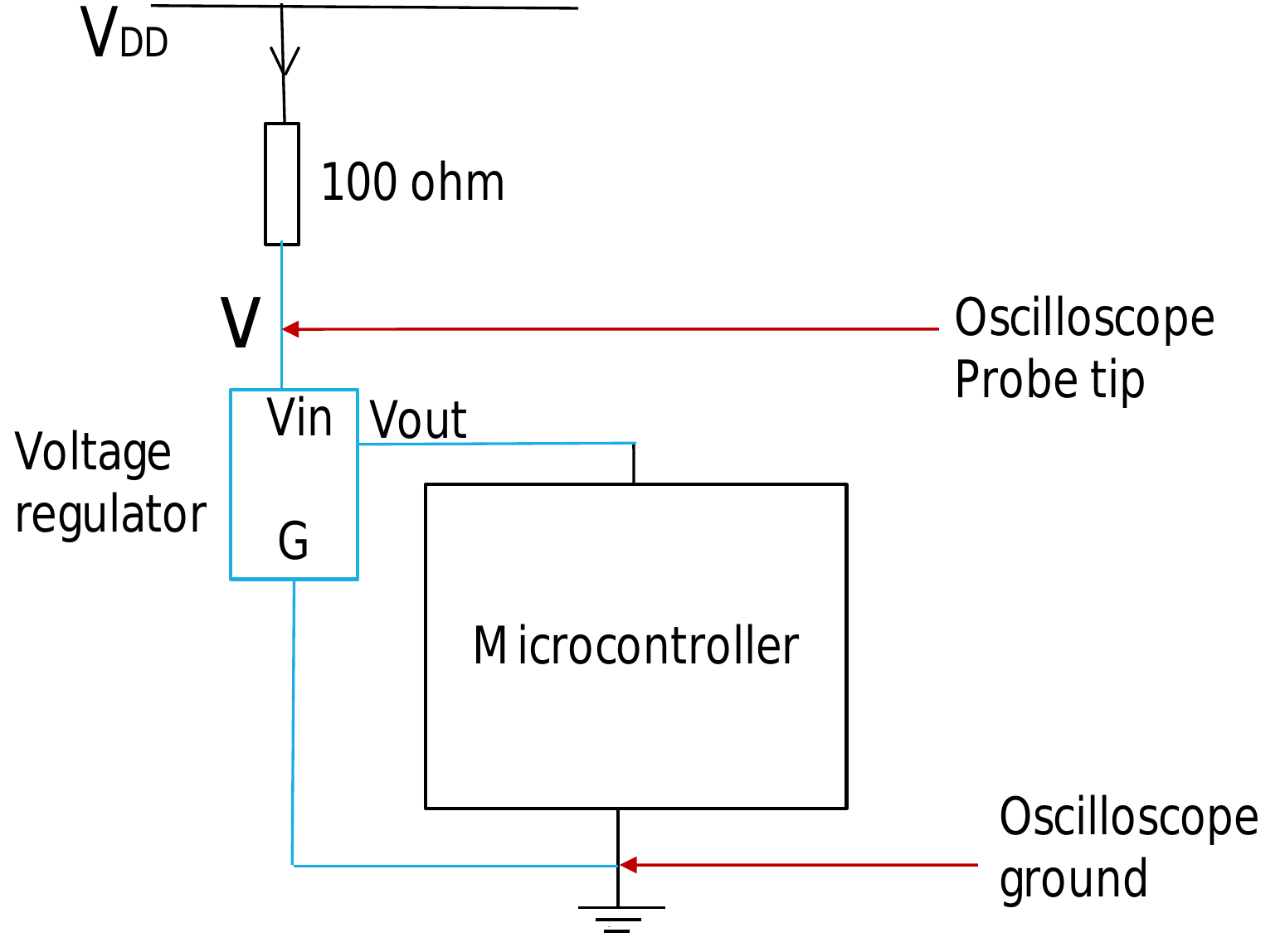}
    	\end{center}
    	\caption{\label{f:voltagereg}Giving power through a voltage regulator}
    \end{figure}

\begin{figure}[htb]
    	\begin{center}
    		\includegraphics[width=9cm]{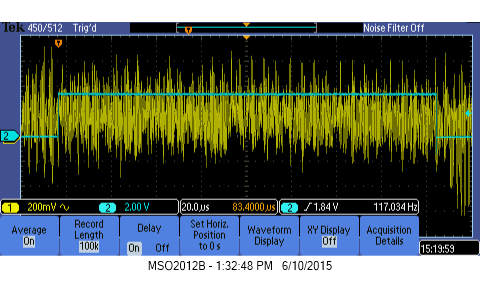}
    	\end{center}
    	\caption{\label{f:voltagereftrace}Power trace when a voltage regulator is used}
    \end{figure}

 A voltage regulator is a device that regulates the voltage. Even when a varying input voltage is provided to a voltage regulator, the output voltage is smooth and constant. 
 \autoref{f:voltagereg} shows how the voltage regulator is connected. Here the voltage regulator is expected to keep the voltage across the microcontroller constant. Therefore we believed that it would smoothen the current drawn by the microcontroller making
 the attack difficult. But unfortunately the power traces were not affected at all. The power trace is shown in \autoref{f:voltagereftrace}.
 When comparing with the power trace without the voltage regulator in \autoref{f:trace2}, it can be seen that the shape of the power trace
 has not been affected even by a little by the voltage regulator. Instead the signal has been amplified. This can be
 observed by the fact that
 the vertical scale of \autoref{f:trace2} is 100 mV and the vertical scale of \autoref{f:voltagereg} is 200 mV.
 Therefore providing a power  through a voltage regulator enhances a power analysis attack rather than being a countermeasure.
 After carrying out the attack it was found out that now 50 traces were more than enough to derive the key.
 
 It seems that even though the voltage regulator keeps the output voltage constant, the current drawn by the microcontroller is still varying 
 as before. Since we are providing an input voltage of about 12 V to the regulator, the voltage variation observed by the oscilloscope is 
 increased causing an amplified waveform. Two types of voltage regulators namely 7805 and AMS1117 were tested but the results were the same.
 
 \subsection{Zener diode}

\begin{figure}[htbp]
    	\begin{center}
    		\includegraphics[width=7cm]{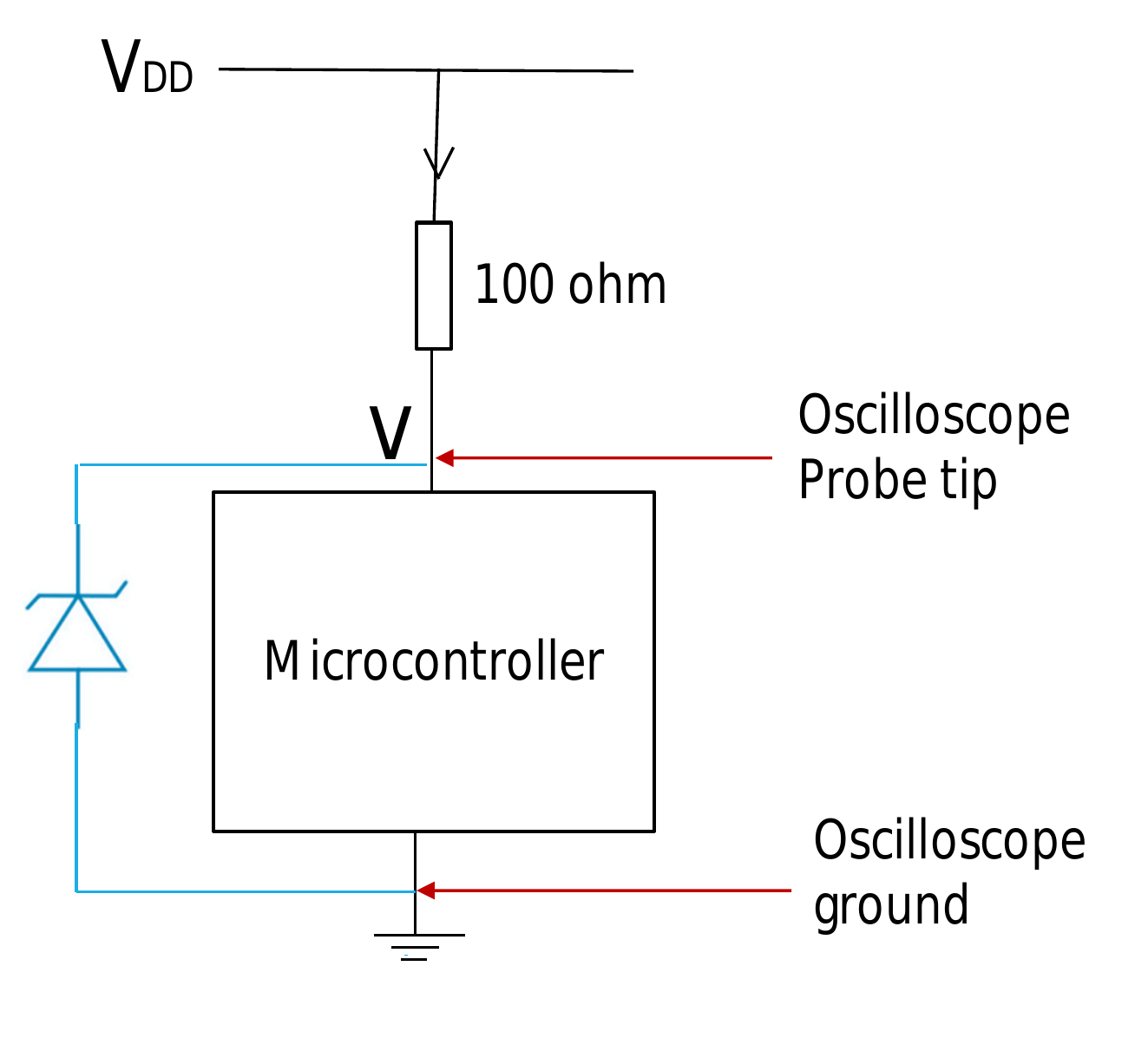}
    	\end{center}
    	\caption{\label{f:zener}Connecting a reversed biased zener diode in parallel to the  microcontroller}
    \end{figure}

\begin{figure}[htbp]
    	\begin{center}
    		\includegraphics[width=9cm]{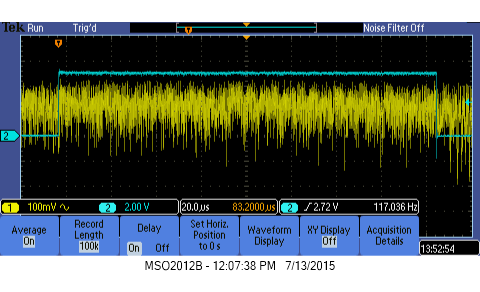}
    	\end{center}
    	\caption{\label{f:zenertrace}Power trace when a zener diode is used}
    \end{figure}

 A reverse biased zener diode keeps the voltage across it constant. Therefore a zener diode with 5 V zener voltage was connected as shown in \autoref{f:zener}. Due to the action of the zener diode, the voltage at V in \autoref{f:zener} should be constant giving a constant
 signal. But the devices are not ideal and hence the observed waveform is as shown in \autoref{f:zenertrace}. 
 Although it is not constant it is smooth than when nothing was used. 

    \begin{figure}[htbp]
    	\begin{center}
    		\includegraphics[width=9cm]{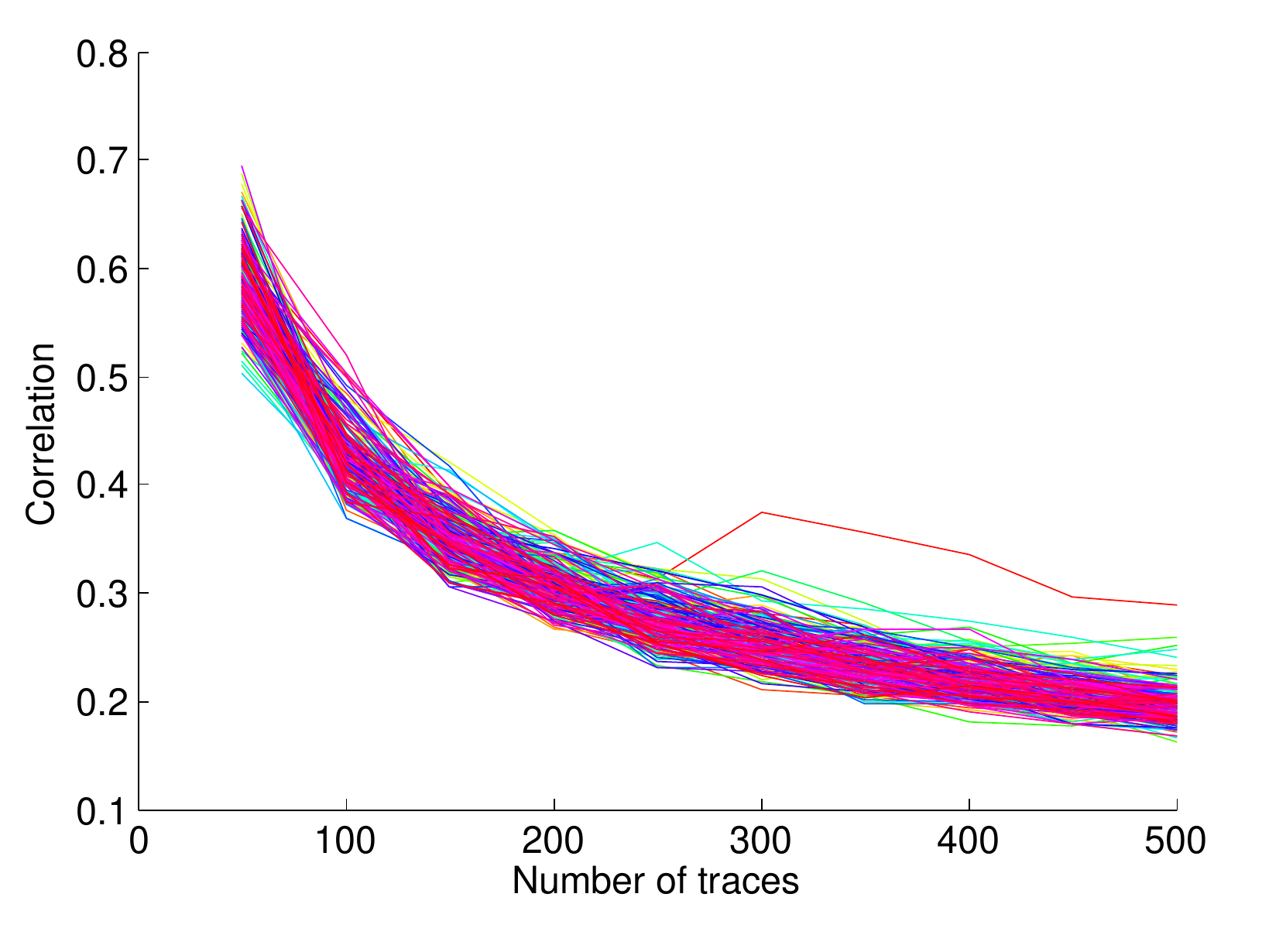}
    	\end{center}
    	\caption{\label{f:zenergraph}Number of power traces needed when a zener diode is used}
    \end{figure}        
    
  When an attack was carried out the number of traces required increased. According to the graph in \autoref{f:zenergraph} 
  at about 300 power traces, the key becomes identifiable. Hence this is not that much effective as a countermeasure.

 \subsection{Operational amplifier} 
 
\begin{figure}[htbp]
    	\begin{center}
    		\includegraphics[width=8cm]{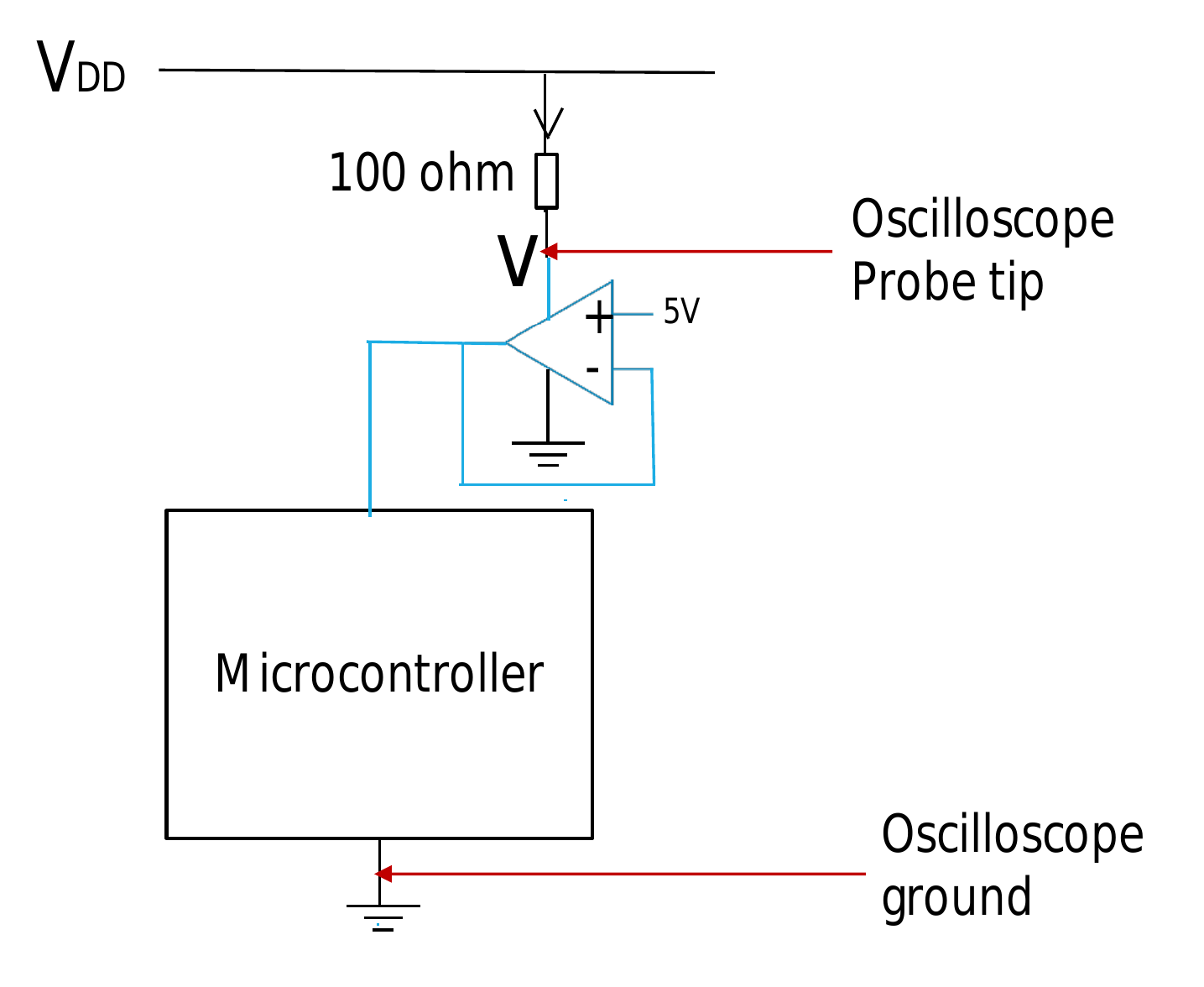}
    	\end{center}
    	\caption{\label{f:opamp}Providing power to the microcontroller through an operational amplifier}
    \end{figure}

\begin{figure}[htbp]
    	\begin{center}
    		\includegraphics[width=9cm]{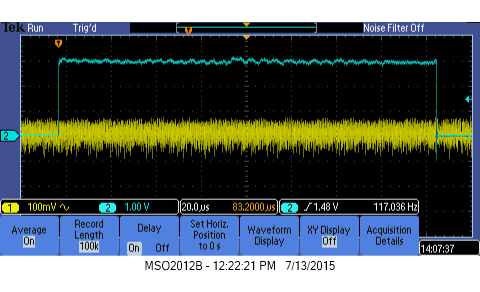}
    	\end{center}
    	\caption{\label{f:opamppic}Power trace when a operational amplifier is used}
    \end{figure}

 An operation amplifier is a non linear device. An operational amplifier (UA741) was connected as shown in \autoref{f:opamp} to give power
 to the microcontroller. In this circuit, the op-amp acts as a buffer / voltage follower. A 5 V input is given to the non-inverting input and
 this same value appears as the output because a negative feedback has been given from the output to the inverting input. 
 Since the operational amplifier is a non linear device, the input - output current relationship would be non linearl as well.
 The causes a change in the power trace as shown in \autoref{f:opamppic}.
 
\begin{figure}[htbp]
    	\begin{center}
    		\includegraphics[width=9cm]{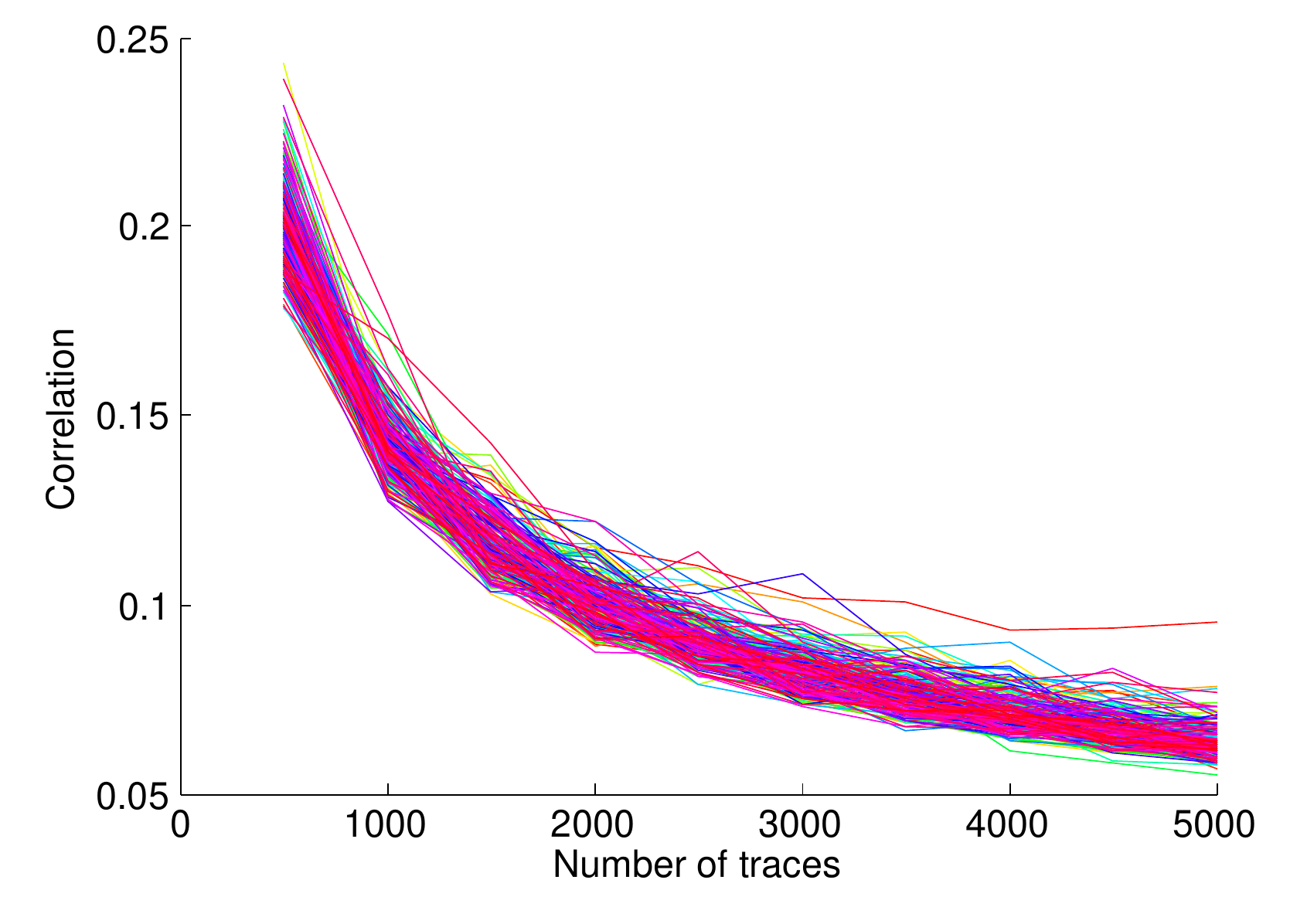}
    	\end{center}
    	\caption{\label{f:opampgraph}Number of power traces needed when an operational amplifier is used}
    \end{figure}

    From \autoref{f:opampgraph} it can be observed that the correct key becomes identifiable only after about 4000 power  traces.
    Therefore this method is somewhat effective. But yet it is not good enough to prevent an attack.
   
 \subsection{Constant current source}
 
\begin{figure}[htbp]
 	\begin{center}
 		\includegraphics[width=9cm]{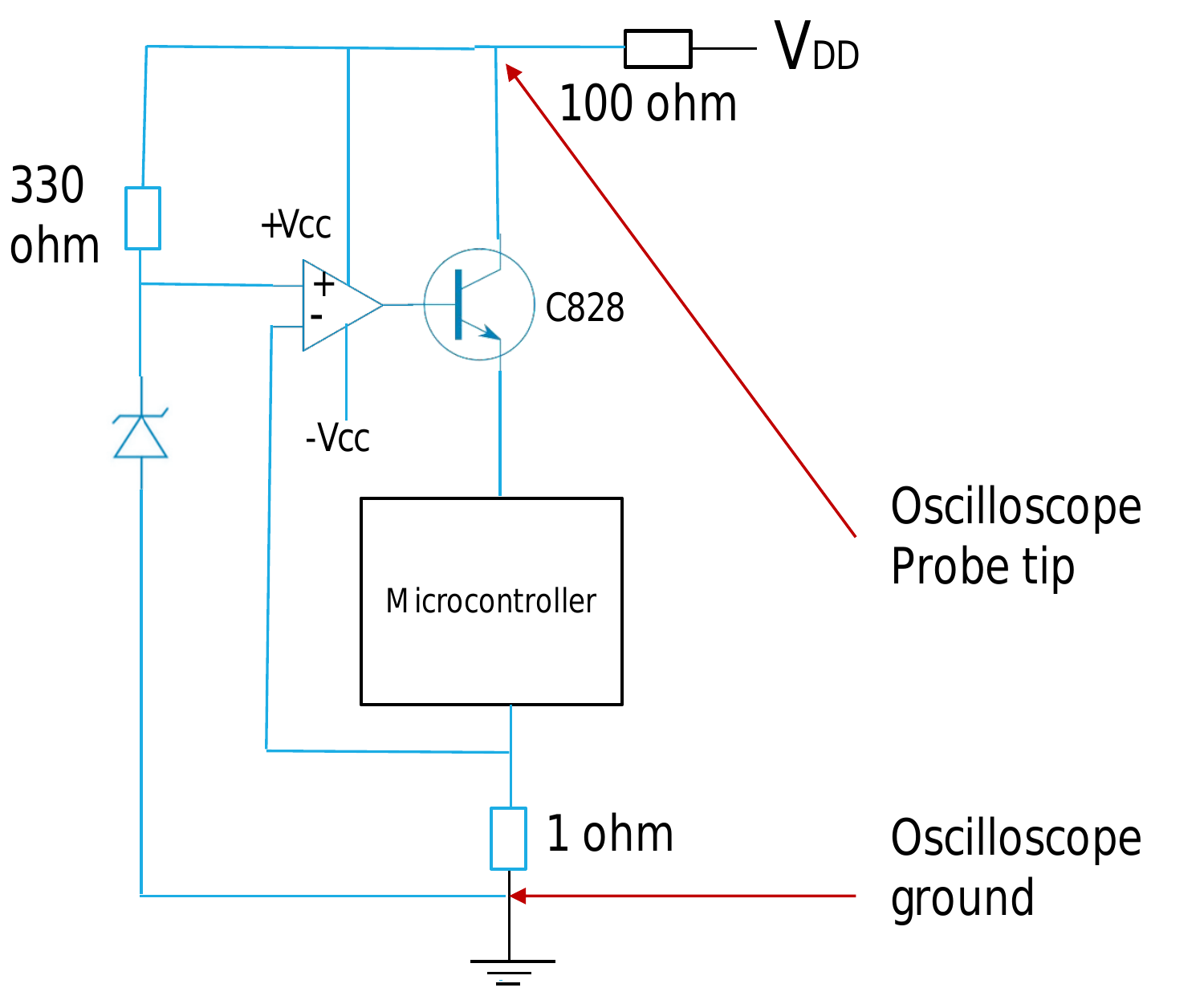}
 	\end{center}
 	\caption{\label{f:constantcurrent}Giving power through a constant current source}
 \end{figure}

 A constant current source is a device that always outputs a constant current independent of the voltage across it. 
 A current source that uses an operational amplifier (UA741), a transistor (C828) and a zener diode (5 V zener voltage) was implemented as shown in \autoref{f:constantcurrent}. The 1 ohm resistor in the circuit acts  as a current sensor which provides a feedback
 to the inverting input of the operation amplifier. Due to this action current through the microcontroller is expected to be constant. 
 But the power trace observed through the oscilloscope was as same as when nothing was used. The reason would have been that though this setup
 is called a constant current source it is not an ideal one and hence current variations in Milli Ampere range still happens.
 Therefore an attack needed only about 50 traces and this method is not useful at all as a countermeasure.
 
 \subsection{An additional microcontroller in parallel}
 
\begin{figure}[htb]
 	\begin{center}
 		\includegraphics[width=8cm]{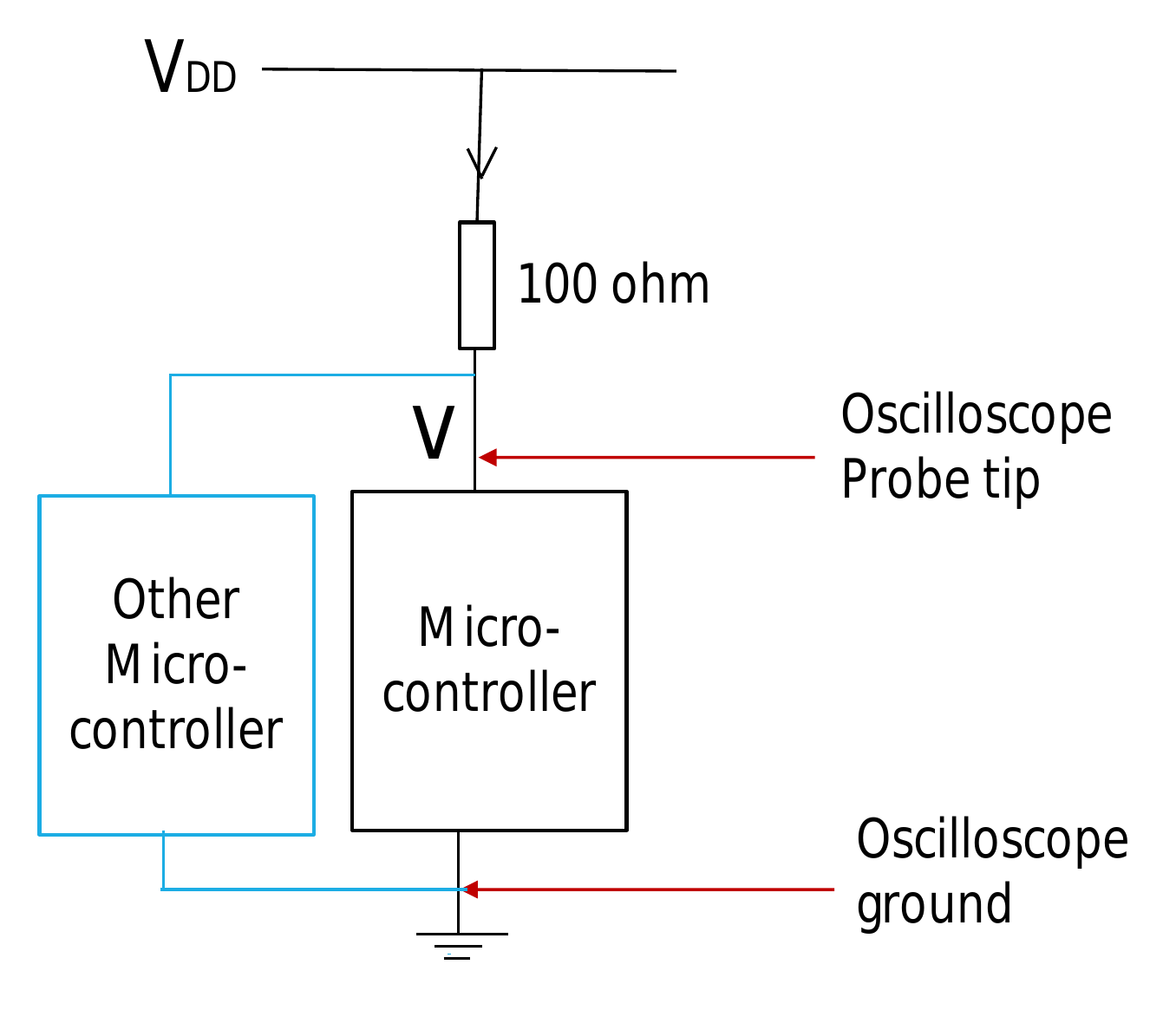}
 	\end{center}
 	\caption{\label{f:dualcore}Connecting another microcontroller in parallel}
 \end{figure}

 In all above cases, only one microcontroller was connected to the power supply and hence the power consumption was solely due to the action of that specific microcontroller. But if another microcontroller that does some thing else is connected in parallel to the power supply, then the power consumption is for both of the devices. In such case, the power consumption of the second microcontroller is considered as noise by the power analysis algorithm. But as SNR is increased, the number of power traces would be increased. This setup can be also considered as an emulation of a dual core chip which is usually powered by a single power supply.
 The circuit diagram is shown in \autoref{f:dualcore}. There, the microcontroller to the right does the real encryption while the one to the left does some other job just to consume false power.

\begin{figure}[htb]
    	\begin{center}
    		\includegraphics[width=9cm]{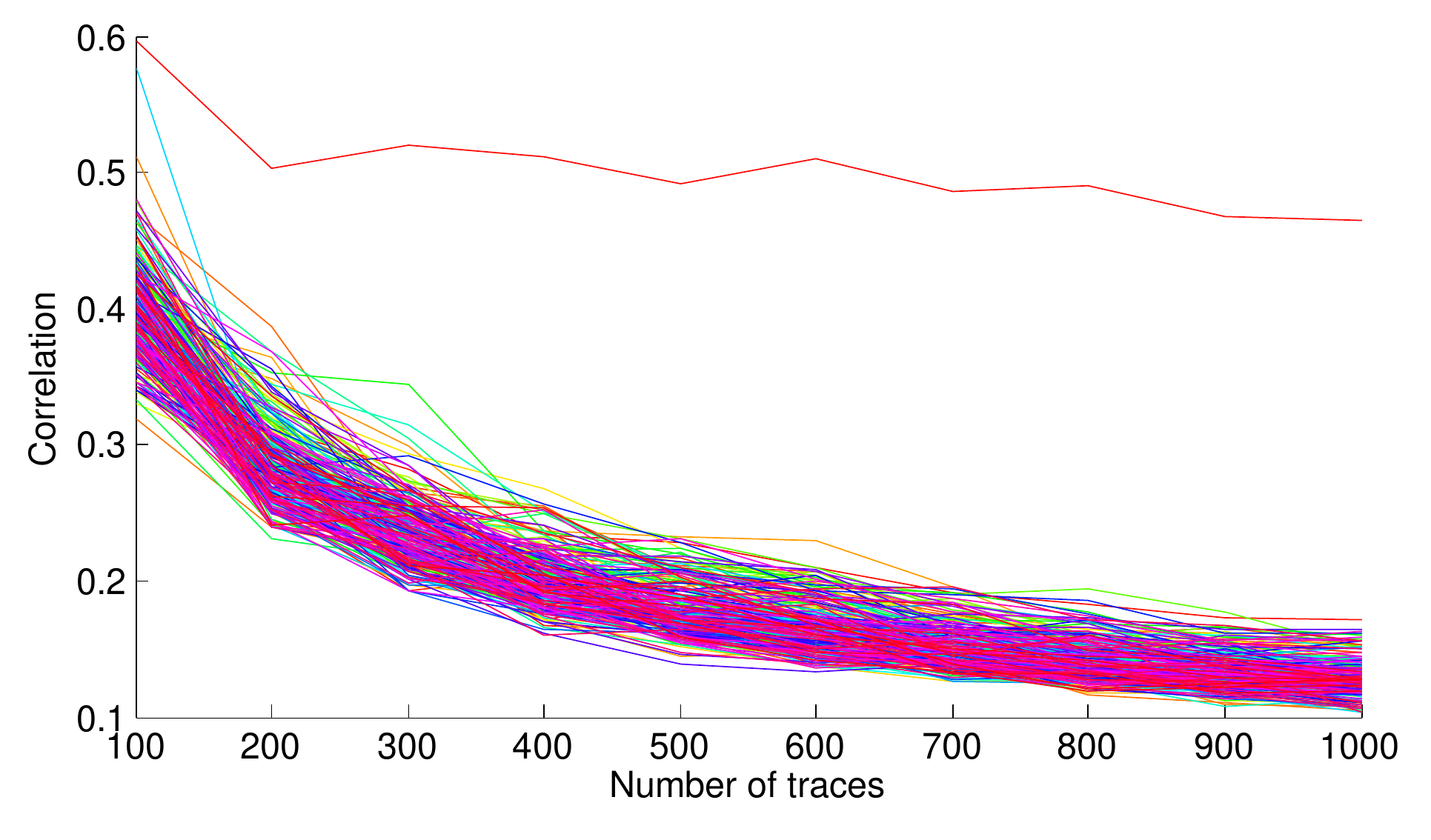}
    	\end{center}
    	\caption{\label{f:dualcoregraph1}Number of power traces needed when another microcontroller that write to the serial port infinitely is connected in parallel to the power supply}
    \end{figure}

 First, we programmed the second microcontroller such that it prints to the serial port in an infinite loop. Power traces were taken when the correct microcontroller was doing the real encryption. The obtained correlation vs number of traces plot is shown in 
 \autoref{f:dualcoregraph1}. But unfortunately it is quite clear that at about 200 power traces, the correct key becomes visible.
 
\begin{figure}[htb]
    	\begin{center}
    		\includegraphics[width=9cm]{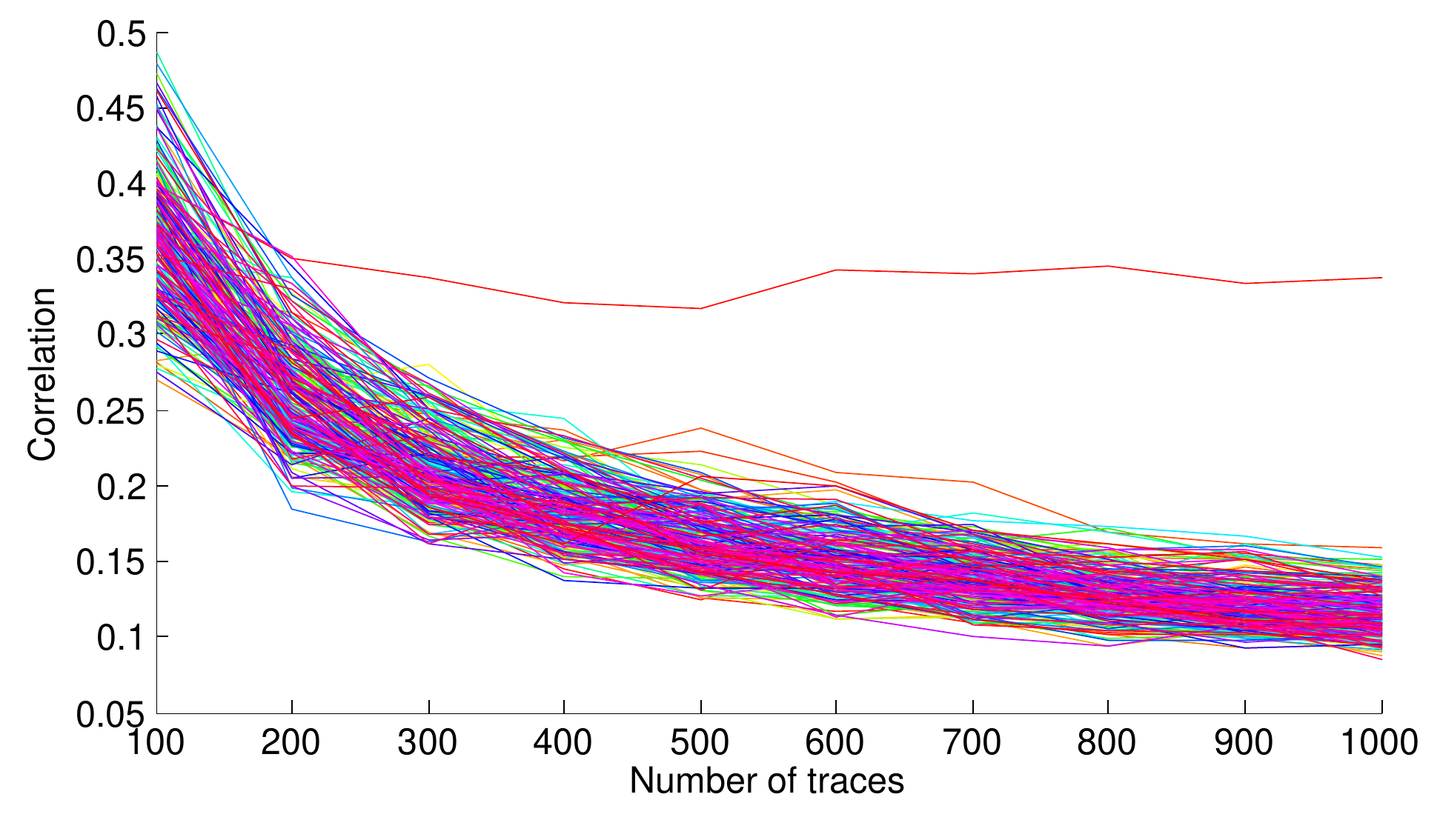}
    	\end{center}
    	\caption{\label{f:dualcoregraph2}Number of power traces needed when another microcontroller that do a mock AES is connected in parallel to the power supply}
    \end{figure}

 When the power traces for the other microcontroller was alone inspected via the oscilloscope, it was noted that the power consumption for serial printing was not that significant. Therefore the introduced noise would not have been significant. Therefore it was programmed to do a mock AES instead of mock serial printing. Then power traces were obtained and analysed to obtain a plot as shown in \autoref{f:dualcoregraph2}.

  But still it is visible in \autoref{f:dualcoregraph2} that at about 300 power traces, the correct key becomes visible. Therefore we concluded that the dual core idea was also not successful for hiding the leakage.

 \subsection{FTDI chip}  
 
 An FTDI FT232RL chip is a USB to Serial UART(Universal Asynchronous Receiver/Transmitter) interface chip manufactured by Future Technology Devices International Ltd (FTDI) \cite{ftdi}. Throughout this chapter we refer to the FTDI FT232RL chip by the short name FTDI. In this subsection the effect of connecting such a chip to the power supply of the microcontroller in parallel is discussed. The reason for trying that out was because an attempt by us to do a power analysis attack on an Arduino board failed apparently due to the effect of the FTDI chip which is explained subsequently.

\begin{figure}[htb]
    	\begin{center}
    		\includegraphics[width=9cm]{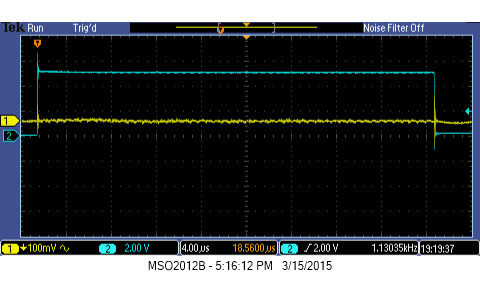}
    	\end{center}
    	\caption{\label{f:arduinouno}An averaged power trace on an Arduino Uno}
    \end{figure}

 Once it was tried by us to do a power analysis attack on an Arduino Uno board. Arduino is an electronic prototyping board based on Atmel microcontrollers \cite{arduino}. The Arduino Uno board was programmed using the Arduino toolkit to perform AES and communicate with the computer similar to the behaviour explained for PIC2550 based testbed in \autoref{c:testbed}. Then the oscilloscope was connected and an average power trace as shown in \autoref{f:arduinouno} was obtained. The power traces was almost null that a power analysis attack would not succeed. Just to verify 100,000 power traces were collected and analysed where it failed. 
 
\begin{figure}[htb]
    	\begin{center}
    		\includegraphics[width=5cm]{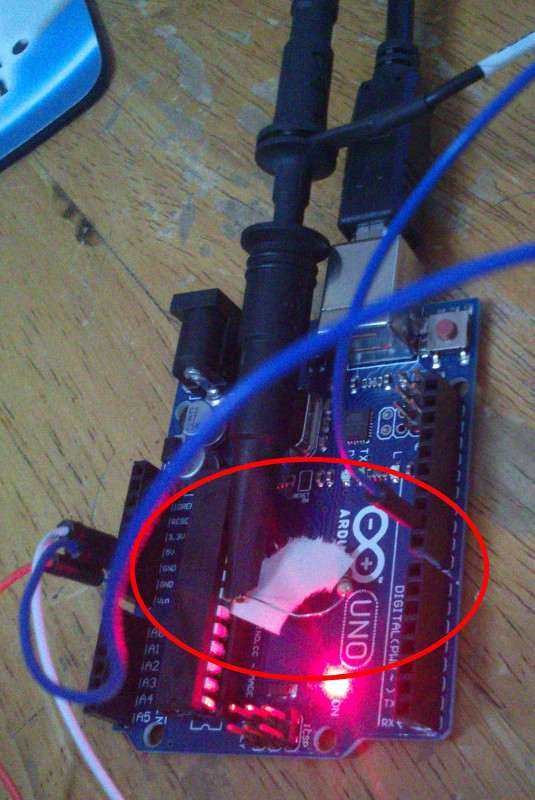}
    	\end{center}
    	\caption{\label{f:arduinohacked}Hacking the Arduino Uno board for a successful power analysis attack}
    \end{figure}

\begin{figure}[htb]
    	\begin{center}
    		\includegraphics[width=9cm]{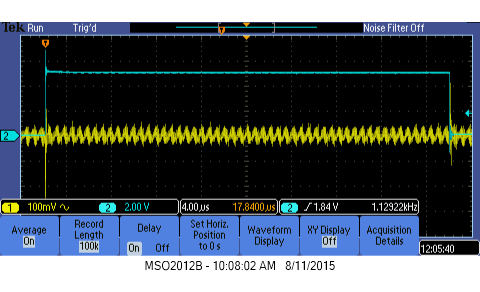}
    	\end{center}
    	\caption{\label{f:arduinounohackedtrace}An averaged power trace on a hacked Arduino Uno}
    \end{figure}

	Two possibilities for the failure can be either the fact Atmel chips are safer or numerous other chips and components on the Arduino board affects the power consumption. To identify the exact cause, the power pin of the Atmel 328P chip was folded upwards and then it was connected to base on the board via a 10 ohm resistor as shown in \autoref{f:arduinohacked}. A 10 ohm resistor is selected in contrast to a 100 ohm one or otherwise the drop across the resistor would be too large that the microcontroller would not get enough voltage. The averaged power trace now looked as shown in \autoref{f:arduinounohackedtrace} where now the power trace is quite clear and an attack should succeed. As expected, at about 2000 traces, the correct key started appearing.    
 
 With that it is clear that Atmel 328P chip is also vulnerable though it not that vulnerable as PIC2550. Therefore the reason for the failure for direct power measurements on the Arduino Uno would have been the extra components on the board. Since Arduino Uno had lot of components we minimized the scope by testing a similar attack on an Arduino Nano. Arduino Nano is also a prototyping board based on an Atmel microcontroller, but which does not have extra components and features as in Arduino Uno. But yet the averaged power traces was similar to one in \autoref{f:arduinouno}, and even at 100,000 power traces the attack became unsuccessful.
 
\begin{figure}[htb]
    	\begin{center}
    		\includegraphics[width=9cm]{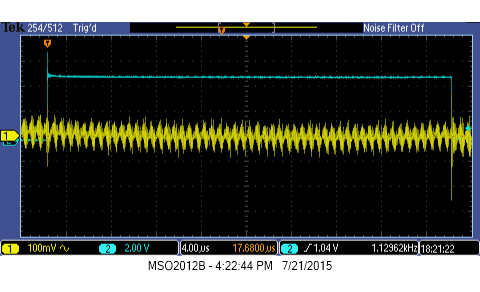}
    	\end{center}
    	\caption{\label{f:minipro}An averaged power trace on an Arduino Mini Pro}
    \end{figure}

 Then we obtained an Arduino Mini Pro which was even more simpler than an Arduino Nano. There, the average power trace was clear as shown in \autoref{f:minipro}. As expected the attack became successful at about 2000 power traces.
 
 Now obviously it should be a component that is present in the Arduino Nano but not in Mini Pro, the cause for the failure of the attack. 
 Schematic diagram for Arduino Nano \cite{arduinonano} and Arduino Mini Pro \cite{arduinomini} was compared to see the difference. As there were few differences, in order to test the success of an attack by eliminating one item by one item, the Arduino Nano was implemented on a breadboard.

    \begin{figure}[htb]
    	\begin{center}
    		\includegraphics[width=16cm]{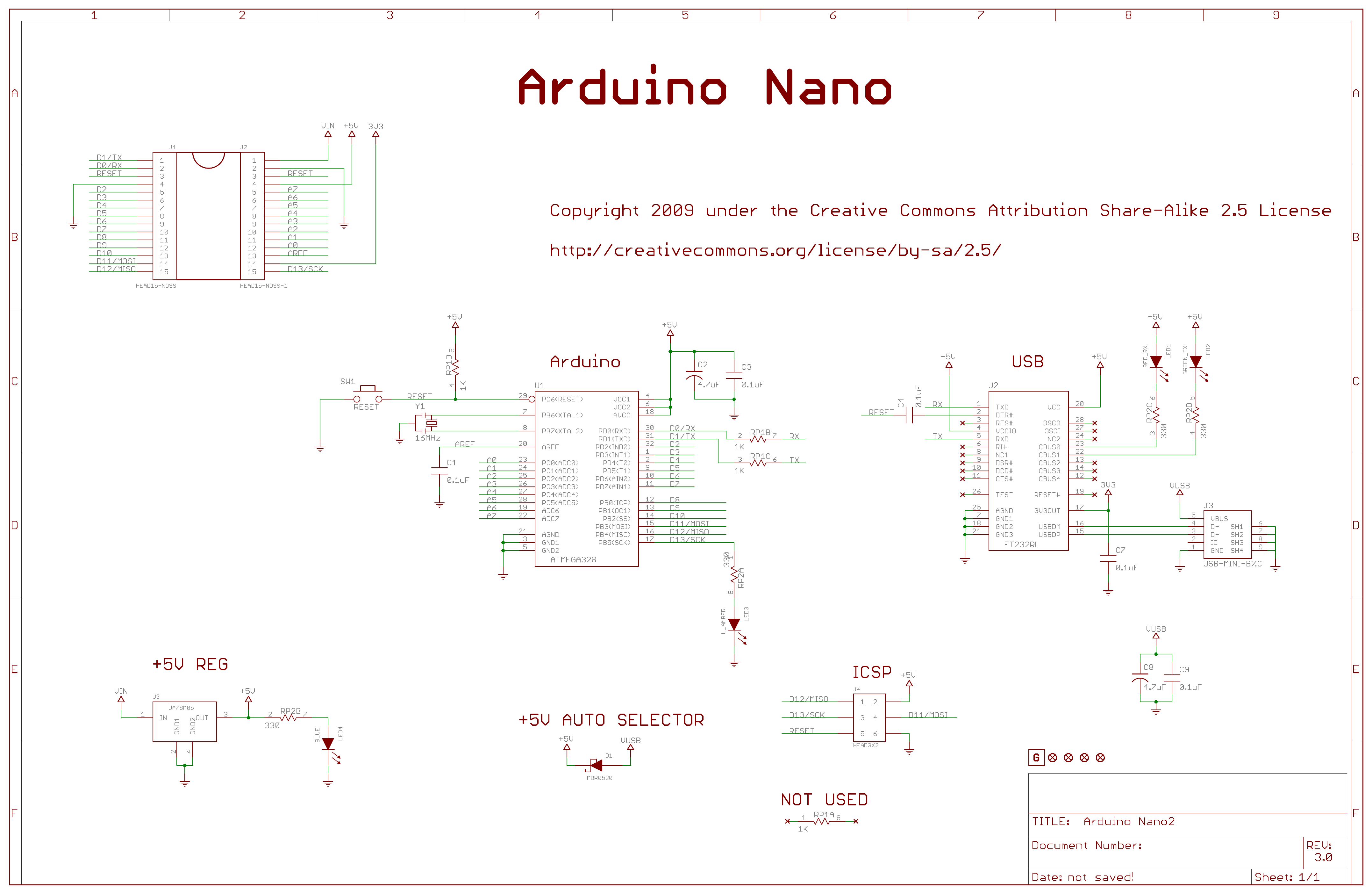}
    	\end{center}
    	\caption{\label{f:schamatic}Schematic of Arduino Nano \cite{arduinonano}}
    \end{figure}  

\begin{figure}[htb]
    	\begin{center}
    		\includegraphics[width=14cm]{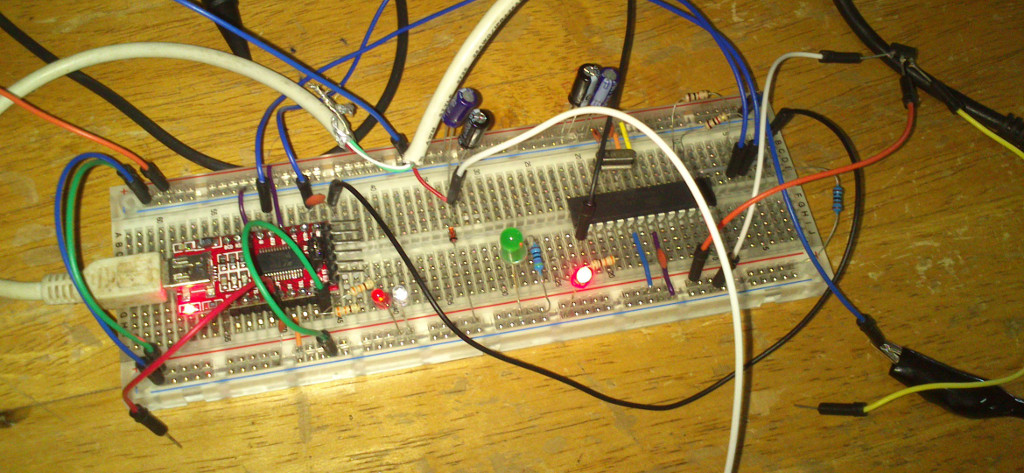}
    	\end{center}
    	\caption{\label{f:adrduinoonbreadboard}Arduino Nano implemented on a breadboard}
    \end{figure}

	The schematic diagram of the Arduino Nano is shown in \autoref{f:schamatic}. There the microcontroller is an Atmel 328 chip. Since the chip on the Arduino Nano is a surface mount one permanently soldered to the board, a removable Atmel 328 chip from an Arduino Uno was used. In \autoref{f:schamatic} an FTDI FT232RL chip is present. It is the chip that interfaces the microcontroller's serial port to USB port of the computer. Such a chip was obtained and was connected on the breadboard. All other components such as voltage regulators, zener diodes, resistors and capacitors were connected on the breadboard exactly according to the circuit diagram in \autoref{f:schamatic}. The Arduino Nano implemented on a breadboard looks like as shown in \autoref{f:adrduinoonbreadboard}.
    
 Then power traces were obtained for the Arduino Nano implementation on the breadboard. But the average power trace was almost null similar to the real Arduino Nano and the attack did not succeed even at 100,000 traces. Then the FTDI chip was eliminated from the circuit and an averaged power trace was observed thinking that FTDI would have been the culprit. 
 The averaged power trace looked very clear similar to \autoref{f:minipro}. Therefore the attack would definitely succeed.
 
 These observation lead us to infer that FTDI chip is the culprit that makes the attack unsuccessful on Arduino Nano. The power consumption by the FTDI chip would have been too noisy that reduces the SNR for the power consumption signal of the microcontroller by very much. This made us think on the possibility to use an FTDI chip connected to the power supply in parallel to the micrcontroller as a countermeasure. Therefore it was decided to test on the PIC2550 microcontroller which was used in earlier evaluations.
 
\begin{figure}[htb]
    	\begin{center}
    		\includegraphics[width=9cm]{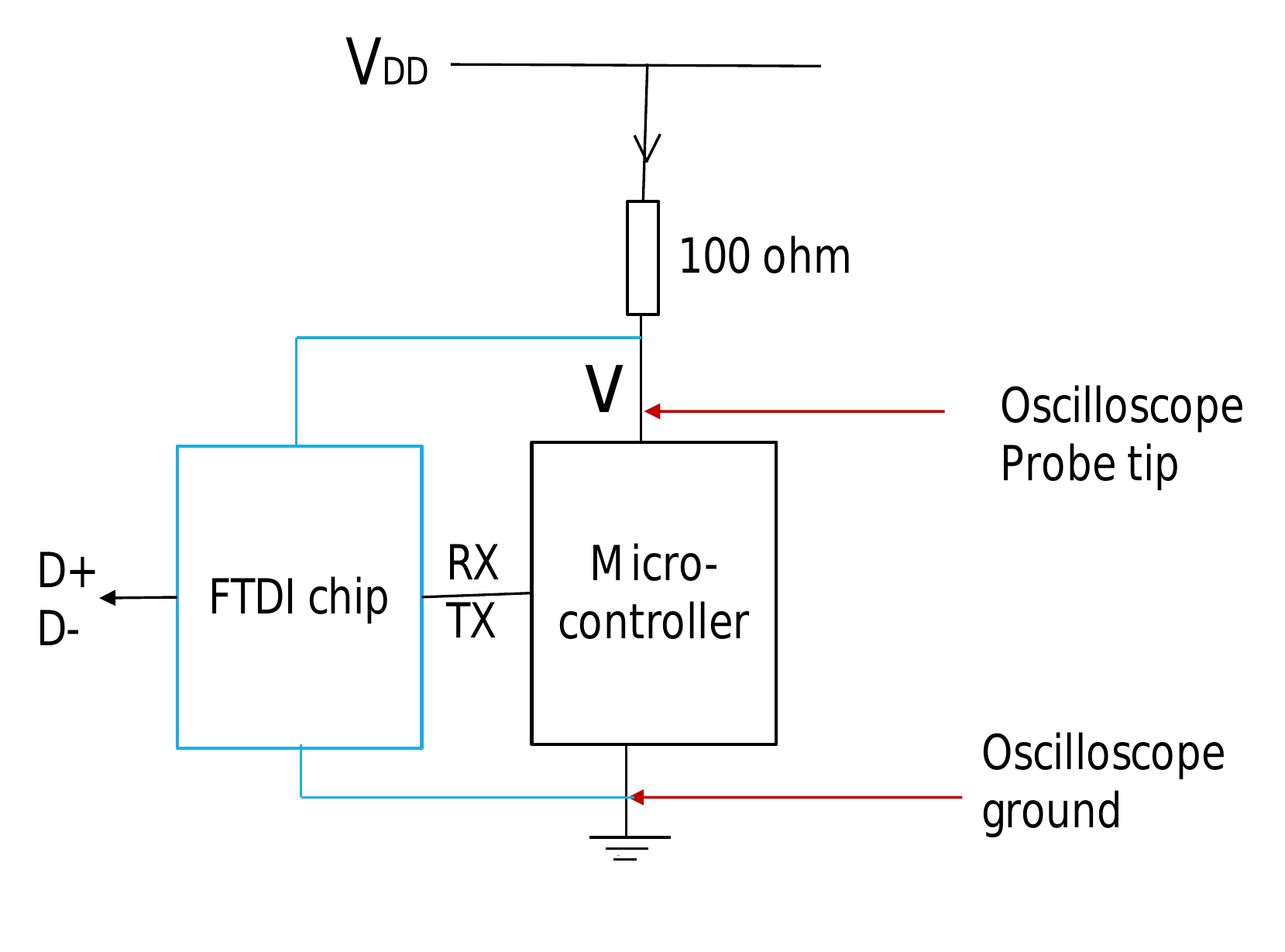}
    	\end{center}
    	\caption{\label{f:ftdicircuit}FTDI chip connected in parallel to the microcontroller}
    \end{figure}

 The FTDI chip was connected to the PIC2550 microcontroller as shown in \autoref{f:ftdicircuit}. Up to now a similar FTDI chip was used for communication between the microcontroller and the PC as it was discussed in \autoref{c:testbed}, but in those cases power to the FTDI chip was provided power separately from the 5 v supply of the PC's USB bus. Now power to the FTDI chip is supplied from the same power supply used to supply power to the microcontroller. 
 
\begin{figure}[htb]
    	\begin{center}
    		\includegraphics[width=9cm]{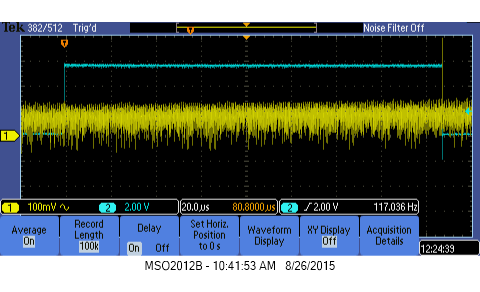}
    	\end{center}
    	\caption{\label{f:ftditrace}Averaged power trace when FTDI chip is connected in parallel to PIC2550}
    \end{figure}

\begin{figure}[htb]
     	\begin{center}
     		\includegraphics[width=9cm]{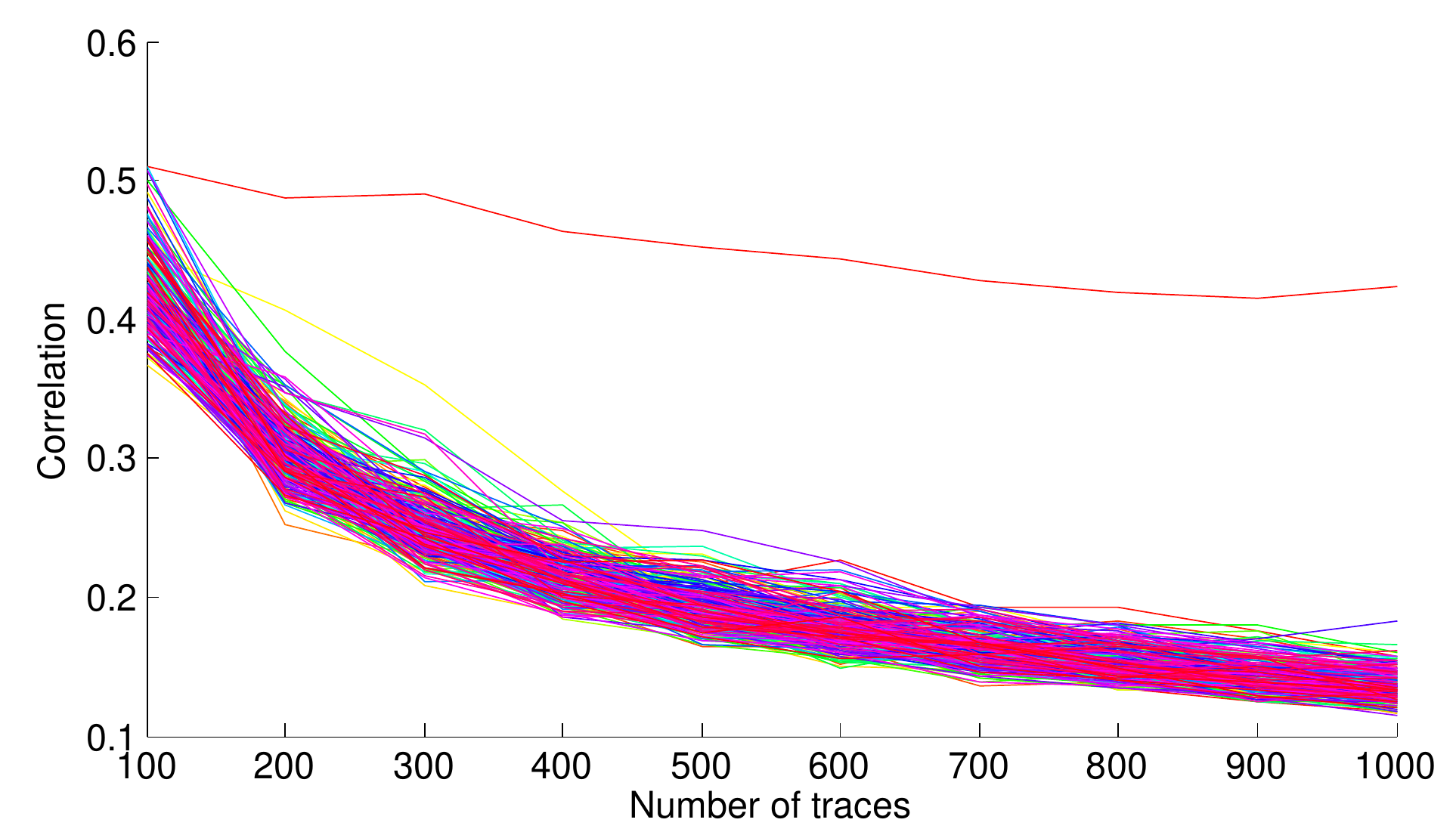}
     	\end{center}
     	\caption{\label{f:ftdigraph}Number of power traces required when FTDI chip is connected in parallel}
     \end{figure}

 An averaged power trace was obtained which is shown in \autoref{f:ftditrace}. Here we can see that the power trace is clear in opposed to null power trace observed for Arduino Nano. Power traces were collected and the correlation vs number of traces plot was plotted and it is shown in \autoref{f:ftdigraph}. The key is clear at just about 200 traces!

   Finally after all those efforts, even though the FTDI made an attack on an Atmel chip infeasible for us, introducing the same FTDI had no effect at all on PIC2550. At the moment the reason why it happens so is unknown to us.

 \subsection{Effectiveness of tested methods}
 
 Comparative results for the ideas we implemented are shown in \autoref{t:other}. Voltage regulator and the current source are not effective at all because they have not made any difference to the number of power traces. Zener diode seems to be better than nothing but yet 20 minutes is too small time. Operational amplifier is somewhat effective but yet it is not better than the second order LC filter tested in \autoref{s:filter}.
 
	\begin{table}[!t]
		\begin{center}
		\begin{tabular}{|l|l|l|}\hline
			\textbf{Method} & \textbf{Approximate minimum} & \textbf{Approximate minimum}\\
			& \textbf{number of traces} & \textbf{time}\\\hline
			Without countermeasures & 50 & 5 minutes\\\hline
			Voltage regulator & 50 & 5 minutes\\\hline
			Current source & 50 & 5 minutes\\\hline
			FTDI chip & 200 & 15 minutes\\\hline
			Zener diode & 300 & 20 minutes\\\hline
			Another microcontroller in parallel & 300 & 20 minutes\\\hline			
			Operational amplifier (UA741) & 4000 & 3.5 hours\\\hline	
				
		\end{tabular}
		\end{center}
		\caption{Results for different methods tried by us}
		\label{t:other}	
	\end{table}

  \chapter[Software countermeasures]{\label{c:software}Software based Countermeasures} 
  
  Software based countermeasures do modifications to the program code that runs on the cryptographic device as it was discussed in \autoref{s:countermeasures_intro}. From many available software based countermeasures, due to reasons that were discussed under \autoref{relatedcounter}, we selected the technique called random instruction injection \cite{ambrose2007rijid} to be practically tested on our testbed. An insight into random instruction method and the results from the carried out attack is explained under \autoref{s:randominsight}. 
  Randomly shuffling Sbox operations is another proposed software countermeasure as mentioned in \autoref{relatedcounter}.
  This method was also implemented and tested by us and the details are elaborated in \autoref{s:sboxshuffle}.
  A problem with these countermeasures is with respect to the randomness. We elaborate a low cost method that 
  can be integrated into any cryptosystem to provide a true randomness. 
  Those improvements are elaborated in \autoref{s:trng}.
  
  \section[Random instructions]{\label{s:randominsight}Random instruction injection}
  
  A power analysis attack requires all the power traces be aligned in time. Random instruction injection destroys the alignment of power traces with the intention of making the attack difficult.  
  Random instruction injection is simple to implement. Further it does not require any additional hardware components and can be implemented on a general purpose microcontroller. Therefore the cost of the countermeasure is not that significant. But as we show in following sections, it provides reasonable immunity against power analysis attacks for its cost and simplicity.
  
  \subsection{\label{s:priciples}Principles}
  
\begin{figure}[htb]
 	\begin{center}
 		\includegraphics[width=16cm]{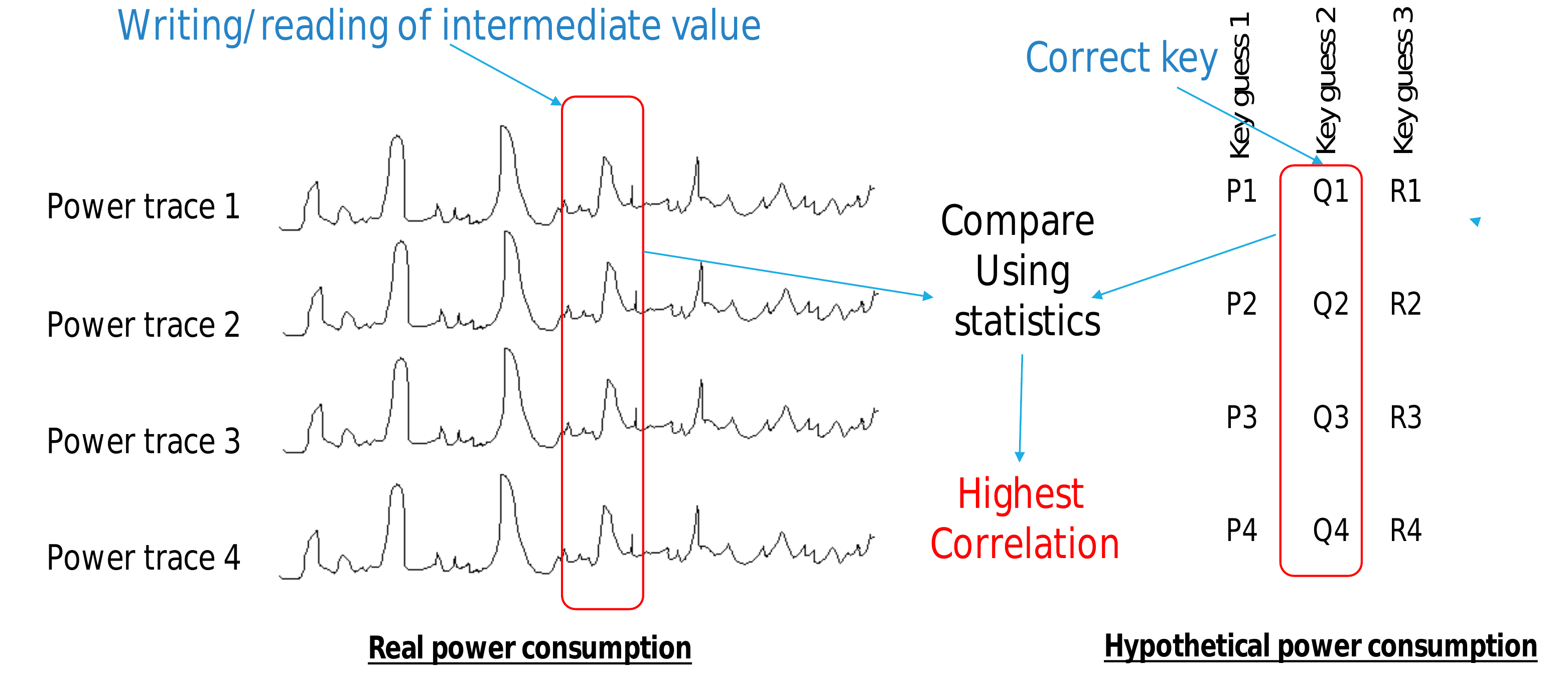}
 	\end{center}
 	\caption{\label{f:align}Alignment requirement of power traces for a successful attack}
 \end{figure}

  In \autoref{s:cpa} how a power analysis attack is done was discussed. 
  \autoref{f:align} is a pictorial recap of the attack. Left side of \autoref{f:align} shows the real power consumption values which are the power traces 
  collected through the oscilloscope.
  Four power traces are shown where each corresponds to the 
  power consumption pattern during encryption of a plain text sample. N number of plain text samples are encrypted and therefore
  N number of such power traces are there even not shown in \autoref{f:align}. The power traces are nicely aligned with respect to time. 
  Therefore, if a certain event like writing of a certain intermediate value during encryption is considered, 
  it should happen on the same time sample of every power trace as marked in the \autoref{f:align}.
  Right side of \autoref{f:align}
  shows the hypothetical consumption values which are the ones calculated using the hamming weight model.
  There, for each combination of key guess and plain text those values are calculated for a specific intermediate value during the encryption.
  During analysis, each vertical column in real power consumption values are compared with each vertical column in hypothetical
  power consumption values using Pearson correlation. When the column at the correct time moment is compared with the correct key 
  as marked in \autoref{f:align}, it would cause the highest correlation. This is how the correct key is identified.
  Note that for this to work, the alignment of the power traces is an important requirement.

\begin{figure}[htb]
 	\begin{center}
 		\includegraphics[width=12cm]{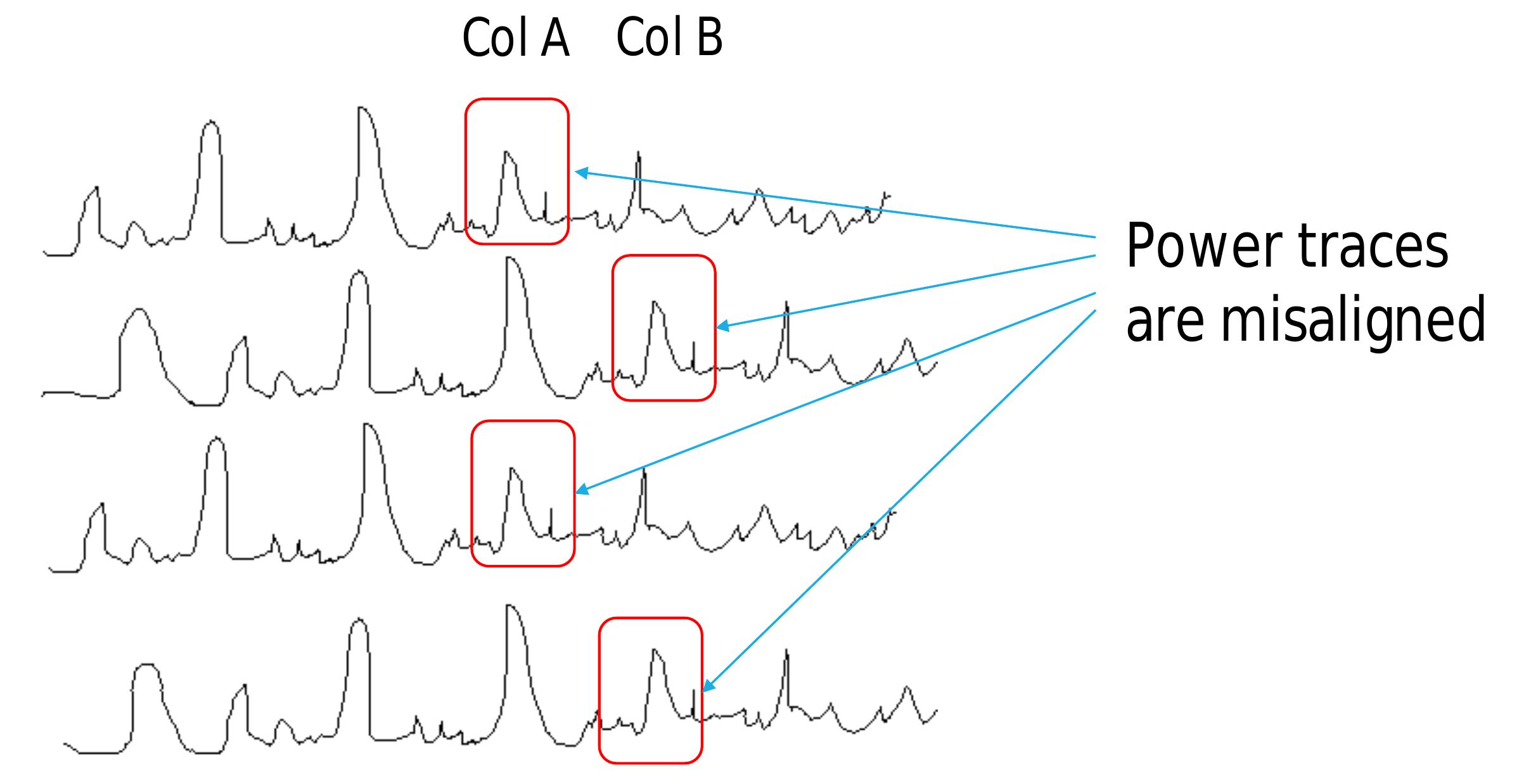}
 	\end{center}
 	\caption{\label{f:misalign}How power traces are misaligned when a single random instruction is inserted}
 \end{figure}

  In random instruction injection, encryption program is written such that additional instructions are randomly inserted and executed at the
  middle of the encryption.  \autoref{f:misalign} shows the same four power traces in \autoref{f:align}, but now with just one random instruction injected. One random instruction means that sometimes no random instruction is injected but sometimes one random instruction is
  injected. The first power trace in \autoref{f:misalign} is same as in  \autoref{f:align} because in this case no extra
  instruction has been injected. The marked position in the trace is the position that corresponds to the
  access to the specific intermediate value that is going to be used for the attack. For the second encryption, a random instruction has been injected at the middle. Therefore now the corresponding point has shifted to the right as marked in \autoref{f:misalign}.
  Because of this phenomena, the power traces are misaligned as shown in \autoref{f:misalign}. 
  Now when you consider a certain vertical column in the power trace set, the specific time sample not only has power consumption values for the intended operations but also for random operations. Therefore the pairwise comparison is affected.
  
 \subsection{Experiments and results}

 Mangard \cite[pp.~200-205]{mangard} mathematically proved that even though the power traces are misaligned still the attack would work when the number of power traces are increased. In the example depicted by \autoref{f:misalign},
 if the randomness was uniformly distributed, half of the power traces would have the corresponding point in column A while the other half would be in column B in \autoref{f:misalign}.
 Therefore if enough power traces are used for the analysis, the power consumptions due to random operations could be considered noise and the attack would still work. 
 
 When the number of random instructions injected is increased, the number of columns which the required operation would fall also increase. If N number of random instructions are injected, any number of random instructions from 0 to N can be injected to the middle. In that case there would be N+1 columns which the necessary moment can fall into. Mangard \cite[pp.~200-205]{mangard} shows that the required number of power traces in such a case increases by (N+1)\textsuperscript{2} times. The increase in power traces lets enough number of correct power consumption points fall into the column as well as to compensate the
 effect on SNR ratio due to the random power consumption points.
 
 The AES implementation in \autoref{c:testbed} was modified to inject random instruction. AES was selected for testing as AES was more vulnerable than Speck and because the most vulnerable is better for testing countermeasures. 
 In AES, Sbox lookup is the point used for the attack.
 Our implementation is such that N number of random instructions are only inserted between the trigger and the first Sbox lookup. As a result the power traces would be shifted to the right from 0 to N number of slots.
 
 \begin{figure}[htb]
 	\centering
 	\subfloat[]{\includegraphics[width=3in]{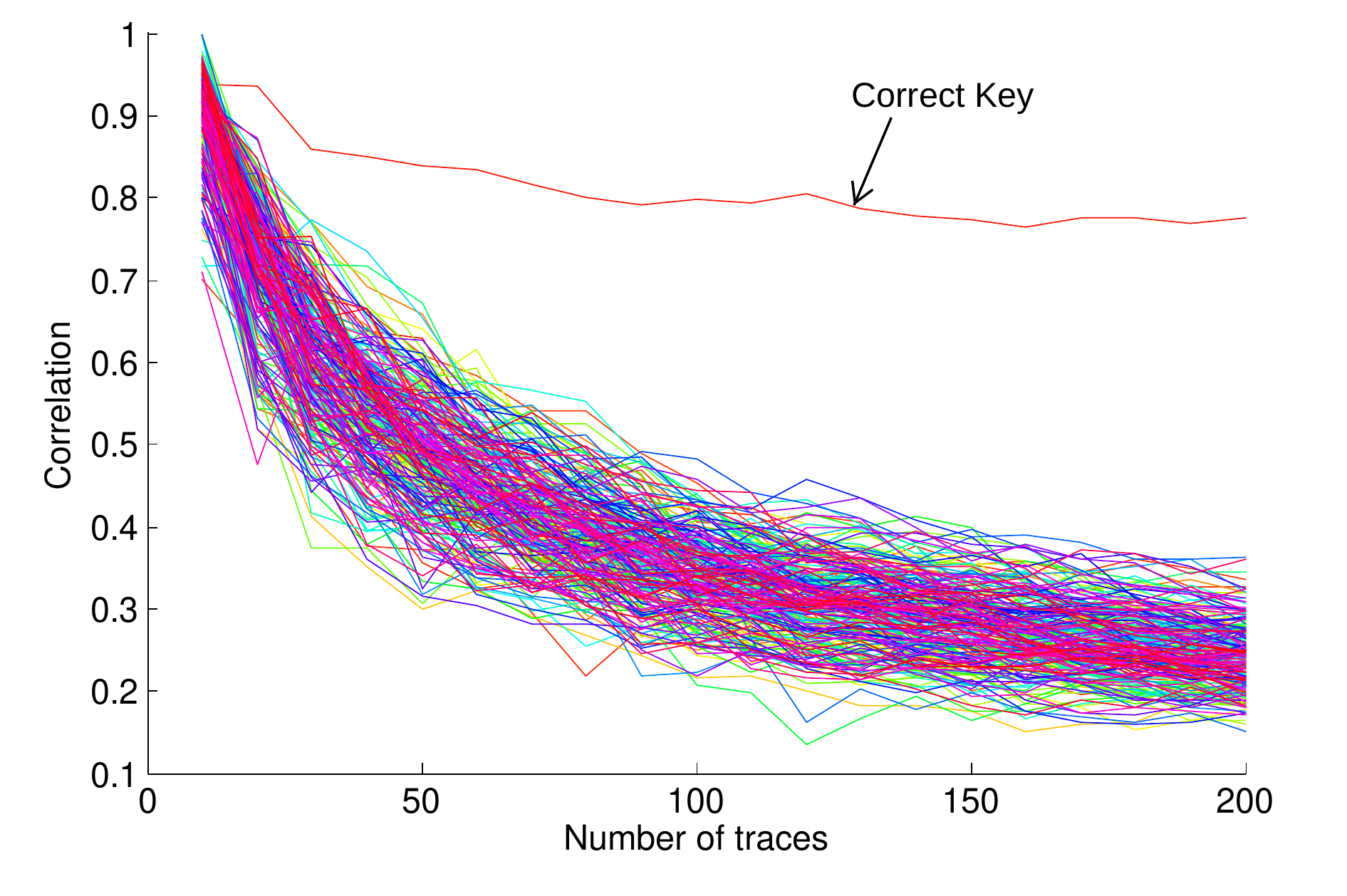}
 		\label{f:random0}}
 	\subfloat[]{\includegraphics[width=3in]{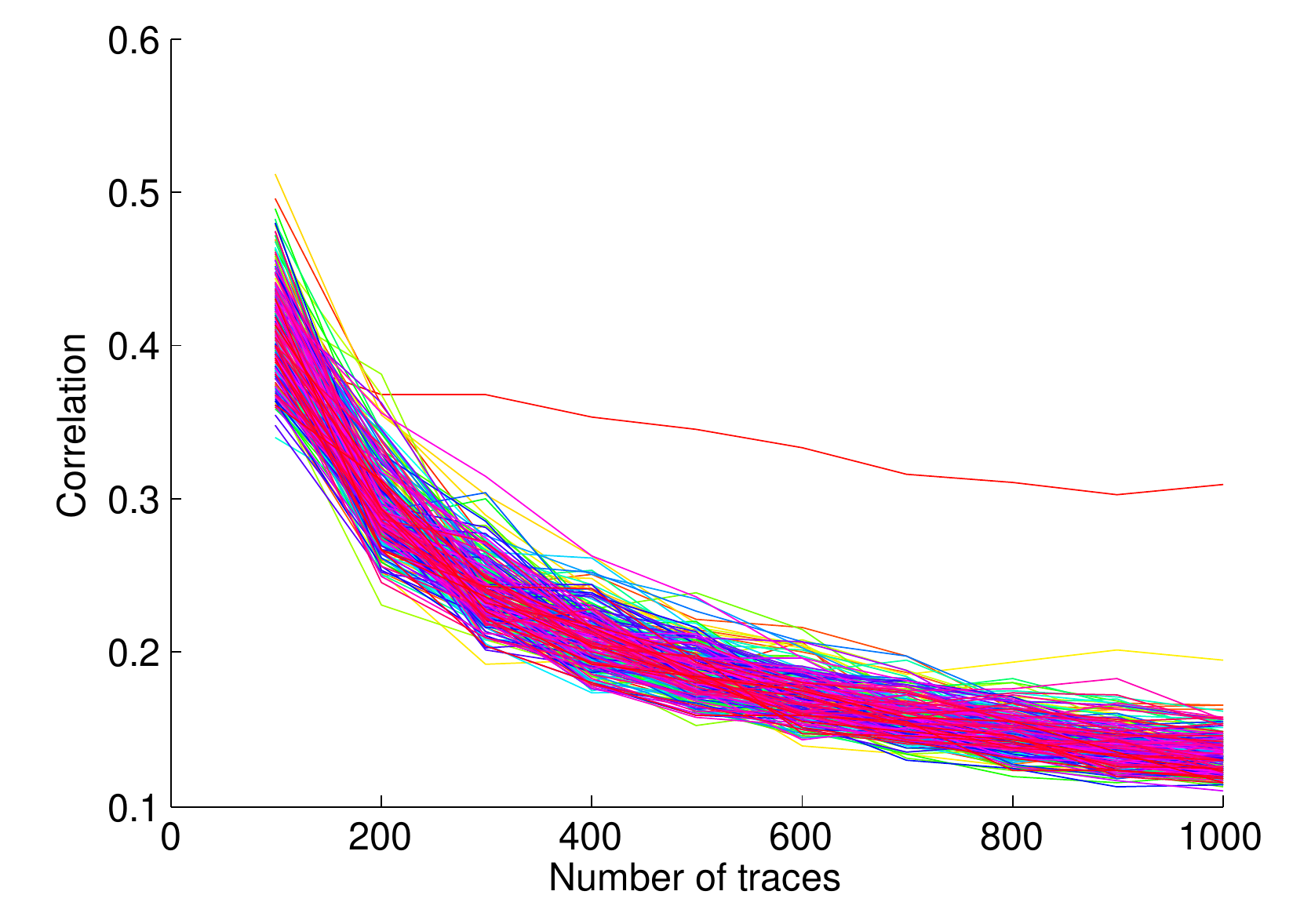}
 		\label{f:random1}}
 	\\
 	\subfloat[]{\includegraphics[width=3in]{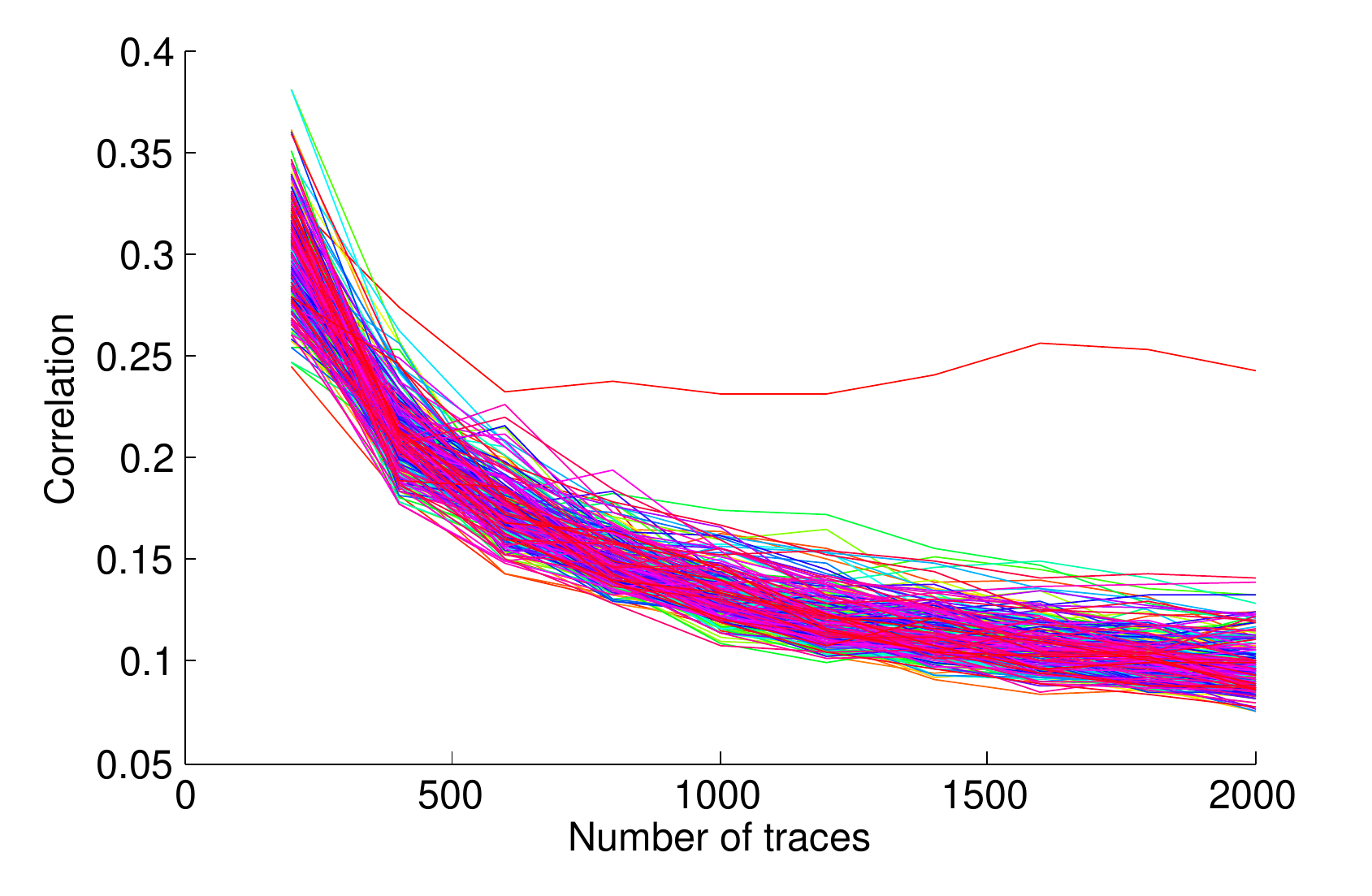}
 		\label{f:random3}}
 	\subfloat[]{\includegraphics[width=3in]{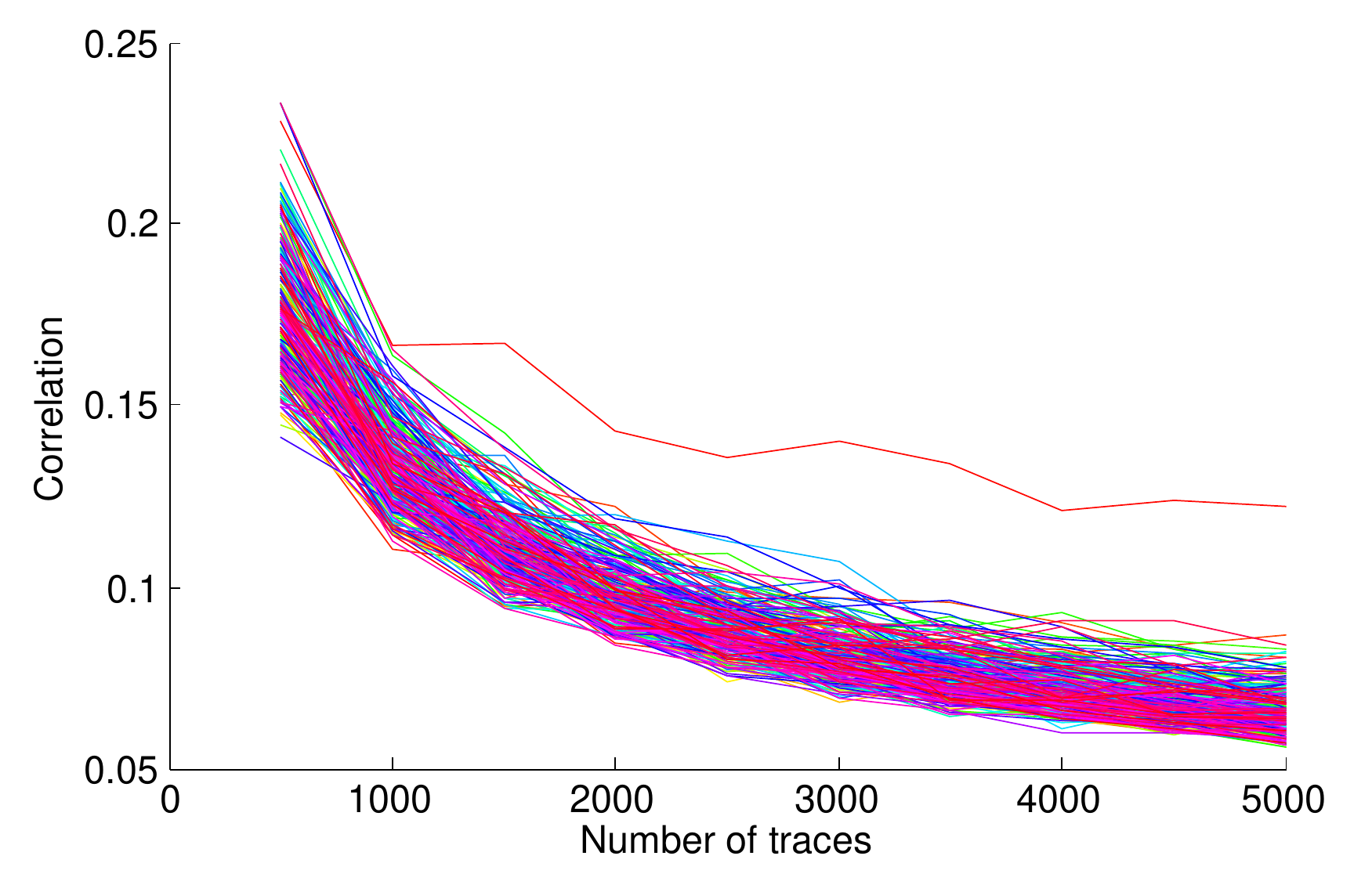}
 		\label{f:random7}} 	
 	
 	\caption{The number of power traces required for injection of (a) no random instructions (b) one random instruction (c) Three random instructions (d) seven random instructions}
 	\label{f:random}
 \end{figure} 
 
 The results for the first keybyte that corresponds to the first Sbox are shown in \autoref{f:random}. How the correlation coefficient changes with the number of power traces for each key is shown for different number of random instructions injected. Figure \autoref{f:random0} is when no random instructions are inserted. Even before 50 traces, the correct key is significant. 
 When one random instruction is injected the number of power traces at which the key becomes significant has increased up to more than 200 according to Figure \autoref{f:random1}. Then according to Figure \autoref{f:random3}, the number of power traces has grown up to more than 500 traces when three random instructions are injected. As shown in Figure \autoref{f:random7}, the number of power traces further increases up to more than 2000 power traces when 7 random instructions are injected.
 
 \autoref{t:randomsummary} summarizes the approximate minimum number of power traces and also the approximate minimum time required to carry out a successful attack for different number of random instructions. When no random instructions are injected, the attack can be done in no time. But with the number of 
 random instruction insertions, the required time as well as the time increase in a quadratic fashion.
  At fifteen random instruction injections it requires more than 40000 power traces where the time required for the attack is more than 45 hours. 
  If the number of traces increase at this rate when about 100 instructions are injected, the required number of traces would be massive. Assuming a quadratic relationship it can be predicted that 200 random instructions would need more than 500,000 power traces which would increase the attack time to more than 25 days.
  Therefore when comparing with the circuit countermeasures tried before in \autoref{c:hardware}, random instruction method appears to be more effective.
 
 	\begin{table}[!t]
 		\begin{center}
 			\begin{tabular}{|p{4cm}|p{4cm}|p{4cm}|}\hline
 				\textbf{Number of random instructions injected} & \textbf{Approximate minimum number of traces} & \textbf{Approximate minimum time}\\\hline
 				0 & 50 & 5 minutes\\\hline
 				1 & 200 & 15 minutes\\\hline
 				3 & 500 & 30 minutes\\\hline
 				7 & 2000 & 2 hours\\\hline
 				15 & 40000 & 45 hours\\\hline			
 			\end{tabular}
 		\end{center}
 		\caption{Number of power traces and time required for different number of random instruction injections}
 		\label{t:randomsummary}	
 	\end{table}

%
 
\begin{figure}[htb]
 	\centering
 	\subfloat[]{\includegraphics[width=3in]{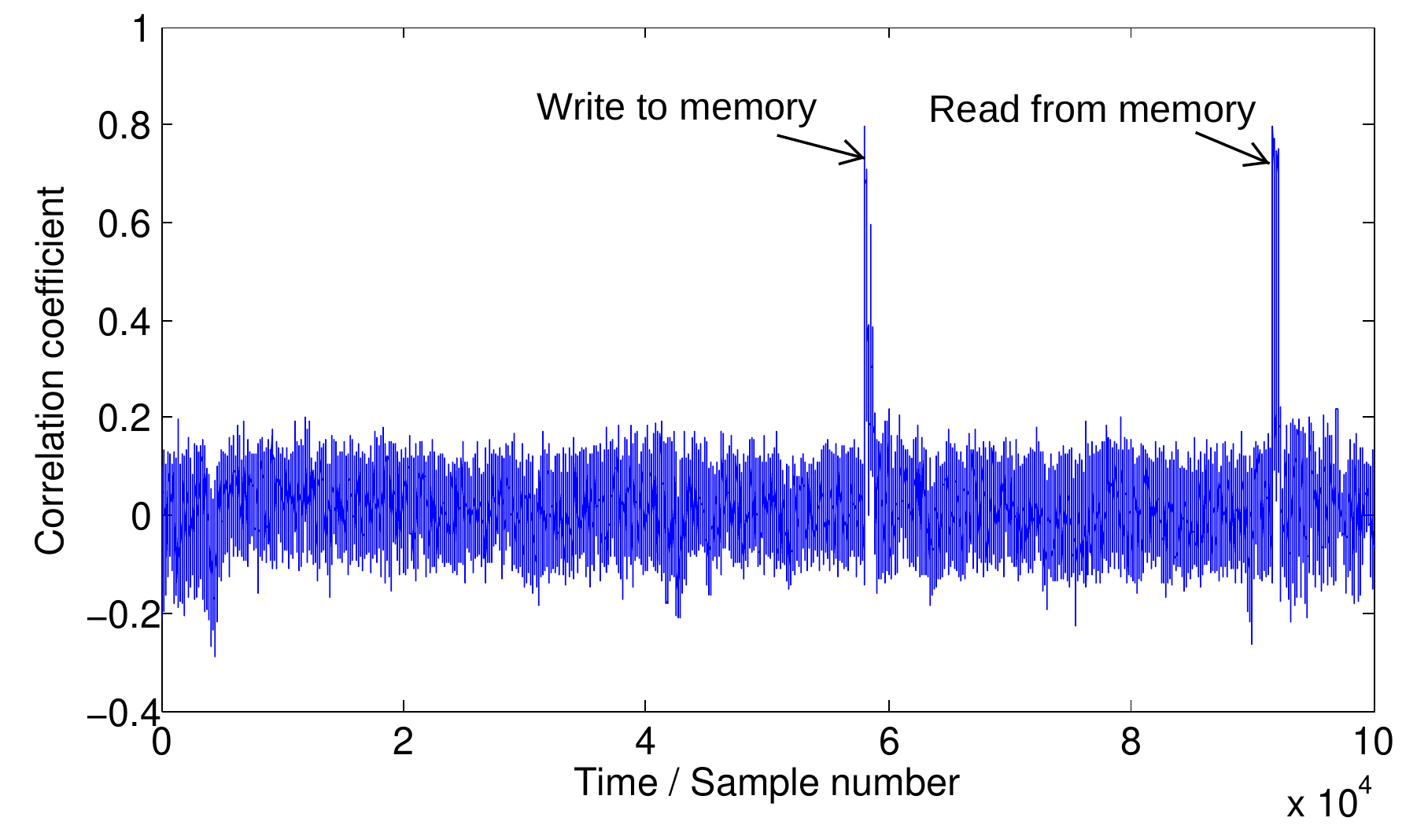}
 		\label{f:peaksshortandwide0}}
 	\subfloat[]{\includegraphics[width=3in]{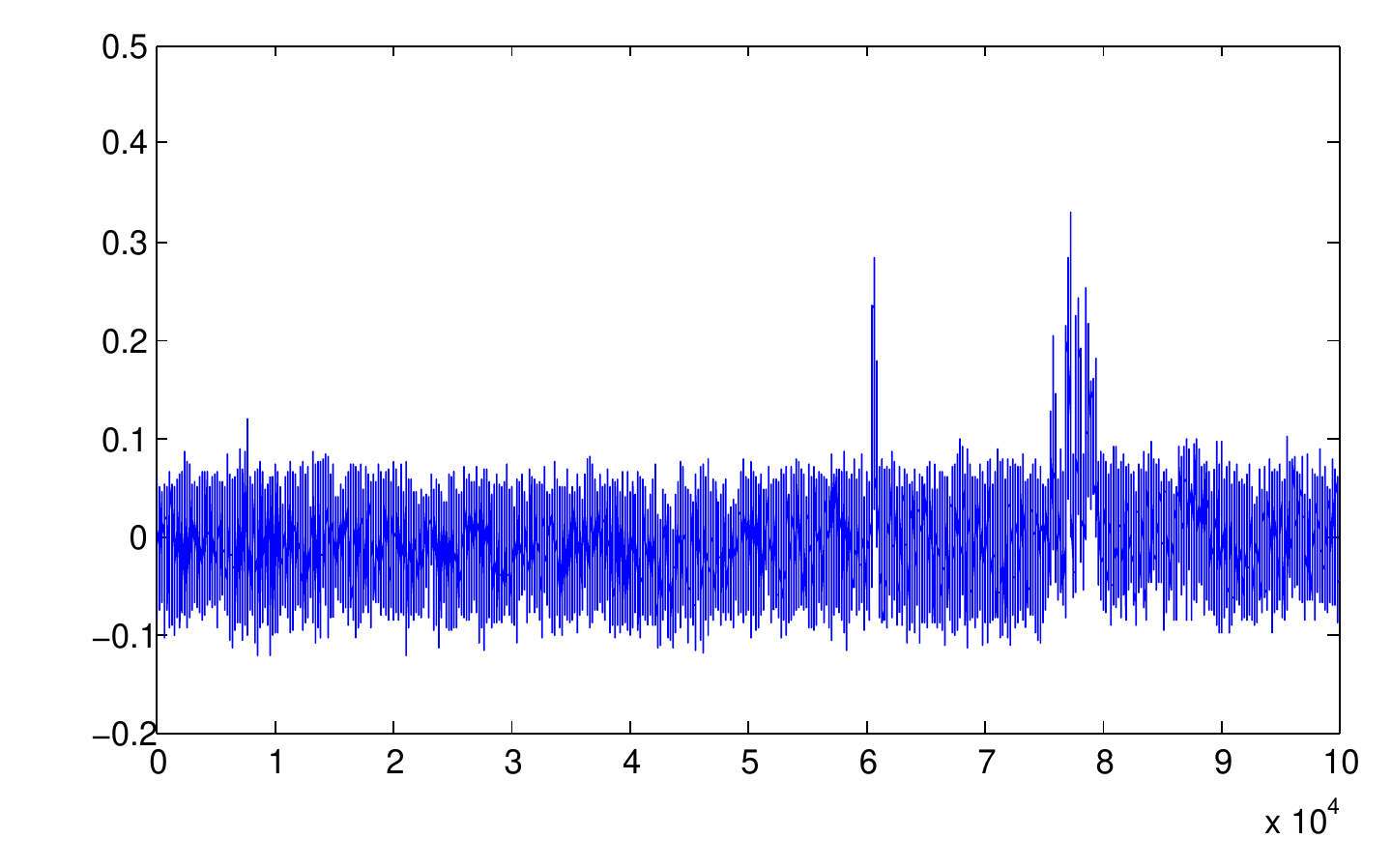}
 		\label{f:peaksshortandwide1}}
 	\\
 	\subfloat[]{\includegraphics[width=3in]{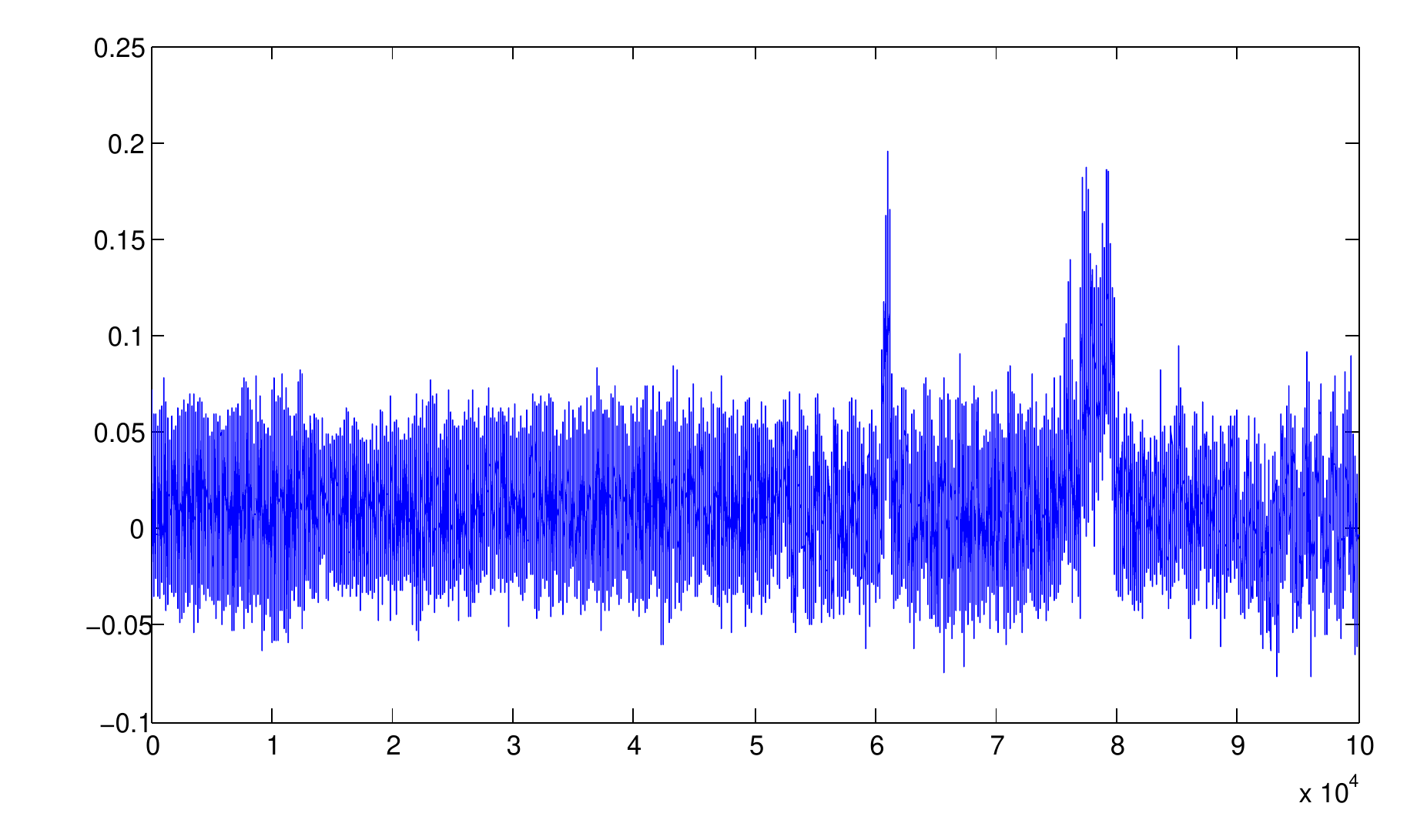}
 		\label{f:peaksshortandwide3}}
 	\subfloat[]{\includegraphics[width=3in]{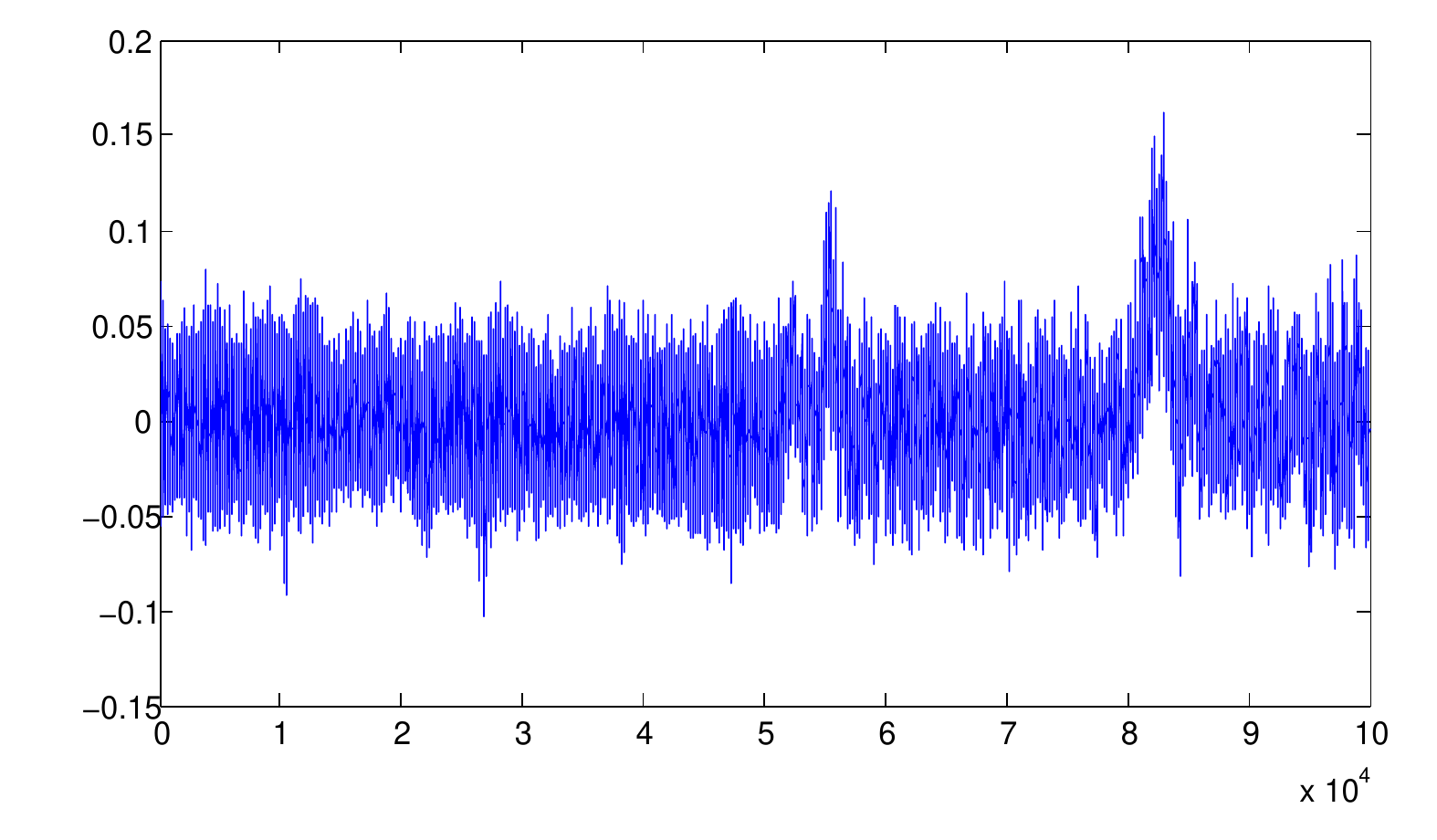}
 		\label{f:peaksshortandwide7}} 	
 	
 	\caption{Correlation peaks for injection of (a) no random instructions (b) one random instruction (c) Three random instructions (d) seven random instructions}
 	
 	\label{f:peaksshortandwide}
 \end{figure}

 An interesting observation could be made in correlation coefficient vs time graphs for the correct keys. \autoref{f:peaksshortandwide} shows those graphs for different number of random instructions injected. The x axis in these graphs denote the time with respect to the sample number of the power traces. According to Figure \autoref{f:peaksshortandwide0} it can be observed that the correlation peaks for zero random instruction injections is high. But the rest of the sub figures show that when the number of random instruction injections increases, the correlations peaks become shorter and wider.
 It was discussed earlier with respect to \autoref{f:misalign} that when the number of random instructions increases, the number of possible slots in time that a correct power consumption point can fall also increases. All these slots will have a certain number of correct power consumption points useful for the attack. Therefore all these slots would cause correlation peaks. As the slots are adjacent we see that the width of the correlation peaks becoming large with the increase in number of random instructions.
 The same fact causes the reduction in the height of correlation peaks as well. If N random instructions are injected there would be N+1 slots and the probability of a correct power consumption point falling on to a certain slot is 1/(N+1), if the randomness is uniform. Therefore, when the number of random instructions increases, the ratio between the correct power consumption points to wrong points for a considered time slot decreases. Wrong power consumptions points are like noise and hence the SNR ratio also decreases. Because of that, the correlation coefficients becomes smaller with
 the number of random instructions. Hence the correlation peaks appear to become shorter. Here note that the correlation peaks becoming short means that the value of the correlation peaks becoming short.  Because the y axis of sub figures in Figure \autoref{f:peaksshortandwide} are not in same scale this is not directly visible unless the scale of the y axis is observed.
 
 Though random instruction injection is a simple but effective method that can be easily implemented on any microcontroller, its main draw back is the wastage of CPU time. More and more random instructions increase the security but yet increase the CPU time as well. This reduces the efficiency of the cryptosystem. Further power is wasted for useless computations. The AES implementation we used for PIC was around 700 instructions. Therefore adding like 100 extra instructions itself would increase the time and power consumed by 1/7 times. But not only them, but also considerable amount of instruction would be added for generating random numbers. This will further increase time and power consumption. 
 
  \section[Sbox shuffling]{\label{s:sboxshuffle}Randomly shuffling the Sboxes}
  
  Randomly shuffling Sboxes is a countermeasures mentioned in \cite{mangard}. Many block ciphers such as AES and DES
  contain a non linear type of operation called substitution boxes (Sboxes). AES algorithm in each round consists of 16 Sbox operations which happen byte wise. These 16 Sbox operations in a certain round are independent to each other and hence the order which they are carried out is not important. Therefore the order in which these Sbox operations happen can be randomized without any problem. This is known as randomly shuffling the Sboxes.   
  
  The principle on which randomly shuffling the Sboxes, works as a countermeasure is extremely similar to the concept explained in 
  \autoref{s:priciples}. The difference is that, misalignment of power traces in \autoref{s:priciples} 
  was due to injected instructions, but here it is due to shuffling the order in which the Sbox operations happen.
  
  If Sbox operations happen in same order for all encryptions, the power consumption
  peaks due to those operations would be nicely aligned in same time moment in all power traces.
  But when they are randomized the alignment is damaged. If all 16 Sboxes are randomized, then for a certain time moment
  there are 16 possible different Sbox operations to happen. Therefore due to the reasons described in \autoref{s:priciples},
  the number of power traces required increases. The same quadratic increase in number of power traces
  occurs here and for N number of shuffled Sboxes, the number of power traces required would increase by N\textsuperscript{2}.
  Note that this is not (N+1)\textsuperscript{2} as in \autoref{s:priciples}. In \autoref{s:priciples}, N was the number of
  random instructions injected and to include the correct instruction itself, it is made N+1. But in Sbox shuffling it 
  is not the case.
  
  An advantage of randomly shuffling the Sboxes is the simplicity. When compared to random instruction injection this method is more efficient as well. In random instruction injection, the extra added random instructions consume extra time and power.
  But in random shuffling of Sboxes, it is the actual work that is shuffled. Therefore no such wastage of time and power occurs.
  
  But the disadvantage is that the maximum possible shuffling is equal to the number of Sboxes. For AES, the maximum possible shuffling possible is 16. Therefore the number of required traces can be increased up to 256 times, but no more.
  On the other hand any number of random instructions can be injected. Depending on the number of injected instructions, the
  required number of power traces would increase. Another approach is to introduce some false Sbox operations so that
  the degree of shuffling can be increased. But also note that randomly shuffling Sboxes can be done together with
  random instruction injection. As Sbox shuffling does not consume extra time and power, it can be used as a way to improve
  the security of a system that injects random instructions by some amount at extra no cost.
  Another apparent issue is that not only Sboxes leak information through power consumption.
  It was discussed in \autoref{s:nummoretraces} that Sboxes are very much vulnerable to power analysis attacks when compared to other operations. But yet we also showed that other operations such as xor can be used for attacking as well. Shuffling
  Sboxes will not reduce leakage due to other operations. But still any set of independent operations can be shuffled in
  a similar fashion. For example in an 8 bit Speck implementation, as xor operations are happened byte wise, they can be shuffled similarly.
  
\begin{figure}[htb]
 	\begin{center}
 		\includegraphics[width=12cm]{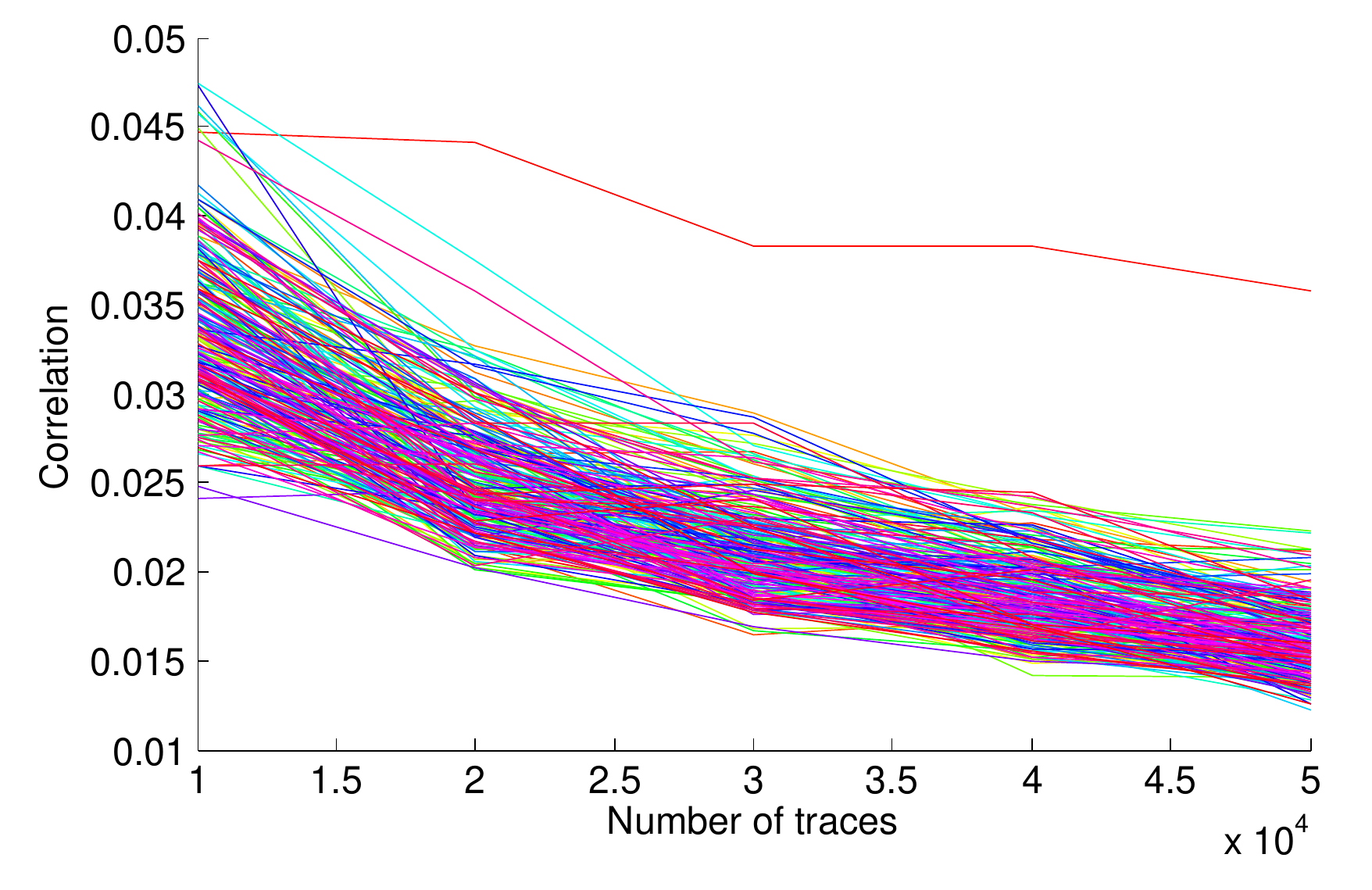}
 	\end{center}
 	\caption{\label{f:sboxtraces}Number of power traces required when Sbox operations are randomly shuffled}
 \end{figure}

\begin{figure}[htb]
 	\begin{center}
 		\includegraphics[width=12cm]{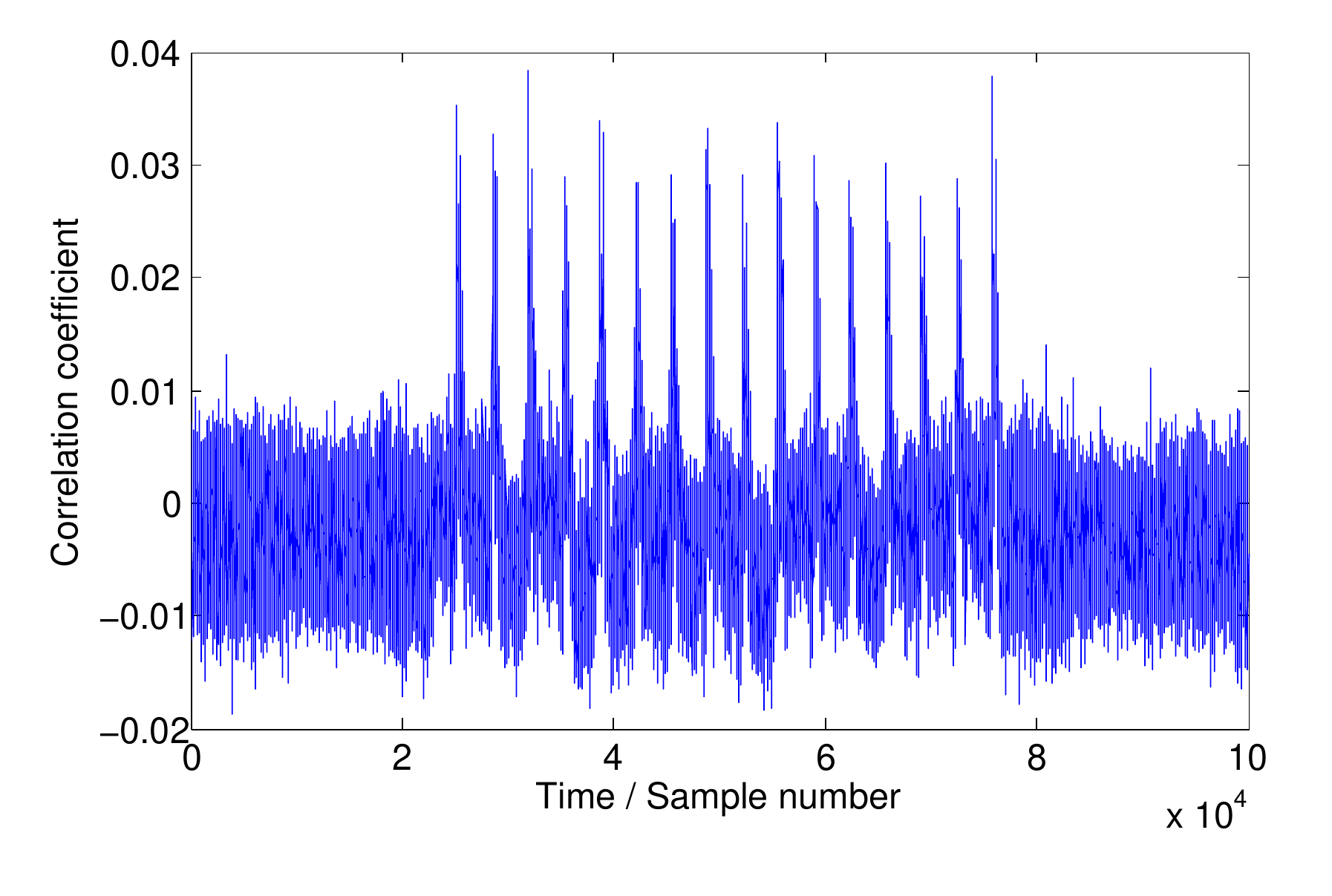}
 	\end{center}
 	\caption{\label{f:sboxtime}Correlation peaks when SBox operations are randomly shuffled}
 \end{figure}

  The AES implementation was modified by us to integrate the random Sbox shuffling functionality. After programming the PIC2550 based testbed, power traces were taken to carry out an attack. After collecting about 20000 power traces, the attack became successful. 
  \autoref{f:sboxtraces} shows how the correlation coefficient varies with the number of power traces. From that it
  is observable that even at about 15000 traces the correct key becomes significantly deviated from the others.
  \autoref{f:sboxtime} shows how the correlation coefficient changes with the time. The plot is only for the time range during which Sbox lookups are done and the results are written to the memory. If Sboxes were not shuffled, only one correlation peak should be visible with respect to the point at which the result of the Sbox operation is written. But here 
  16 peaks are visible. Since all 16 Sbox lookups are shuffled randomly, a given Sbox operation can fall on to one of 16 possible time moments. When large number of power traces are taken and if the randomness is uniformly distributed, each time slot would 
  have approximately the same number of the specific Sbox that is currently under attack. These cause approximately equal correlation peaks.  But note that the height of correlation peaks here are lesser than 0.04, where in \autoref{f:sboxcorrect} when no Sbox shuffling was done, the peaks were larger as 0.8. This reduction of height of correlation peaks is due to the low
  SNR caused by the shuffling.

  \section[Improvements]{\label{s:trng}Hardware random seed generator as an improvement for random instruction generation}
   
  The security of the random instruction injection method as well as randomly shuffling the Sboxes depends on the randomness. The time taken for a certain instruction can be easily calculated using the clock frequency of the device. If the attacker knows the random number sequence determined by the device,
  he can easily realign the collected traces. Random numbers in software are generated using pseudo random generators \cite{random}. Pseudo random generators are algorithms which generate a sequence of random numbers. But the random number sequence depends on an initialization value called the seed. Therefore, pseudo random generators are also known as deterministic random number generators. If the seed is the same, the generated random number sequence will be always the same
  for a given pseudo random algorithm. 
  
  If the seed is hard coded in the program of the cryptosystem, it would be static and once an attacker figures it out by dissembling the program or a similar method, there is no use of that countermeasure any more. In a computer, the candidate for the seed for a random number is
  usually the current time given in milliseconds. As milliseconds flow quite fast it would be a challenge for an attacker to figure out the seed. But embedded cryptographic devices are made of microcontrollers and these microcontrollers usually do not have clocks that keep track of the current time (real time clocks). Therefore some other method for generating a random seed is required.
  We introduce a technique to improve the security of random instruction injection by providing the seed for the pseudo random generator using a true hardware random generator. 
  
\begin{figure}[htb]
 	\begin{center}
 		\includegraphics[width=10cm]{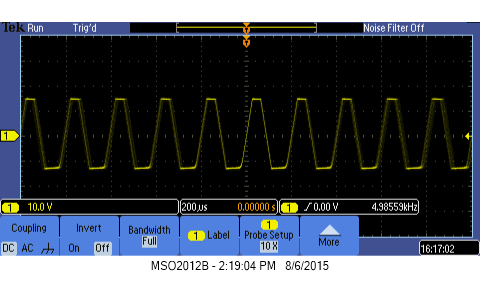}
 	\end{center}
 	\caption{\label{f:saturated}Amplification of thermal noise via a resistor through an op amp caused saturation}
 \end{figure}

  Natural phenomena such as thermal noise across a resistor, nuclear decay of atoms, signal produced by base of a reversed biased transistor
  and photoelectric effect are unpredictable using theories in physics \cite{trng}. Therefore such processes are considered true random.
  Hardware that uses such processes to generate random numbers are known as true random generators \cite{random}. Amplifying
  thermal noise across a resistor or the signal produced by base of a reversed biased transistor provides an easy way to generate a random signal \cite{trng}. We first attempted amplifying thermal noise across a one mega ohm resistor via an operational amplifier, but the 60 Hz noise from the power lines made the operational amplifier go to saturation causing waveform similar to the one shown in \autoref{f:saturated}. But amplifying the signal produced by base of a reversed biased transistor, gave good results and hence it was used to implement the true random generators discussed in the following subsections.
   
   \subsection{\label{s:setup1}Setup 1}
   
   A circuit for a random number generator based on noise generated by a reverse biased transistor is found in \cite{rand1}. 
   We implemented that method on a breadboard to check how good it is as a hardware random number generator.
   
 \begin{figure}[htb]
 	\begin{center}
 		\includegraphics[width=12cm]{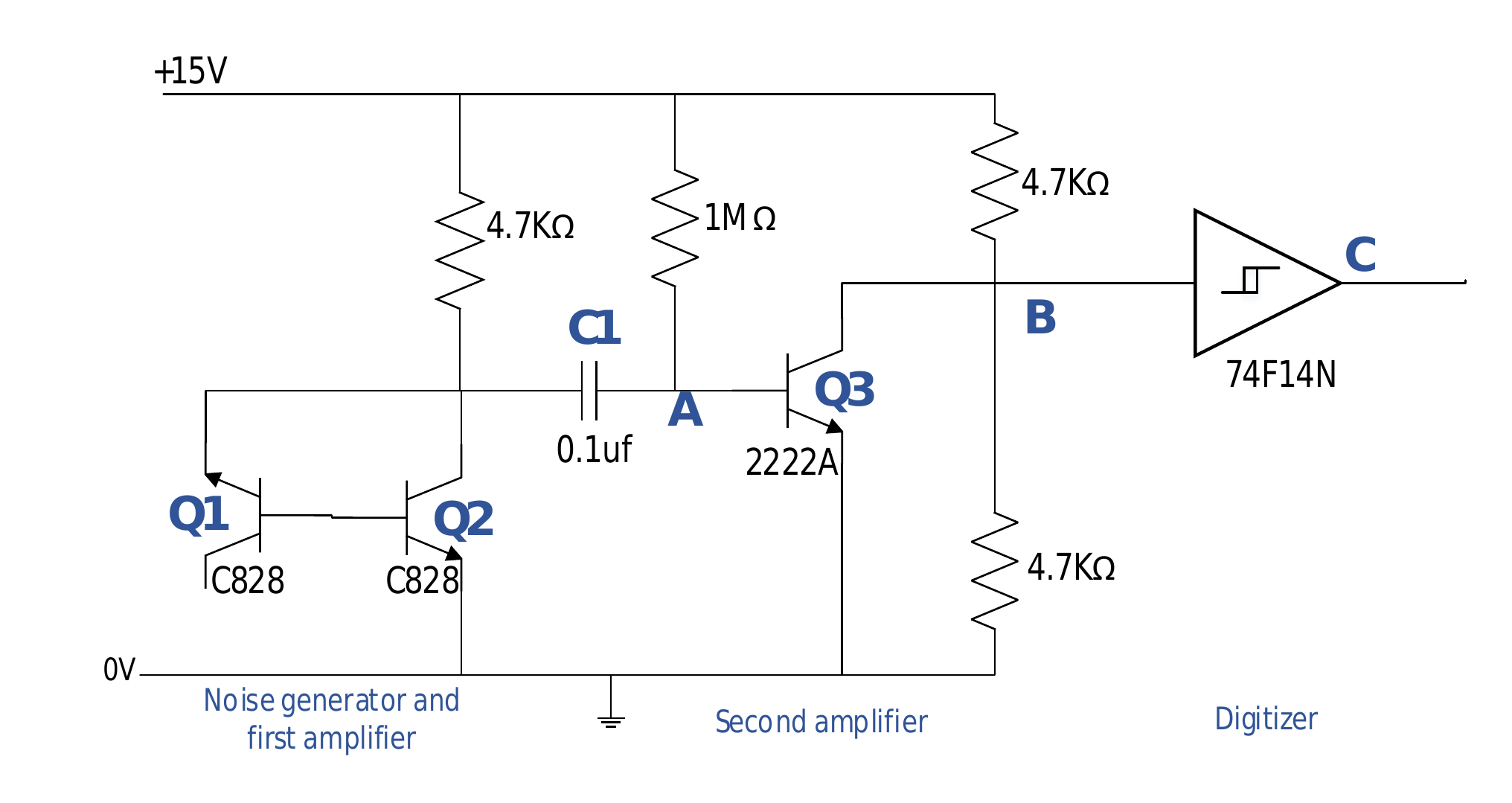}
 	\end{center}
 	\caption{\label{f:rand1}Hardware random number generator that generates random pulses}
 \end{figure}      
   
\begin{figure}[htb]
 	\begin{center}
 		\includegraphics[width=9cm]{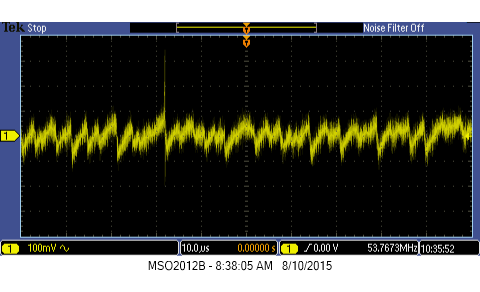}
 	\end{center}
 	\caption{\label{f:rand1A}Amplified reversed biased noise}
 \end{figure}

\begin{figure}[htb]
 	\begin{center}
 		\includegraphics[width=9cm]{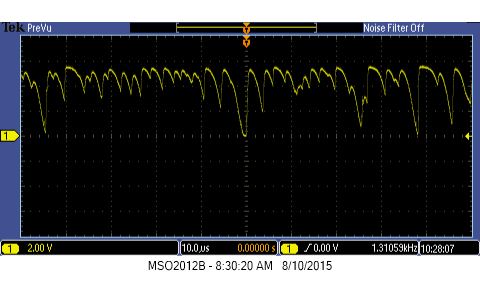}
 	\end{center}
 	\caption{\label{f:rand1B}Further amplified reversed biased noise}
 \end{figure}

\begin{figure}[htb]
 	\begin{center}
 		\includegraphics[width=9cm]{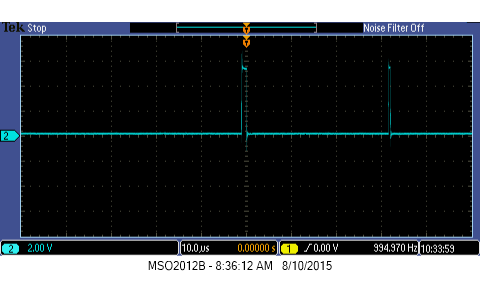}
 	\end{center}
 	\caption{\label{f:rand1C}Digitized random pulses}
 \end{figure}

   \autoref{f:rand1} is the circuit diagram of the circuit that we implemented. 
   The collector of the transistor Q1 is not connected. The base-emitter junction of Q1 is reversed biased. This causes a noise signal
   on the base of the Q1 transistor. Transistor Q2 which is in the active region amplifies the noise signal.The amplified noise signal observed
   at position marked as A in \autoref{f:rand1} is given in \autoref{f:rand1A}. Then the transistor Q3 in \autoref{f:rand1} 
   further amplifies the signal. The signal observed at position B in \autoref{f:rand1} is shown in \autoref{f:rand1B}.
   The signal is an analog signal and it is finally sent through a schmitt trigger to  create a digital signal with random pulses.
   The signal observed at the position marked as C is as shown in \autoref{f:rand1C}. The problem here is that, though pulses appear randomly, value 0 is much more present than value 1. This issue is later solved using software. 
   For Q1 and Q2, C328A transistors and for Q3, 2222A transistor was used by us. The selection was due to availability reasons and
   note that any transistor with a proper gain would work. 15V have been provided to reverse bias the base emitter junction of Q1 to 
   a level that produces enough noise. Any voltage value between 12V to 20V is suitable for the components used here.
   As the Schmitt trigger, we have used the IC 74F14N but even a not gate such as 74LS04 would be suitable.

   The point C in \autoref{f:rand1} is connected to a digital pin of the microcontroller of the cryptosystem. The digital pin is configured
   as an input pin. When the program running on the cryptosystem requests for a random seed, the signal on the digital pin
   is sampled repeatedly. As the signal on the pin is a random bit pulse as in \autoref{f:rand1C}, one sample gives a bit. 
   Therefore, if an n bit random seed is required, n samples have to be taken. By shifting each collected bit appropriately and then
   doing an or operation among them, an n bit random number can be formed. But the wave form in \autoref{f:rand1C} is zero biased as it was mentioned earlier. Therefore most of the bits in a generated random number would be zero. This would make the random generator
   biased to generate certain values only, which is not suitable. 
   To remove the bias, several random numbers generated in the above fashion were xored \cite{removebias}. 
   
\begin{figure}[htb]
 	\begin{center}
 		\includegraphics[width=14cm]{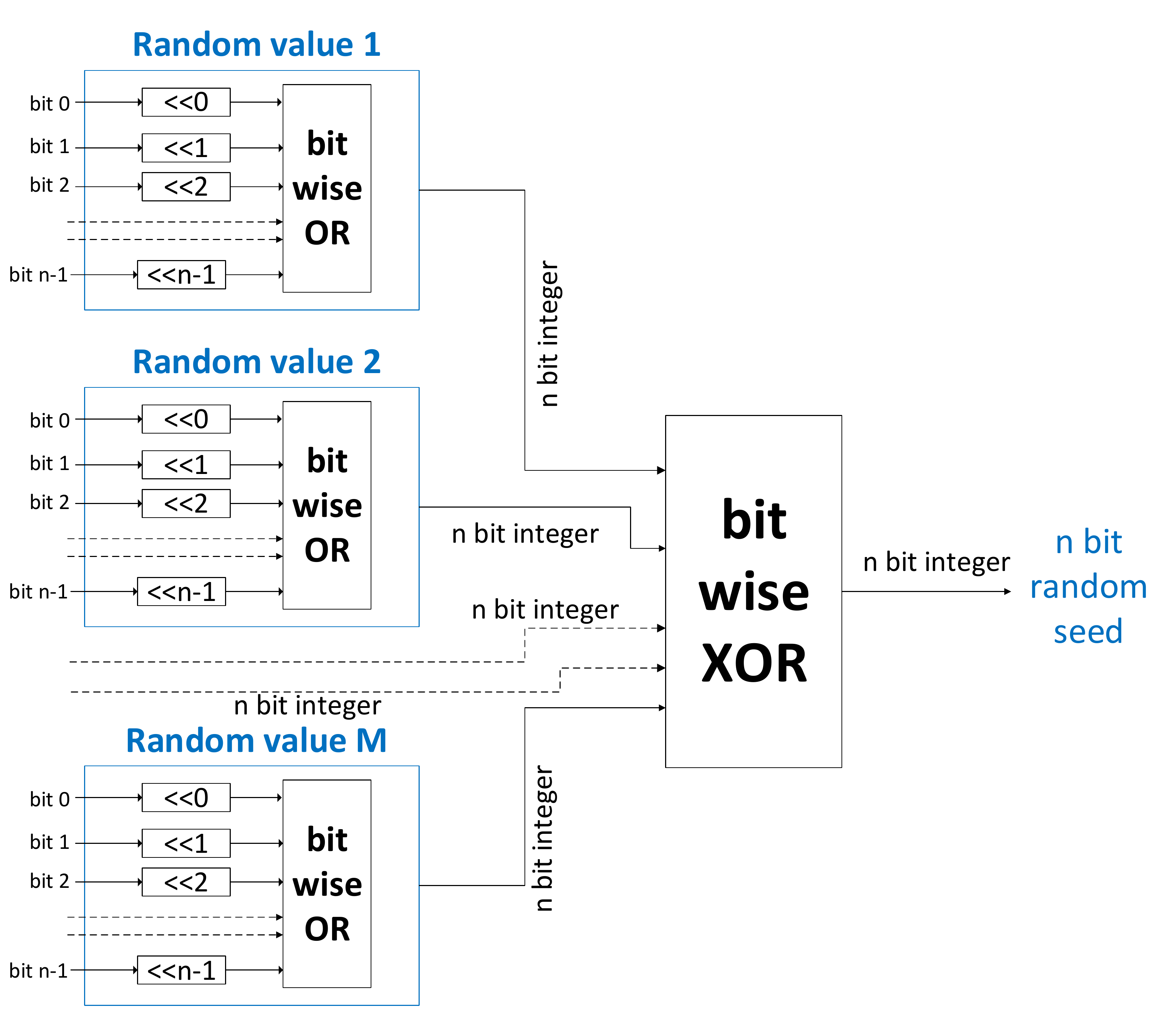}
 	\end{center}
 	\caption{\label{f:bitwiserandomseed}The process of generating an n bit random seed using digitized random pulses}
 \end{figure}

   The above process is pictorially elaborated in \autoref{f:bitwiserandomseed}. Here the final outcome is an n bit random seed. The value on the digital pin of the microcontroller is sampled n number of times to obtain bit 0, bit 1, ... bit n-1 shown in \autoref{f:bitwiserandomseed}. By shifting them as shown in \autoref{f:bitwiserandomseed} and then doing a bitwise or, the random value 1 is generated. But due to the 0 bias such generated values would not be uniformly distributed. Therefore the process is repeated M number of times to get M number of random values. Those are subjected to bitwise xor to obtain an unbiased n bit random seed as shown in \autoref{f:bitwiserandomseed}.

\begin{figure}[htbp]
 	\begin{center}
 		\includegraphics[width=9cm]{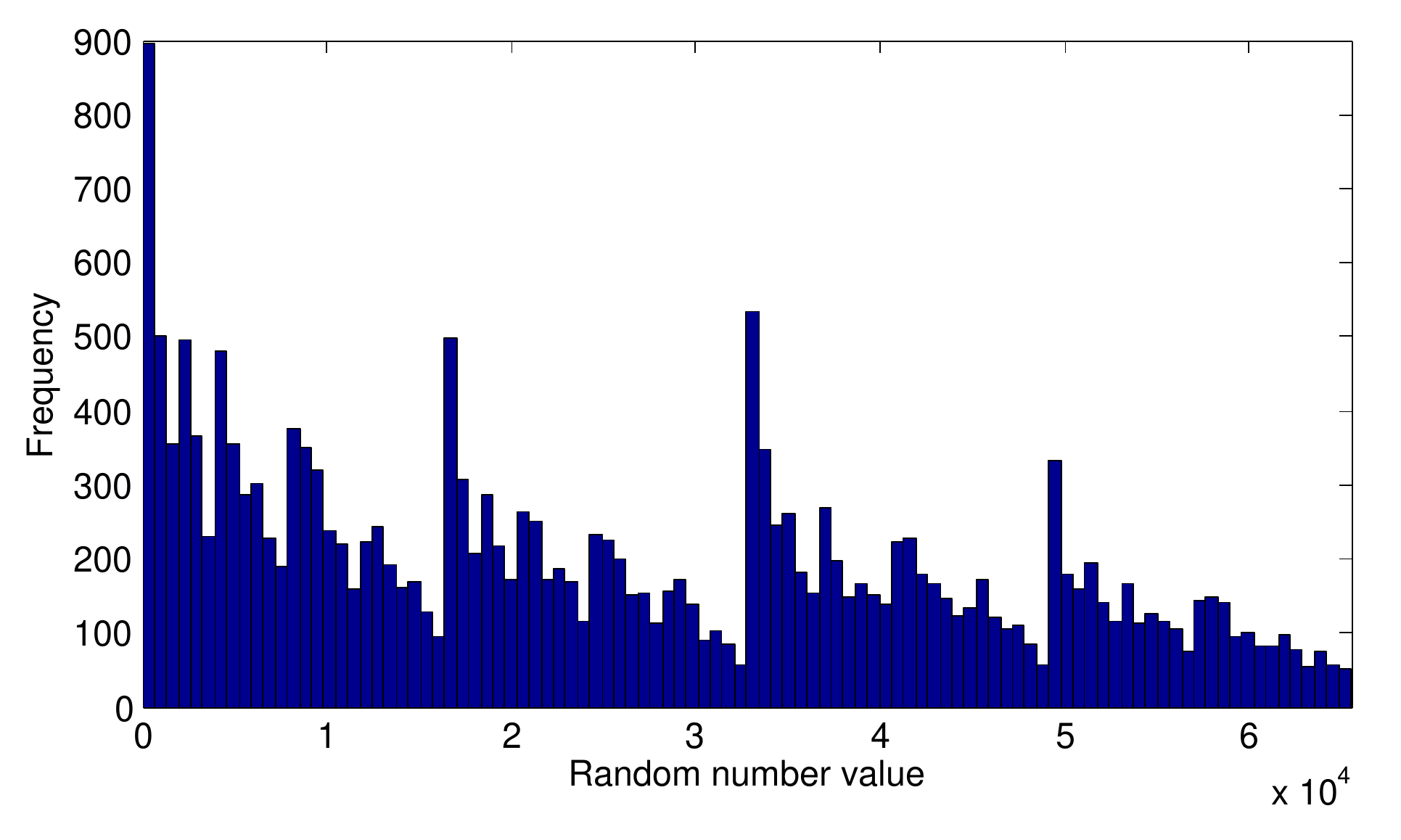}
 	\end{center}
 	\caption{\label{f:badhisto}Histogram for random numbers generated by xoring 100 values}
 \end{figure}

   The effectiveness of a hardware random generator must be definitely practically tested. 16 bit unsigned integers were formed
   in the above described method. Hundred such numbers were xored together to form a single random number. Large number of such random
   numbers (20000) were collected and the frequency distribution (histogram) was drawn to obtain \autoref{f:badhisto}. The x axis represents the numerical value of the random numbers and y axis is the frequency count. Since the range of 16 bit unsigned integers span from 0 to 65535 the axis has that range. The histogram contains 100 bins in the x axis. It is
   quite visible that the histogram is biased to certain values. The reason is that the number of xors done was not enough to remove the zero bias.
   
\begin{figure}[htbp]
 	\begin{center}
 		\includegraphics[width=9cm]{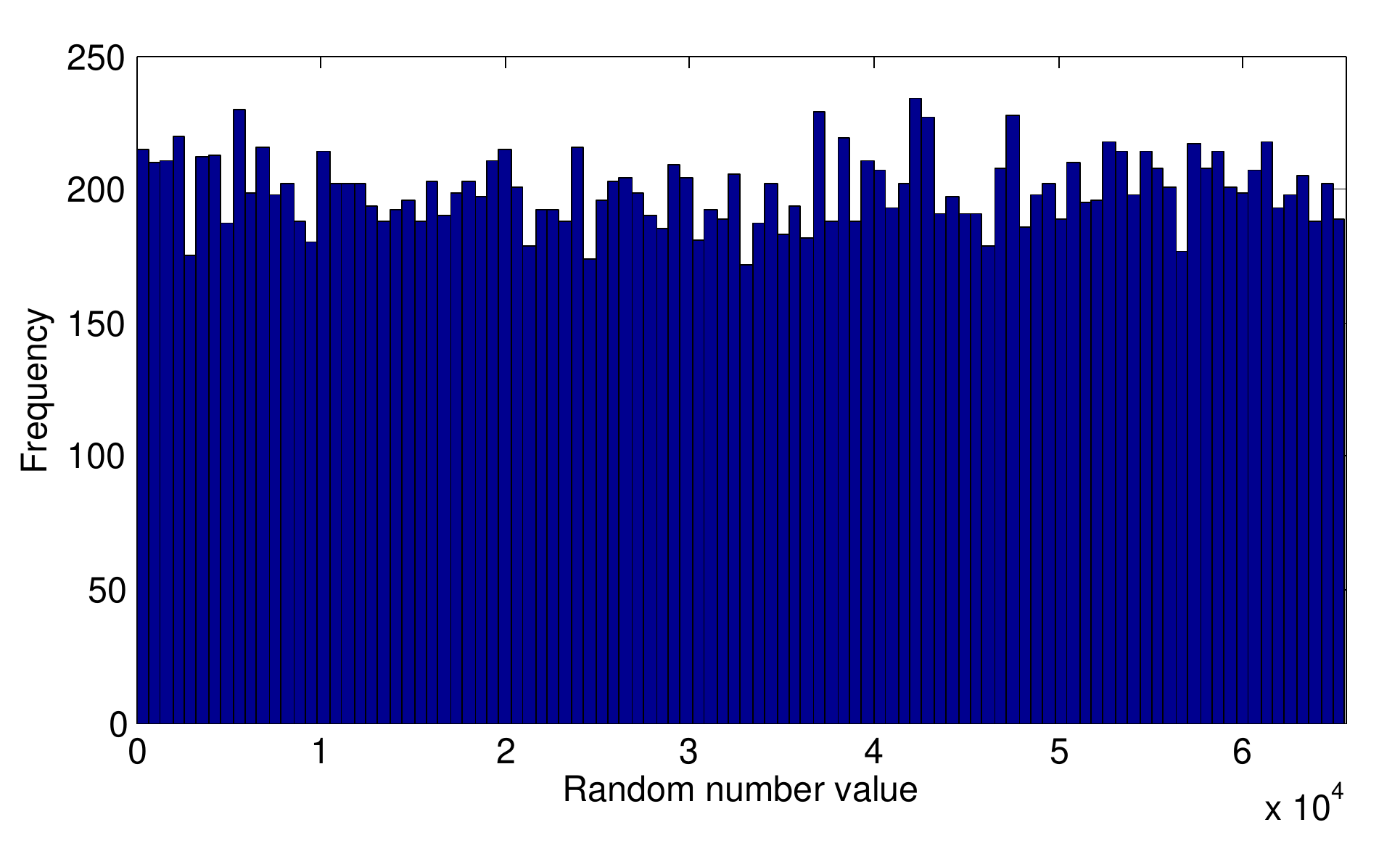}
 	\end{center}
 	\caption{\label{f:goodhistoran1}Histogram for random numbers generated by xoring 1000 values}
 \end{figure}

\begin{figure}[htbp]
 	\begin{center}
 		\includegraphics[width=9cm]{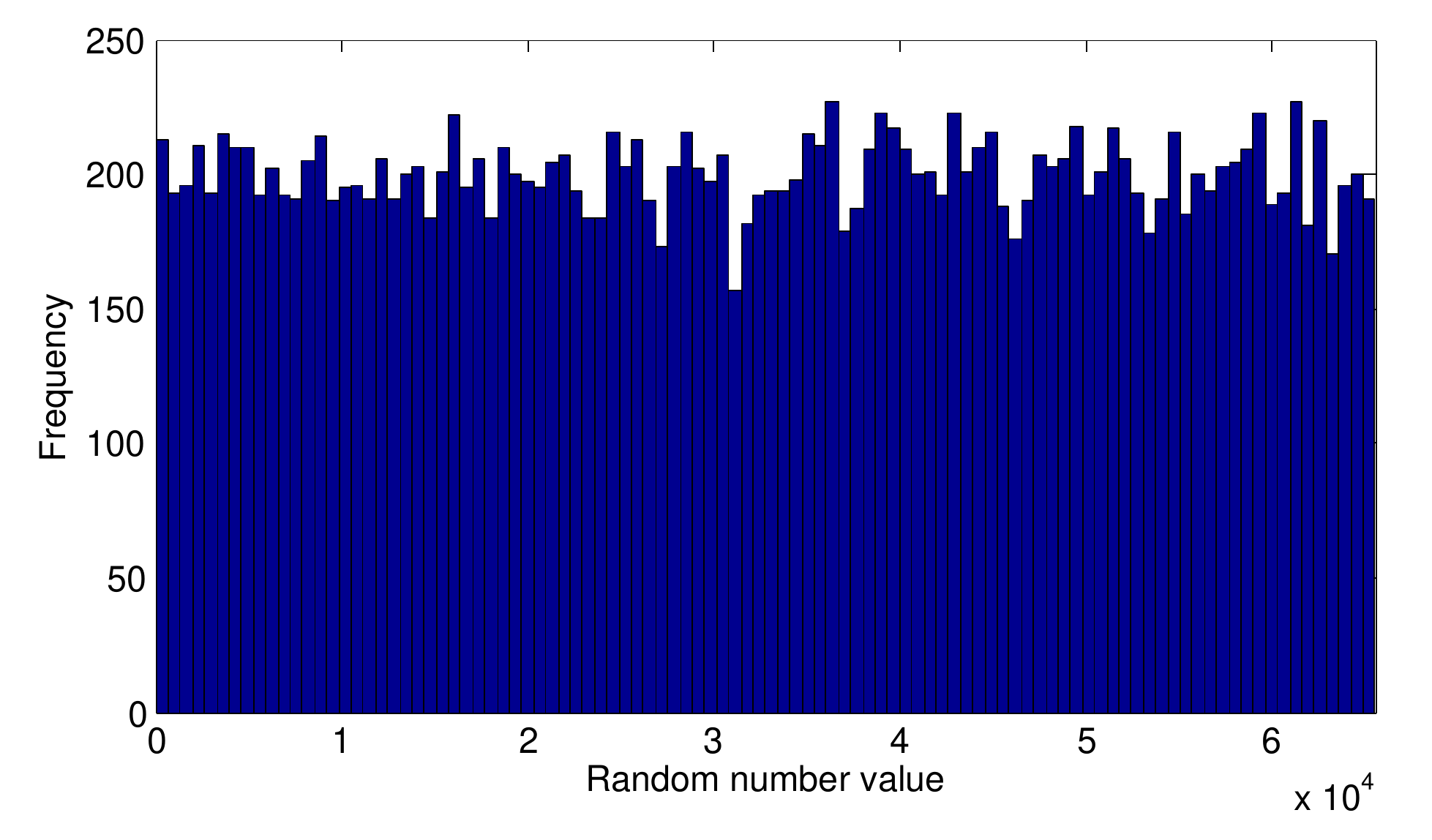}
 	\end{center}
 	\caption{\label{f:histomatlab}Histogram for random numbers generated by Matlab}
 \end{figure}

   When the number of values xored was increased 1000, a histogram as shown in \autoref{f:goodhistoran1} was obtained. Now an almost uniform distribution is seen which is good enough for a random number generator. In order to verify, a histogram was drawn by generating 20000 random numbers
   using the random number generator in Matlab. That histogram is shown in \autoref{f:histomatlab}. By comparing the histogram of our random number generator with the histogram for Matlab, it can be inferred that it is good enough.

 Another factor to consider when evaluating the effectiveness of a random number generator is whether the sequence generated 
 is periodic. A sequence with a uniform distribution of values is not  random if the sequence is periodic.
 Periodicity of an can be evaluated by taking the Fourier transform \cite{DSP}. A periodic signal would have a discrete
 frequency spectrum opposed to a continuous one. 
 
\begin{figure}[htbp]
 	\begin{center}
 		\includegraphics[width=9cm]{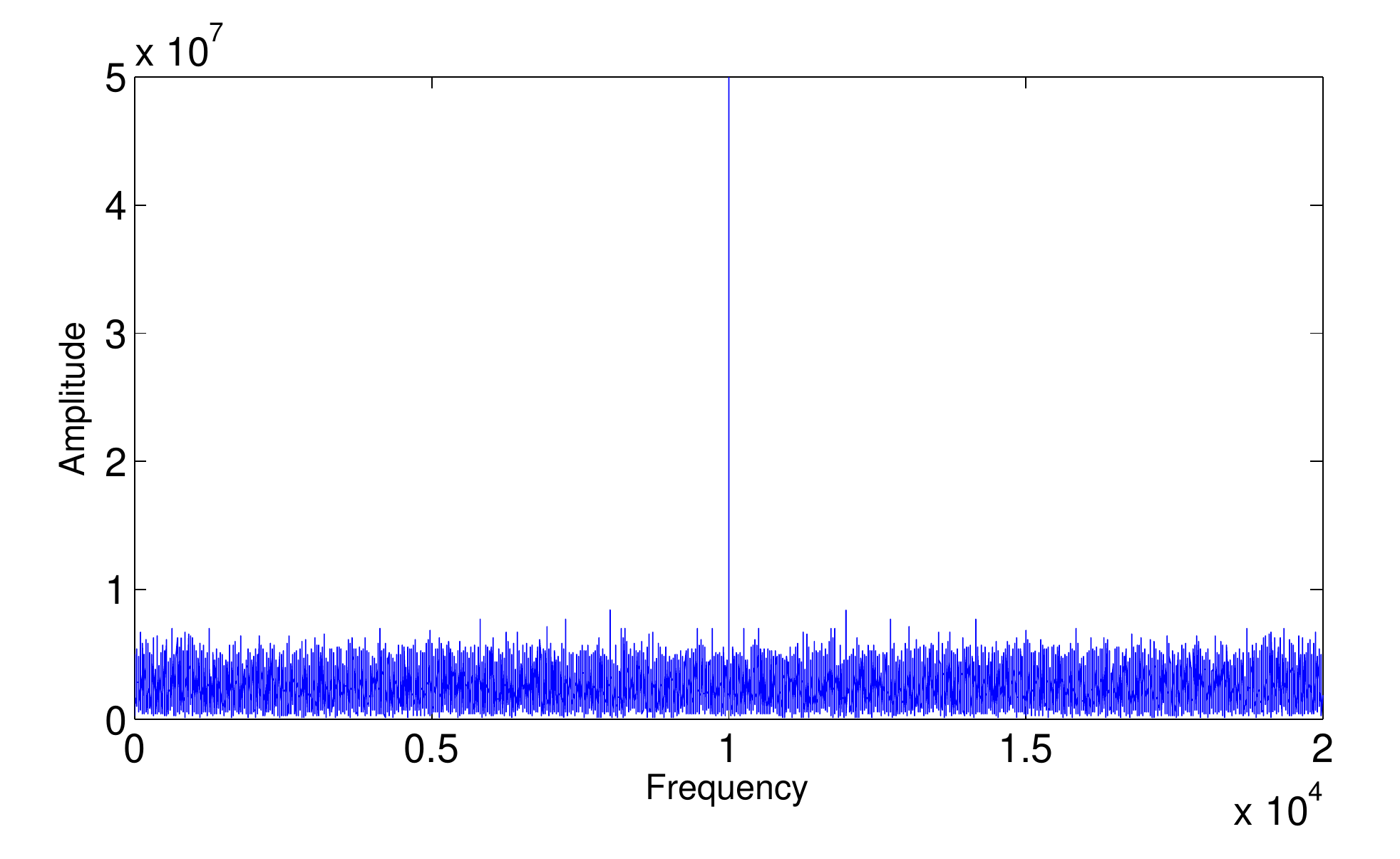}
 	\end{center}
 	\caption{\label{f:fft1}Frequency spectrum of the signal formed by the hardware random number generator}
 \end{figure}

\begin{figure}[htbp]
 	\begin{center}
 		\includegraphics[width=9cm]{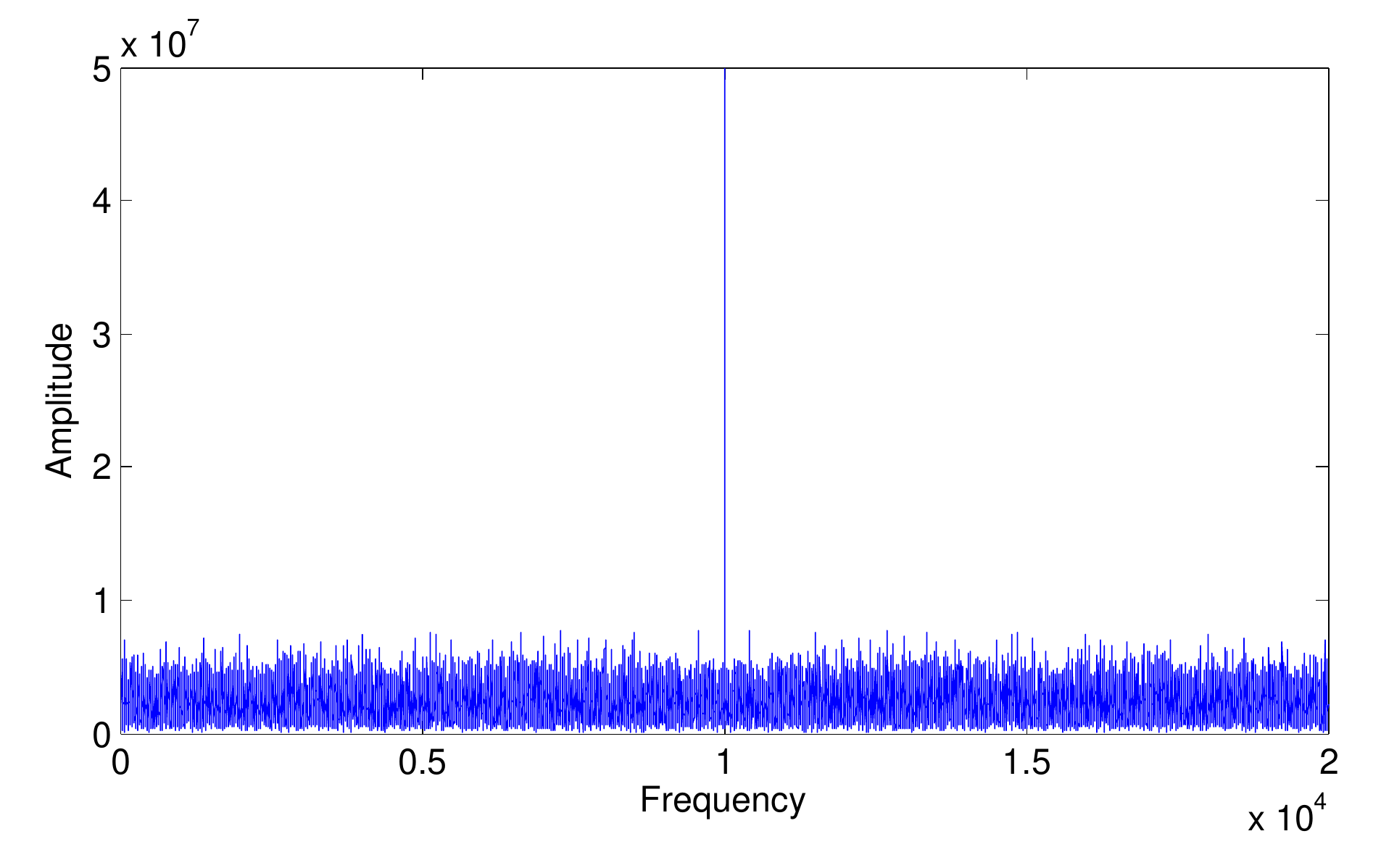}
 	\end{center}
 	\caption{\label{f:fftmatlab}Frequency spectrum of the signal formed by Matlab random number generator}
 \end{figure}

 The random number sequence we are trying to evaluate is a discrete signal. For such discrete signals,
 Fast Fourier Transform (FFT) \cite{DSP} can be used. Therefore, using Matlab, the frequency spectrum for the 20000 random numbers collected before was obtained. The obtained frequency spectrum is shown in \autoref{f:fft1}. The x is the frequency in terms of bins. The y axis is the amplitude of the frequency components. It is clearly visible that the 
 spectrum contains all the frequency components. Except the DC value (0 frequency) rest have similar amplitudes as well. The higher amplitude for the DC frequency component is because all the random generated values are positive values
 and hence the signal is a DC shifted signal. As all frequency components are there in the signal, we can infer that the signal is not periodic. In order to verify, a similar test was done for the random generator in Matlab and the obtained frequency spectrum is shown in \autoref{f:fftmatlab}. Frequency spectrums are almost similar, establishing the fact that our random number generator is good enough.
   
   \subsection{Setup 2}
   
\begin{figure}[htbp]
 	\begin{center}
 		\includegraphics[width=12cm]{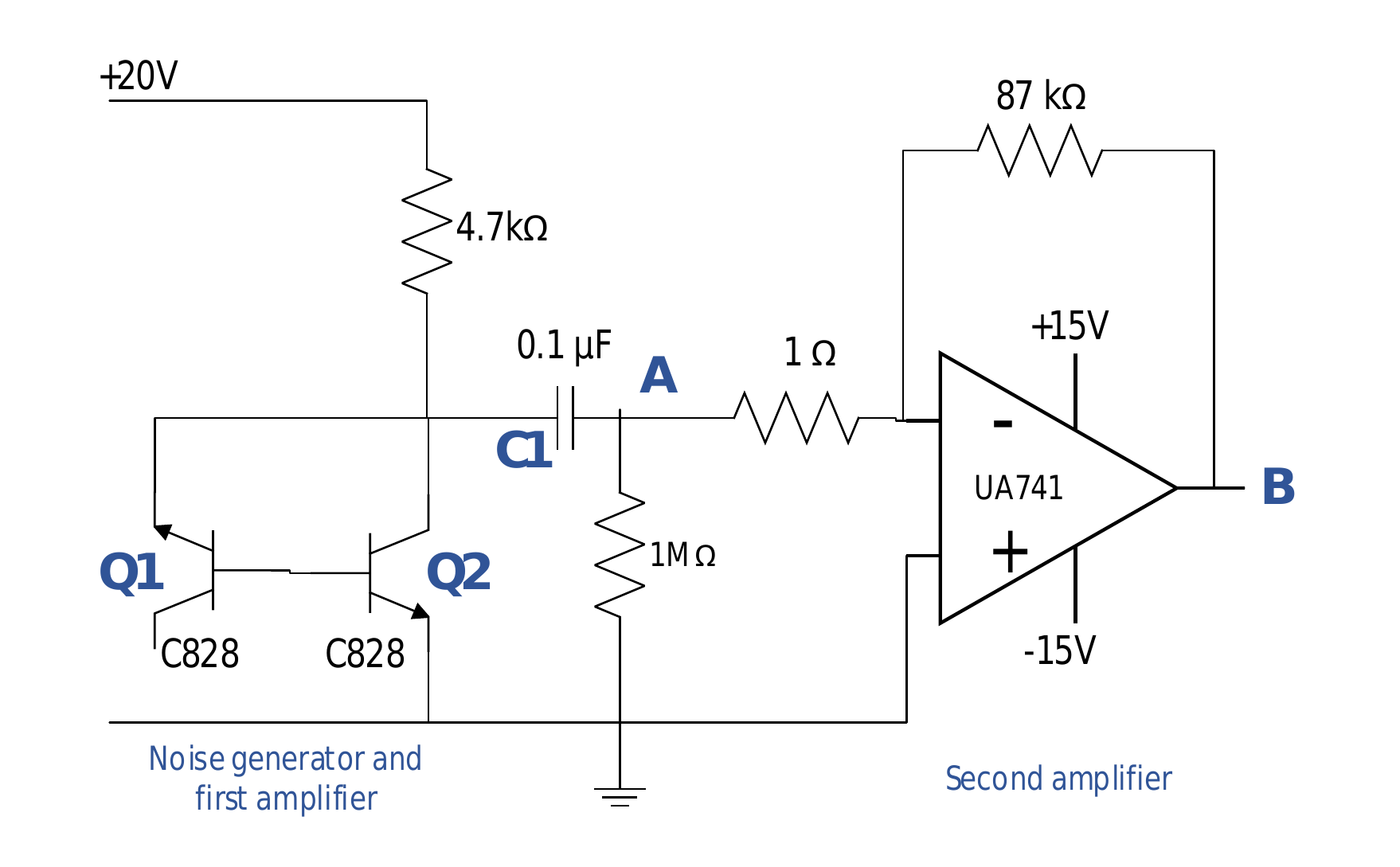}
 	\end{center}
 	\caption{\label{f:rand2}Hardware random number generator that generates an analog waveform}
 \end{figure}

  The randomness for the generated seed by the setup in \autoref{s:setup1} is good enough, but collecting and doing xor operations for 1000 values, where each value requires multiple number of samples, takes some time (about 1 second).
  Since generating the seed is not that frequent 1 second delay is not a much issue, but the process can be made much faster by doing modifications to the random number generator setup. The circuit diagram for the new setup after modifications is given in \autoref{f:rand2}
   
\begin{figure}[htbp]
 	\begin{center}
 		\includegraphics[width=9cm]{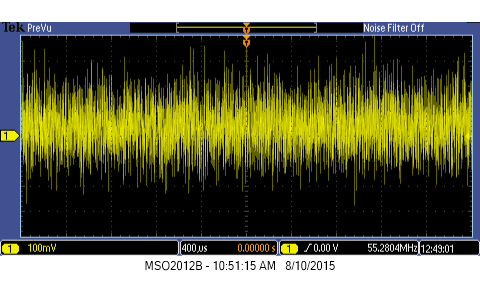}
 	\end{center}
 	\caption{\label{f:rand2A}Amplified reversed biased noise}
 \end{figure}

\begin{figure}[htbp]
 	\begin{center}
 		\includegraphics[width=9cm]{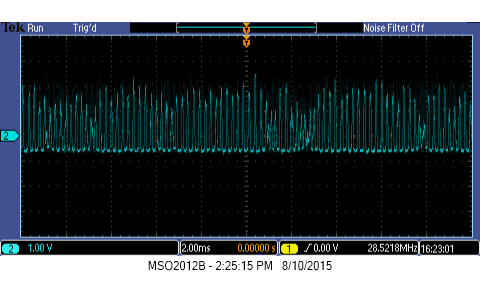}
 	\end{center}
 	\caption{\label{f:rand2B}Further amplified reversed biased noise}
 \end{figure}

 The noise generator and the first amplifier is exactly same as in the previous setup in \autoref{f:rand2}. The difference is that the second amplifier is now an operational amplifier instead of a transistor. An operation amplifier has a better gain than a transistor. Then the digitizer found in the earlier setup is not present here.  The amplified noise signal from the first amplifier is passed through the AC coupling capacitor C1  to remove the DC offset. The signal at place marked as A in \autoref{f:rand2}, observed via the oscilloscope is shown in \autoref{f:rand2A}. This signal is fed to the operational amplifier configured as an inverting amplifier with a large gain of 87000 times. The amplified signal by the operation amplifier at place marked as B is as shown in \autoref{f:rand2B}. This signal is expected to be an amplified version of the signal in \autoref{f:rand2A} but the deformation is due to the bandwidth limitation of the operational amplifier where higher frequency components have been attenuated. Anyway the signal is a random analog signal. 
 The output of the setup (place marked b) is connected to an analog input pin of the microcontroller of the cryptosystem.
 This is why the negative values of the signals have been clipped to zero in \autoref{f:rand2B}. This flattening causes a zero bias in the random signal which has to be unbiased in software.
 
 When a random seed is requested by the program, the voltage value on the analog input pin is converted via the analog to digital converter (ADC) of the microcontroller. The inbuilt ADC of general microcontrollers are limited to about 10 bit resolution. If the seed is required to be a number with large number of bits, several 10 bit numbers can be taken and then can be shifted and subjected to or operation to form large values. To remove the zero bias mentioned earlier,
 several such generated values can be subjected to xor. 
 
\begin{figure}[htbp]
  	\begin{center}
  		\includegraphics[width=14cm]{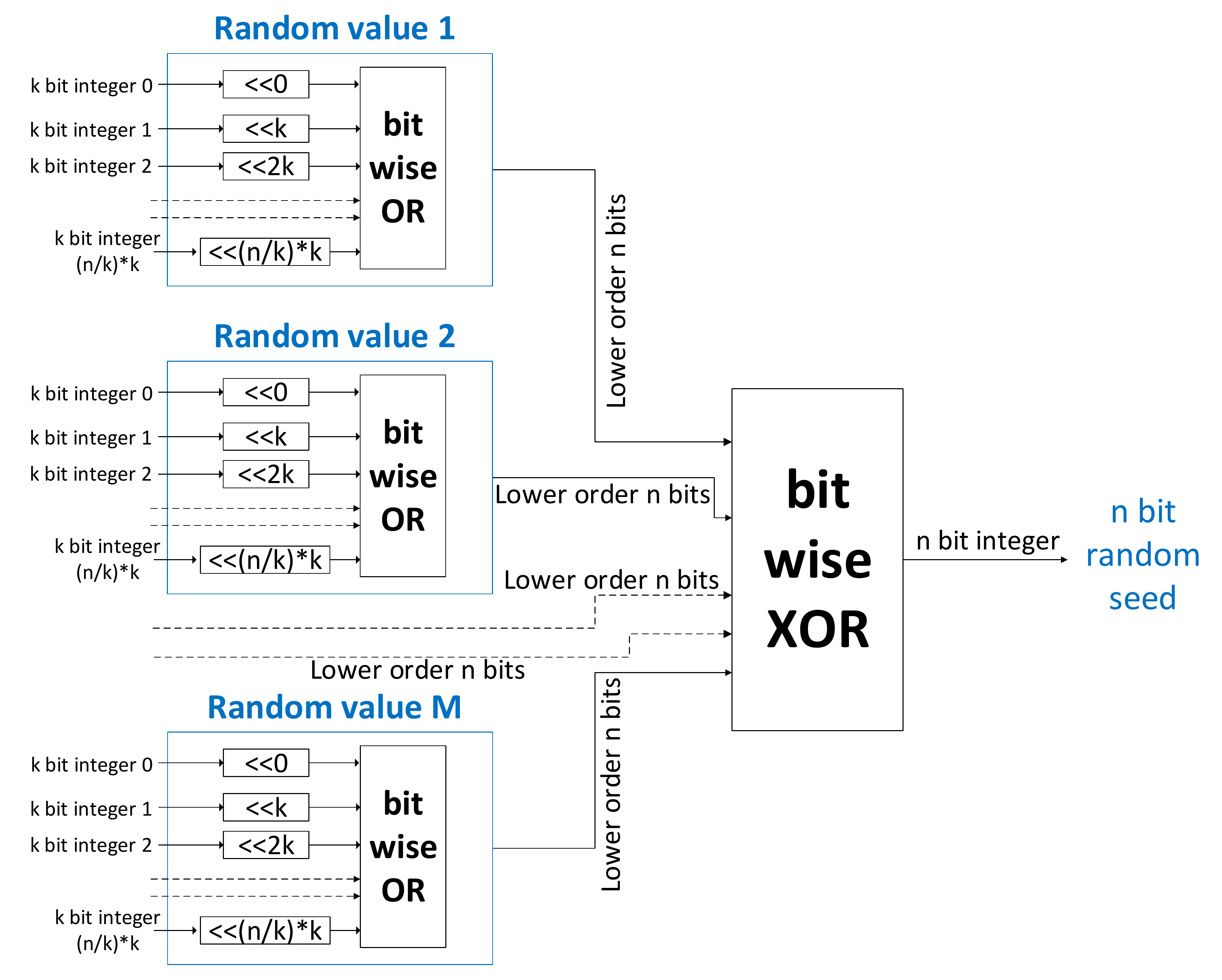}
  	\end{center}
  	\caption{\label{f:analohnbit2}Process of generating a random seed via the analog to digital converter by sampling the amplified analog noise signal}
  \end{figure}

 The approach above is graphically elaborated in \autoref{f:analohnbit2}. Objective is to generate an n bit random seed. The signal on the ADC is sampled to obtain a k bit integer 0. In PIC2550 or similar device ADC is 10 bit and hence k is 10. If n is larger than k, then more values are required. Hence (n/k)*k+1 number of such samples are taken and they are shifted appropriately and subjected to bitwise or as shown in \autoref{f:analohnbit2}. Here note that the division symbol refers to integer division. The output would be n bits if k is a multiple of k, but otherwise the output would be more than n bits. In that case the lower order n bits are taken. To remove the 0 bias, the process is repeated M times to obtain M random values. Those M values are xored to obtained the n bit random seed as shown in \autoref{f:analohnbit2}.
 
 The zero bias of the waveform in \autoref{f:rand2B} is much low when compared with the waveform in the earlier setup (\autoref{f:rand1C}). Therefore now collecting and xoring 10 values which had to be 1000 in the earlier case was enough.
 This makes the setup such efficient than before. 
 
\begin{figure}[htbp]
 	\begin{center}
 		\includegraphics[width=9cm]{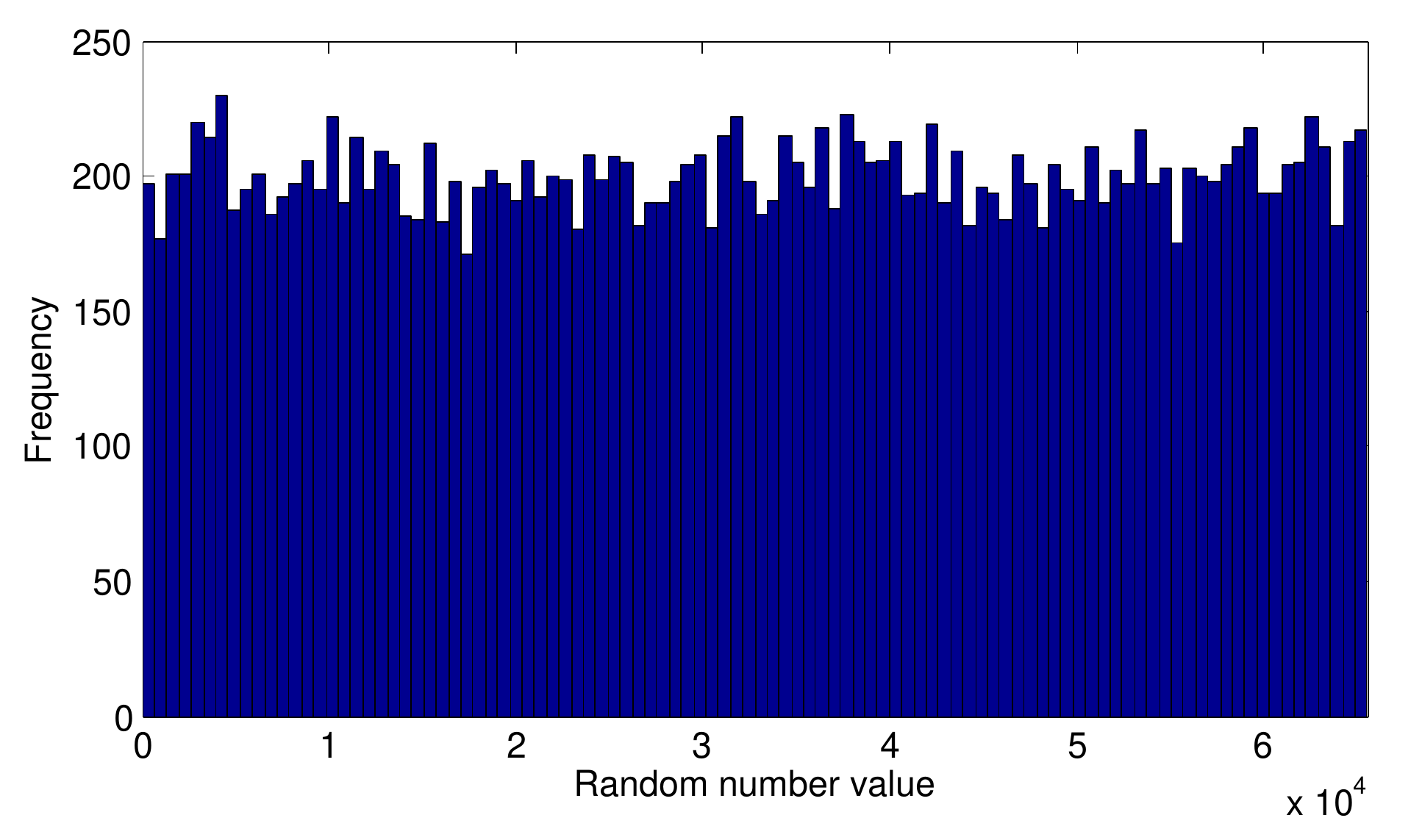}
 	\end{center}
 	\caption{\label{f:goodhistoran2}Histogram for random numbers generated by xoring 10 values}
 \end{figure}

\begin{figure}[htbp]
 	\begin{center}
 		\includegraphics[width=9cm]{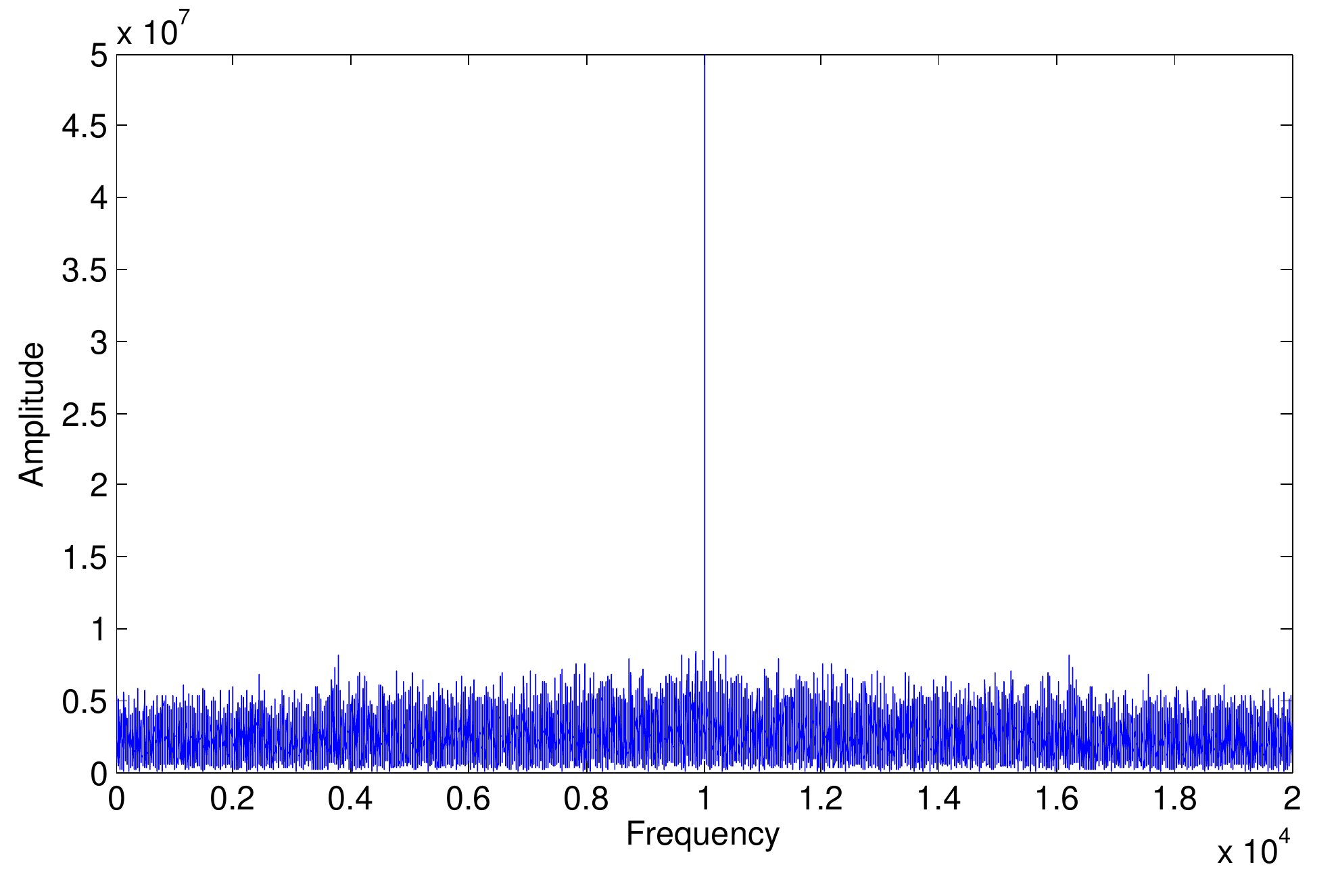}
 	\end{center}
 	\caption{\label{f:fft2}Frequency spectrum of signal formed by generated random numbers}
 \end{figure}

  The randomness was verified by collecting a large number of random numbers and then drawing a histogram and a frequency spectrum as it was done for the previous case. Obtained histogram is given in \autoref{f:goodhistoran2} and the frequency spectrum is given in \autoref{f:fft2}. Based on the arguments elaborated in \autoref{s:setup1} it can be inferred that the random number generator is good enough.
 
  \chapter{\label{c:conclusion}Conclusion}
  
  Power analysis attacks are a type of side channel attacks that has become a popular method of attacking 
  an encryption algorithm due to the fact that time taken is significantly less than what is needed for 
  other types of attacks. Power analysis involves measuring the power consumption of a cryptographic device
  during its operation and then analysing it using statistical techniques such as Correlation Power Analysis in order to derive the key.
  
  \section[Testbed]{Phase 1 : Bulding the testbed}
  
  In this project as the first phase we built a testbed for power analysis attacks. The testbed involved a PIC microcontroller
  based cryptosystem that could communicate with a PC via USB, circuit for measuring power during encryption, software needed
  to automate the power capturing process and the analysis software based on the statistical approach, Correlation Power Analysis . This was the most time consuming phase specially because debugging problems with respect
  to USB and power measurement needed lot of time and research. Currently according to best of our knowledge there is no step by step guide for building  
  such a testbed and there is no commercially available testbed either. Therefore we believe that having an already built testbed
  would save the time and effort of anyone who is interested in doing power analysis based research so that they can directly
  start on topics such as countermeasures rather than trying to build a testbed from the beginning. Also by
  following the steps described in this report one can easily replicate the testbed we created.
  
  For testing the testbed AES was used as the encryption algorithm where it helped greatly to identify and fix 
  the bugs and problems in our testbed. Here AES was selected as enough material were available about power analysis done
  on AES. Finally after doing few optimizations to handle noise, using the created testbed 128 bit AES could be broken 
  even in 10 minutes time.
  
  \section[Speck]{Phase 2 : Attacking Speck encryption algorithm}
  
  As the second phase, the latest encryption algorithm called Speck introduced by NSA in 2013 was tested against power analysis.
  Since an implementation of Speck for PIC microcontrollers was not available it had to be implemented our selves.
  Algorithm which originally required 64 bit unsigned integers to perform the encryption, was 
  bit challenging to implement on an 8 bit microcontroller which did not support such large integers,
  but the target was achieved successfully.
  
  As the Speck algorithm had lot of differences with AES, the power analysis method for AES could not be directly used here
  and hence new approached had to be found. As currently no one had attempted a power analysis attack on 
  Speck those approaches had to be newly formed. Though the first attempts of the attacks on Speck using power analysis failed, 
  with the help of observations found by attacking AES algorithm, the mechanisms and methods were identified to realize the attack. 
  Finally using the mechanisms and methods we identified, on our testbed Speck could be broken in less than 1 hour.
  
  With those findings we conclude that though Speck is a very recent algorithm introduced specially for
  microcontroller based embedded systems where embedded systems are anyway the target of power analysis attacks,
  still it is vulnerable
  to power analysis attacks. Therefore just using Speck in embedded systems without any countermeasures against power analysis
  thinking that
  a recent algorithm would be less vulnerable to such attacks is dangerous.
  
  \section{Phase 3 : Countermeasures}
  
  The third phase of the project was working on countermeasures. Countermeasures are the techniques implemented to safeguard cryptosystems from power analysis attacks. Countermeasures fall into two categories namely hardware countermeasures and software countermeasures. Hardware countermeasures can be further categorized into two more categories as microarchitecture level countermeasures and circuit level counters. Due to limitations in the laboratory we only worked on circuit level countermeasures and software countermeasures.
  
  Implementing power line filters is a proposed circuit based countermeasure which had not been practically tested so far according to best of our knowledge. After implementing different electronic passive filters on our testbed it was found out that, even a power line filtered by a second order LC filter can be broken in less than 5 hours. Then some of our own ideas were tested. From various methods and components we tested such as voltage regulator, zener diode, operational amplifier, constant current source and dual chip, the best one yet could be broken in less than 4 hours. With that we inferred that circuit level countermeasures are not that worth. Further once it was attempted by us to do a power analysis attack on an Arduino Uno board where it failed even at 100,000 power traces. After a thorough investigation it was found out that the FTDI chip was the reason behind the power trace corruption. But unfortunately that approach when tested with our PIC microcontroller based testbed did not succeed as a countermeasure due to some unknown reason to us at the moment.
  
  Two existing software countermeasures were tested by us. The method called random instruction injection which destroys the alignment of power traces had been only tested on a simulator so far. We tested it practically on a testbed and to observe that it is much better than any of the circuit based countermeasures we tried. The number of required power traces as well as the time needed for the attack quadratically increased with the number of random instructions injected. With that it could be predicted that when the number of random instructions injected is about 100, time required would be more than 20 days. Another existing method called randomly shuffling Sboxes was tested on our testbed. The security provided by this method is limited by the number of Sboxes in the encryption algorithm. But not only Sboxes but any other independent operations can be shuffled. Further, this shuffling method can be used together with random instruction injection to elevate the time taken for an attack. When compared to circuit countermeasures, software countermeasures therefore gave promising results. Software countermeasures also have the advantage that they can be implemented on any existing microcontroller without any additional hardware.
  
  A problem that we identified in both random instruction injection and random shuffling was regarding the randomness. Usually a pseudo random algorithm would be used for generating the randomness but in microcontrollers usually there is nothing like a real time clock to generate a suitable seed for the algorithm. Hence we implemented a method that uses an amplified noise signal formed across a reverse biased transistor to generate a true random seed. Two setups have been introduced by us, where one set up requires a digital pin of the microcontroller to sample a digitized noise signal, while the other method requires an analogue pin to sample an analogue noise signal. Each has its own advantages and disadvantages with respect to the components required and the time needed to generate a seed. The randomness of the seeds generated by both methods were verified by collecting several thousands of samples and then plotting frequency distributions and Fourier spectrum.
 
  \cleardoublepage
  \bibliographystyle{IEEEtran}
  \bibliography{extra}

\end{document}